\documentclass[review]{elsarticle}
\usepackage{graphicx}
\usepackage{subfig, float}
\usepackage{lineno,hyperref,url}
\usepackage{amsmath, amssymb,amsfonts, mathtools, mathrsfs, bbm} 
\newtheorem{theorem}{Theorem} 
\journal{Journal of \LaTeX\ Templates}

\begin{document}
\tableofcontents
\begin{frontmatter}

\title{Decoding the double trouble: A mathematical modelling of co-infection dynamics of SARS-CoV-2 and influenza-like illness} %\ template\tnoteref{mytitlenote}

%\author{Suman Bhowmick\fnref{myfootnote}}
%\fntext[myfootnote]{Corresponding author}

\author[sumanaddress]{Suman Bhowmick\corref{mycorrespondingauthor}}
\author[sumanaddress,2]{Igor M. Sokolov}
\author[3]{Hartmut H. K. Lentz}
\cortext[mycorrespondingauthor]{Corresponding author}
\address[sumanaddress]{Institute for Physics, Humboldt-University of Berlin, Newtonstra{\ss}e 15, 12489 Berlin, Germany}
\address[3]{Friedrich-Loeffler-Institut, Federal Research Institute for Animal Health,  Institute of Epidemiology, S\"udufer 10, 17493 Greifswald, Germany}
\address[2]{IRIS Adlershof, Zum Gro{\ss}en Windkanal 6, 12489 Berlin, Germany}

\begin{abstract}
After the detection of coronavirus disease 2019 (Covid-19), caused by the severe acute respiratory syndrome coronavirus 2 (SARS-CoV-2) in Wuhan, Hubei Province, China in late December, the cases of  Covid-19 have spiralled out around the globe. 
Due to the clinical similarity of Covid-19 with other flulike syndromes, patients are assayed for other pathogens of influenza like illness.  
There have been reported cases of co-infection amongst patients with Covid-19. 
Bacteria for example \textit{Streptococcus pneumoniae, Staphylococcus aureus, Klebsiella pneumoniae, Mycoplasma pneumoniae, Chlamydia pneumonia, Legionella pneumophila} etc and viruses such as influenza, coronavirus, rhinovirus/enterovirus, parainfluenza, metapneumovirus, influenza B virus etc are identified as co-pathogens.
\par
In our current effort, we develop and analysed a compartmental based Ordinary Differential Equation (ODE) type mathematical model to understand the co-infection dynamics of Covid-19 and other influenza type illness.
In this work we have incorporated the saturated treatment rate to take account of the impact of limited treatment resources to control the possible Covid-19 cases.
\par
As results, we formulate the basic reproduction number of the model system.
Finally, we have performed numerical simulations of the co-infection model to examine the solutions in different zones of parameter space. 
\end{abstract}

\begin{keyword}
Covid-19\sep Co-infection\sep ODE \sep Sensitivity Analysis \sep Invasion Reproductive Number
\end{keyword}

\end{frontmatter}

%\linenumbers

\section{Introduction}

Coronavirus belongs to a group of enveloped virus with a single-stranded RNA and viral particles bear a resemblance to  a crown from which the name originates.
It belongs to the order of Nidovirales, family of Coronaviridae, and subfamily of Orthocoronavirinae \cite{doi:10.1164/rccm.2014P7}. 
It can infect the mammals, including humans, giving rise to mild infectious disorders, sporadically causing to severe outbreaks clusters, such as those brought about by the “Severe Acute Respiratory Syndrome” (SARS) virus in 2003 in mainland China \cite{v12020135}.\par

After the first reported case of  new coronavirus disease outbreak in Wuhan, Hubei province, People’s Republic of China, the virus has progressively disseminated to different countries in the world \cite{LAI2020505}.
WHO declared it as a global pandemic on 11 March, 2020 \cite{Cucinotta2020}.
This disease can spread from person-to-person through the breathing in of respiratory droplets from an infected person or having the direct contact with contaminated surfaces \cite{10.1001jama.2020.2565}.\par

The ending season and the culminating severity of the current Covid-19 pandemic wave are still unknown and unsettled issue.
Meanwhile, the influenza season has collided with the current pandemic that could pave the way for more challenges.
This possesses a larger threat to public health domain.
It is still uncertain how the seasonal influenza-like illness (ILI) will have an impact on the long-term effects on the course of Covid-19 pandemic.
Both viruses share similarities in transmission characteristics and alike clinical symptoms. 
ILI and Covid-19 have been reported to cause respiratory infection. 
The interplay between ILI and Covid-19 have been a major concern \cite{doi:10.1177/2324709620934674}.
An emerging study from England has shown that fatality amongst the people infected with both ILI and Covid-19 are twice as that of someone infected with the new coronavirus only \cite{Iacobuccim3720}.
An investigation conducted by the Public Health England (PHE) has shown that people infected with the two viruses were having higher risk of severe illness during the period from January to April 2020.
According to the same analysis, most cases of co-infection occurred in older people, and the mortality rate was high.
Reports from the USA point out that co-infection between Covid-19 and other respiratory pathogens are notably more common compared to the initial data what suggest such an interplay is rare found in China \cite{10.1001/jama.2020.6266, chenz}.
The authors in \cite{Hazra} report the co-infections between Covid-19 and other respiratory pathogens at a large urban medical centre in Chicago, Illinois.
A study \cite{Burrel} reveals that $7\%$ of SARS-CoV-2-positive patients share the burden of co-infection with other respiratory viruses.
According to that study the detection of other respiratory viruses in patients during this pandemic assumes the Covid-19 co-infection.
The authors in \cite{Bandar} note that high prevalence of influenza-Covid-19 co-infection in their study. 
In \cite{Balraj}, the authors describe the cases of influenza and Covid-19 co-infection.
During the last winter and autumn, the co-circulation of influenza-Covid-19 has taken a toll on the health of the patients and taxing the intensive care capacity \cite{10.1093/cid/ciaa1810, Belongia1163}.
The authors in the study \cite{Bai} mention that co-infection with influenza A virus enhances the infectivity of Covid-19.
This is undoubtedly, a significant threat that co-infection of ILI and Covid-19 possess.\par

Proper and appropriate nursing and treatment processes can significantly reduce the outcome of epidemics in society.
In traditional epidemiological modelling assumption, the treatment rate is hypothesised to be constant or correspond to the number of infected people and the recovery rate reckons on the available medical resources such as test kits, ventilators, nursing facilities, efficiency of treatment etc. 
In the course of the ongoing pandemic, we have noticed how this pandemic has stretched the healthcare systems of different countries around the globe while registering high mortality \cite{10.1001/jamahealthforum.2020.0345, sam, Gai}. 
In the classical epidemiological models, it is very common to use the treatment function as $T(I) = \xi I, I \geq 0$, where 
$\xi$ is positive but in the situation of sudden epidemic or pandemic when the infected population is very large then it is not always possible to provide such a type of treatment which is proportional to the infected number of individuals and $I$ is the number of infected people. 
To circumvent the crux, the authors in \cite{WANG2004775} have introduced a constant treatment rate  of the form and demonstrated different bifurcations. 
But to sustain such a constant treatment rate might be plausible when there are a small number of infected individuals as the medical resources are limited \cite{Griffinl72}. 
Following this, the authors in \cite{ZHOU2012312, Suo, ZHANG2008433} have modified the treatment rate by taking account of  Holling type II functional response as given below: $T(I) = \frac{\eta I}{1+\zeta I}$ and have explored the model dynamics to understand the importance of limited medical resources and facilities in the dissemination of infectious diseases. 
It is important to note that this function is clearly an increasing function of $I$ and is bounded above by the least upper bound $\eta/ \zeta$.
This functional form of saturated treatment can provide a better rationale for different disease outbreaks such as SARS, Dengue etc\cite{ZHOU2012312} in a new region because we know from our ongoing Covid-19 pandemic experience that in the beginning of an outbreak there is a lack of effective treatment due to either negligence or lack of knowledge about the disease.
But afterwards the treatment is being increased with the gain of knowledge about the disease as well scaling up medical facilities \cite{Pepita}.
Eventually, the treatment rate is outstretched to its maximum given the boundedness of medical resources of any country \cite{Bene}.
Here, $\eta$ represents cure rate and $\zeta$ denotes the extent of the effect of the infected individuals being delayed for treatment \cite{Suo}. \par
Epidemiological models are quite beneficial to examine the co-infection dynamics and to estimate the treatment facilities. 
There are several studies concentrating on the coexistence of two infectious agents in the susceptible hosts \cite{BLYUSS2005177, MALLELA2016143, ASADUZZAMAN20151, Til}. 
In \cite{GAO2016171}, the authors have proved a sufficient condition for coexistence of  two infectious diseases.
In the same vein, the authors in \cite{Tang}have proposed a new mathematical model Zika-dengue model while describing coupled dynamics of Zika-dengue. 
The authors in \cite{MERLER2008499} have asserted that pandemic outbreaks can possibly be controlled by co-infection with other acute respiratory infections that enhances the transmissibility of influenza virus.
These suggest that co-infection can potentially change the course of current ongoing pandemic and induce multiple waves \cite{Cacci, Fisayo332}.\par

The basic reproduction number ($R_0$) is a critical metric in epidemiology and it is used to determine whether a disease will persist in a population or not\cite{doi:10.1098/rsif.2009.0386}. 
Although it is a very useful threshold, it comes with some restrictions too.
Many diseases are endemic to certain regions around the globe and henceforth, giving rise to the cases of co-infection in the communities. 
Therefore, it becomes necessary to  study invasion reproductive numbers (IRNs) \cite{OLAWOYIN201944, NUNO200720, 1531-3492_2009_2_279}. 
It is defined as the number of secondary infections produced by an infected individual in a population where one or more other pathogens are endemic. 
This critical threshold has similar behaviour as of $R_0$ i.e. if the IRN of an infectious agent is larger than $1$, then the pathogen can transmit in a population infected with other diseases.\par

Since the first pneumonic case of Covid-19, reported in Wuhan City, Hubei Province, China, the world is witnessing the ravaging impact of this global pandemic.
It is acknowledged that Indian Government has implemented timely control measures to mitigate the spread of this pandemic. 
However, in the current situation, medical resources are in a critical need and especially for an ethnically diverse country like India.
This has in fact left an influence in the pattern of Covid-19 dissemination.
The interplay of Covid-19 and other ILI, is of utmost importance as the cases of co-infection induces a higher cases of mortality.
From the perspective of public health control, it is significant to quantify and analyse the burden of co-infection on the medical resources and consequently help to curb the cases of Covid-19.
To address this major issue, we have proposed a deterministic dynamic model featuring the co-infection in a saturated medical facility. 
The primary goal of this current endeavour is to investigate the impact of co-infection in the dynamics of Covid-19. 
Additionally, we derive the analytical expression of IRN to understand the influence of ILI on the potential spread of Covid-19.
We also perform the sensitivity analysis to figure out the essential parameters while taking account of limited medical resources in the modelling framework.

%%%%%%%%%%%%%%%%%%%%%%%
\section{Model formulation and description}

\begin{figure}[H]
\centering
\includegraphics[width=0.8\textwidth]{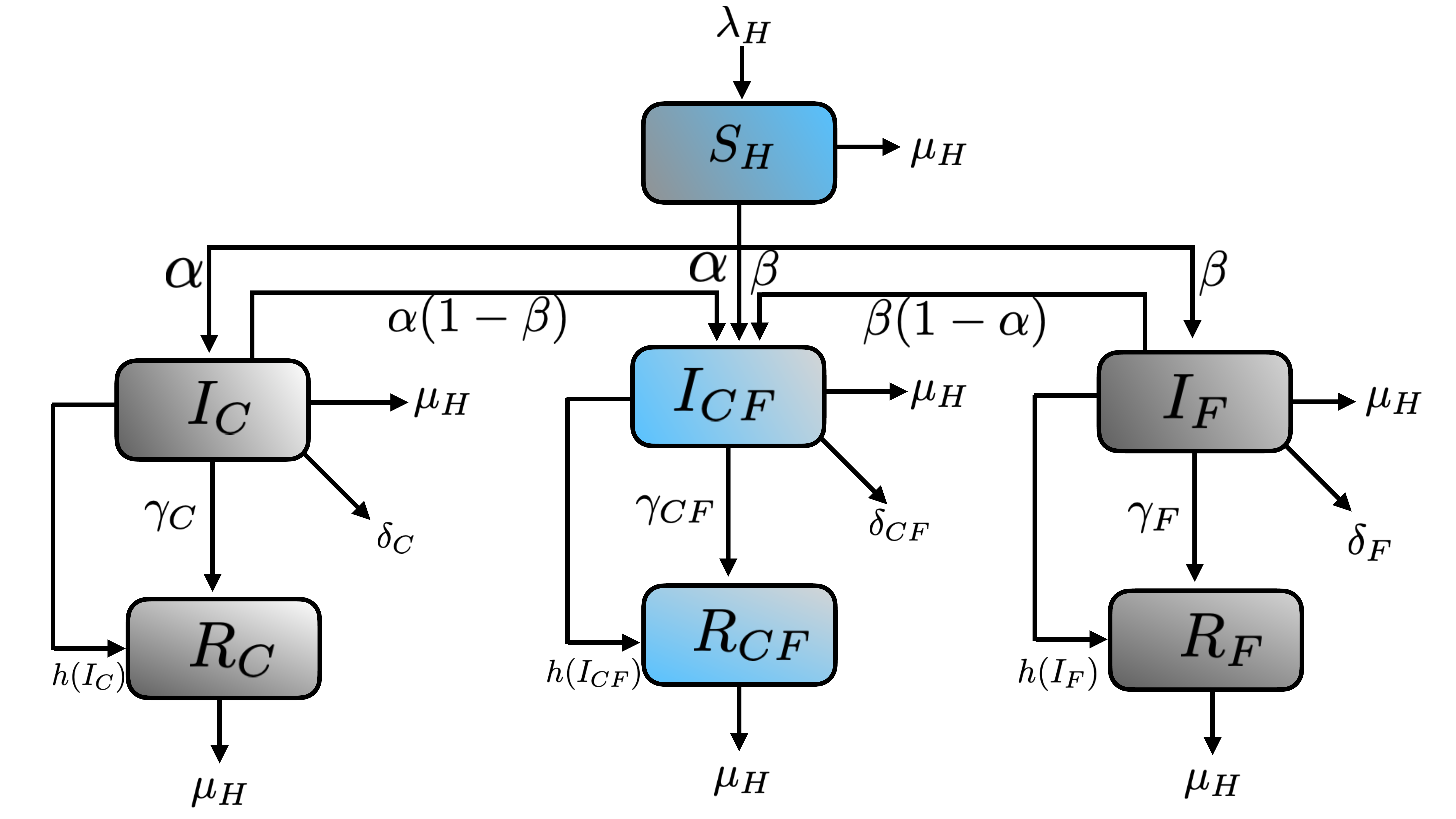}   
\caption{Flowchart of a SARS-CoV-2 and influenza-like illness  co-infection model. $S_H$, $I_C$, $I_F$ and $I_{CF}$ represent the fractions of the susceptible population, infected with Covid-19, 
Influenza-like illness and infected with both infectious agents, respectively.
Details about the model parameters are mentioned in the Table:\ref{Table1}.
}
\label{Fig:1}
\end{figure}

The host population is $N$ and we divide the population into the following classes according to their health status:
susceptible ($S_H$), infected with Covid-19 only ($I_C$), infected with ILI only ($I_F$), co-infected (infected with Covid-19 and ILI simultaneously) ($R_{CF}$), recovered from Covid-19 ($R_C$), recovered from ILI ($I_F$) and recovered from co-infection ($R_{CF}$).
In our model $\lambda_H$ represents the constant recruitment rate, $\mu_H$ is the natural death rate, $\alpha$ and $\beta$ denote the effective transmission rates with which susceptible individuals become infected with Covid-19 and ILI, respectively.
We also take account of the transmission rates to be infected only with Covid-19, ILI or both diseases after having the contact with a co-infected individuals as:  $\alpha (1-\beta)$, $(1-\alpha)\beta$  and  $\alpha \beta$.
We also include the disease induced death rates into our model. 
$\delta_{C}$, $\delta_{F}$ and $\delta_{CF}$ represent the death rates for the Covid-19, ILI and co-infected individuals, respectively.
Natural recovery rates for the Covid-19, ILI and co-infected individuals are denoted as $\gamma_{C}$, $\gamma_{F}$ and 
$\gamma_{CF}$.
We further consider a saturated treatment function rate as $h(I_J) = \frac{\eta I_J}{1+\zeta I_J}$, where $J=C, F$ and $CF$ respectively.
We have not taken account of reinfection and henceforth it is of SIR type model.
We present the schematic diagram of the model \eqref{Co-infModelSys} in the Figure: \ref{Fig:1}.

%\section{Model Formulation}

The model equations are following:

\begin{eqnarray}\label{Co-infModelSys}
\frac{dS_H}{dt} &=& \lambda_H -\mu_H S_H- [\alpha(1-\beta)+\beta(1-\alpha)+\alpha\beta]S_HI_{CF} - \alpha S_HI_C -\beta S_H I_F \\ \nonumber 
\frac{dI_C}{dt} &=& \alpha S_HI_C+\alpha(1-\beta)S_HI_{CF} - (\mu_H+\delta_C +\gamma_C) I_C  -\beta(I_F+I_{CF})I_C  - h(I_C) \\ \nonumber 
\frac{dR_C}{dt} &=& \gamma_C I_C + h(I_C) - \mu_H R_C\\ \nonumber 
\frac{dI_F}{dt} &=& \beta S_HI_F+\beta(1-\alpha)S_HI_{CF} - (\mu_H +\delta_F + \gamma_F) I_F  -\alpha(I_C+I_{CF})I_F  - h(I_F) \\ \nonumber 
\frac{dR_F}{dt} &=& \gamma_F I_F + h(I_F) - \mu_H R_F\\ \nonumber 
\frac{dI_{CF}}{dt} &=& \alpha\beta S_HI_{CF}+(\alpha+\beta)I_F I_C+(\alpha I_F+\beta I_C )I_{CF} - (\delta_{CF}+\gamma_{CF}+\mu_H)I_{CF}-h(I_{CF}) \\ \nonumber 
\frac{dR_{CF}}{dt} &=& \gamma_{CF} I_{CF} + h(I_{CF}) - \mu_H R_{CF}\nonumber 
\end{eqnarray}

In this paper, we further explore the SEIR epidemic model with saturated treatment function as $h(I_J) = \frac{\eta I_J}{1+\zeta I_J}$, where $J=C, F$ and $CF$ respectively. 
%$\left[0.5,  1.5\right]$
\begin{table}[H]
  \centering
  \begin{tabular}{c c c  c} 
  Parameter &  Description  &  Value or range  &  References\\ 
 \hline
  $\pi_H$  & Recruitment rate. &  Varies over states   & \cite{SARDAR2020110078} \\ 
  $\mu_H$ & Natural mortality rate &  Varies over states   & \cite{SARDAR2020110078} \\ 
  $\gamma_C$ & Recovery rate of Covid-19 &  Varies over states   & \cite{SARDAR2020110078} \\
  $\delta_C$ & Average case fatality rate of Covid-19&  Varies over states   & \cite{SARDAR2020110078} \\
  $\alpha$ & Covid-19 transmission rate &  $\left[0.09,  0.89\right]$   & \cite{MAHAJAN2020110156} \\
  $\gamma_F$ & Recovery rate of ILI &  $\left[0.0714 0.1998\right]$   & \cite{Blyuss} \\
  $\delta_F$ & Average case fatality rate of ILI&  $\left[0.01, 0.021\right]$  & \cite{Blyuss} \\
  $\beta$ & ILI  transmission rate &  $\left[0.01,  0.78\right]$   & \cite{Kharis_2018} \\
  $\gamma_{CF}$ & Recovery rate of co-infection &  $\left[0.033,  0.1428\right]$   & Assumed \\
  $\delta_{CF}$ & Average case fatality rate of co-infection &  $\left[0.0189, 0.0332\right]$  & \cite{NADIM2020110163} \\
  $\eta$  & Cure rate & $\left[0.2, 0.95\right]$& \cite{Sezer}\\
  $\zeta$  & Extent of saturation & $\left[0.008, 0.25\right]$& \cite{Sezer}\\
  \hline
\end{tabular}
\caption{Parameters used in the model \eqref{Co-infModelSys}.}
\label{Table1}
\end{table}

\section{Qualitative Analysis}
\subsection{Infectious model with Covid-19 only}
In this subsection, analytical findings are made taking into consideration SARS Covid-19 only.
The model equations are following:
\begin{eqnarray}\label{2}
\frac{dS_H}{dt} &=& \lambda_H -\mu_H S_H - \alpha S_HI_C \\ \nonumber
\frac{dI_C}{dt} &=& \alpha S_HI_C - \mu_H I_C -\delta_C I_C -\gamma_C I_C - h(I_C) \\ \nonumber
\frac{dR_C}{dt} &=& \gamma_C I_C + h(I_C) - \mu_H R_C\\ \nonumber
\end{eqnarray}

\subsubsection{Invariant Region}
\par Total population is $N_C=S_H+I_C+R_C$. Differentiating w.r.t time t and substituting the required expressions, we get:

\begin{equation}\label{3}
\frac{dN_C}{dt}\leq \lambda_{H}-\mu_{H}N_C.
\end{equation}

\par After solving the equation \eqref{3} provides the invariant region as,\\

$ \Psi = {(S_H,I_C,R_C)\in \mathbb{R}^3:0\leq N_C \leq\frac{\lambda_{H}}{\mu_{H}}}$\\

Hence, the solution set is bounded in $\Psi$.

\subsubsection{Positivity of solution}

Utilising the first equation of system \eqref{2} and after rewriting it as,\\

$\frac{dS_{H}}{dt} \geq S_{H}(-\mu_{H}-\alpha I_{C})$.\\

On evaluating, it yields,\\

$S_H \geq S_{H}(0) \exp^{-\int(\mu_H+\alpha I_C)dt}.$\\

Similarly, the result follows for other equations proving the positivity of solutions.

\subsubsection{Disease free equilibrium (DFE) and basic reproduction number $R_{0_{C}}$}

On substituting $I_C=0$, system \ref{2} possesses the disease free equilibrium point $(\frac{\lambda_H}{\mu_H},0,0)$. 
Thereby, $R_C$ is evaluated using the next generation matrix method approach as,\\

$R_{0_{C}}=\frac{\alpha \lambda_H}{\mu_H (\mu_H+\delta_C+\gamma_C+\eta)}$.

\subsubsection{Local Stability of DFE}

\begin{theorem}
The DFE of system \ref{2} is locally asymptotically stable if $R_{0_{C}} < 1$ and unstable if $R_{0_{C}} > 1$.
\end{theorem}
Proof: The variational matrix governing the system at DFE is;\\
                                                                   $\mathrm {J_C} =
                                                                   \begin{bmatrix}
                                                                   -\mu_H & \frac{-\alpha \lambda_H}{\mu_H} & 0\\
                                                                   0 & \frac{\alpha \lambda_H}{\mu_H}-\mu_H-\delta_C-\gamma_C-\eta & 0 \\
                                                                   0 & \gamma_C+\eta & -\mu_H \\
                                                                   \end{bmatrix}$.\\

The eigenvalues are given by $\lambda _1=-\mu_H$, $\lambda _2= -\mu_H$ with $\lambda _3=\frac{\alpha \lambda_H}{\mu_H}-\mu_H-\delta_C-\gamma_C-\eta = R_{0_{C}}-1$ having real negative part if and only if $R_{0_{C}} < 1$ which proves the theorem.

\subsubsection{Global Stability of DFE}
\begin{theorem}
The DFE of system \ref{2} is globally asymptotically stable if $R_{0_{C}} < 1$.
\end{theorem}
Proof: Define Lyapunov function as;\\

$L=a.I_C$,\\

$\frac{dL}{dt}=a\alpha S_HI_C-a(\mu_H+\delta_C+\gamma_C+h(I_C))$\\

$\Rightarrow \frac{dL}{dt}= (\frac{\alpha S_H}{\mu_H+\delta_C+\gamma_C+\eta}-1)I_C$,\\

$\frac{dL}{dt} < 0 $ if $R_{0_{C}} \leq 1$.\\

Hence, the result follows.

\subsubsection{Endemic equilibrium analysis}
The endemic equilibrium is $E^*_{P}=(S^*_H,I^*_C,R^*_C)$ where,

$S^*_H = \frac{\lambda_H}{\mu_H+\alpha I^*_C}$,
$R^*_C = \frac{\eta I^*_C}{1+\zeta I^*_C}$
and $I^*_{C}$ is a positive root of the quadratic equation,
$I^{*^{2}}_C+b_1I^*_C+b_2 = 0$,
where $b_1 = \mu_H\zeta+\alpha+\frac{\eta \alpha-\lambda_H\zeta\alpha}{\mu_H+\delta_C+\gamma_C}$ and $b_2 = (1-R_{0_{C}})\frac{\mu_H(\mu_H+\delta_C+\gamma_C+\eta)}{\alpha(\mu_H+\delta_C+\gamma_C)}$.

\begin{theorem}
The endemic equilibrium is globally stable in region $\phi$ if,\\
%\begin{enumerate}
  I.  $\mu_H(\mu_H+\delta_C+\gamma_C-\frac{\eta}{1+\zeta I^*_C})> \frac{\alpha^2(S^*-I^*_C)^2}{4}$,\\
  II. $\mu_H[\mu_H(\mu_H+\delta_C+\gamma_C-\frac{\eta}{1+\zeta I^*_C})-\frac{1}{4}(\gamma_C+\eta-\frac{\eta}{1+\zeta I^*_C})^2]>\frac{\alpha (S_H^*-I^*_C)}{2}[\alpha \mu_{H}\frac{(S_H^*-I^*_C)}{2}]$
%\end{enumerate}
\end{theorem}
Proof: Define $V=\frac{1}{2}(S_H-S_H^*)^2+\frac{1}{2}(I_C-I^*_C)^2+\frac{1}{2}(R-R^*)^2.$\\
Differentiating with respect to time t, we get,\\
$\frac{dV}{dt}=(S_H-S_H^*)\frac{dS_H}{dt}+(I_C-I_C^*)\frac{dI_C}{dt}+(R-R^*)\frac{dR}{dt}$,\\

On substituting the required expressions and simplifying, we get,\\
$(S_H-S_H^*)^2[-\mu_H-\alpha I^*_C]+(I_C-I_C^*)^2[-\mu_{H}+\alpha S_{H}-\delta_C-\gamma_C-\frac{\eta}{1+\zeta I_C}+\frac{\eta}{1+\zeta I^*_C}]-\mu_H(R_C-R_C^*)^2+(I_C-I_C^*)(R_C-R_C^*)[\gamma_C+\frac{\eta}{1+\zeta I_{C}}-\frac{\eta}{1+\zeta I^*}]+(S_H-S_H^*)(I_C-I_C^*)[-\alpha(S_H^*-I_C^*)]$\\

$\frac{dV}{dt}$ is negative definite in the region $\phi$ subject to the fulfilment of conditions I and II.

\section{Co-infection model system}
In this section we find the basic reproduction number, $R_0$ of the entire model system as defined in \eqref{Co-infModelSys} after following the Next Generation Matrix (NGM) method as described in \cite{doi:10.1098/rsif.2009.0386}.
$R_0$ characterises the average number of new reported cases of an infection caused by one infected individual, in a completely susceptible population.
Entries of NGM relate to the numbers of newly infected individuals in the various categories in consecutive generations.\par
To apply the NGM approach, we notice that the variables associated with Covid-19, ILI and Co-infected stages are $(I_C, I_F, I_{CF})$ and $S_H$ as the susceptible compartments.
Denoting $Y_I = (I_C, I_F, I_{CF})$ and $Y_S = S_H$, we can rewrite the associated system as the difference between the new-infection terms (inflow) and outflow terms then we have, 
\begin{equation}\label{R0Eqn1}
\frac{dY_I}{dt} =  \mathrm{F_{CF}}(Y_S, Y_I)-\mathrm{V_{CF}}(Y_S, Y_I)
\end{equation}

$
\mathrm{F_{CF}} =
\begin{bmatrix}
                              \alpha S_H I_C+(\alpha(1-\beta)S_HI_{CF})-(\beta(I_F+I_{CF})I_C)\\
                              \beta S_HI_F+(\beta(1-\alpha)S_HI_{CF})-(\alpha(I_C+I_{CF})I_F)\\
                              \alpha \beta S_H I_{CF}+(I_{CF}(\alpha I_F+\beta I_C))+((\alpha+\beta)I_CI_F)

\end{bmatrix}
$
\\
$
\mathrm{V_{CF}} =
\begin{bmatrix}
                             \mu_H I_C+\delta_C I_C+\gamma_C I_C+\eta I_C/(1+\zeta I_C)\\
                             \mu_H I_F+\delta_F I_F+\gamma_F I_F+\eta I_F/(1+\zeta I_F)\\
                             \mu_H I_{CF}+\delta_{CF}I_{CF}+\gamma_{CF} I_{CF}+\eta I_{CF}/(1+\zeta I_{CF})
\end{bmatrix}
$\\
So, after computing the Jacobians of $\mathrm{F_{CF}}$ and $\mathrm{V_{CF}}$ at the disease free equilibrium (DFE) point, i.e. $\mathcal{F_{CF}}$ and $\mathcal{V_{CF}}$ respectively, we have 
the $\mathrm{NGM} =  \mathcal{F_{CF}}\mathcal{V_{CF}}^{-1}$.
The matrices for  $\mathcal{F_{CF}}$ and  $\mathcal{V_{CF}} $ are described below:\\
$
 \mathcal{F_{CF}} =
     \tiny{\begin{bmatrix}
    S_H\alpha -\beta \left(I_{CF}+I_F\right) & -I_C\beta  & -I_C\beta -S_H\alpha \left(\beta -1\right)\\ 
    -I_F\alpha  & S_H\beta -\alpha \left(I_C+I_{CF}\right) & -I_F\alpha -S_H\beta\left(\alpha -1\right)\\ 
    I_{CF}\beta +I_F\left(\alpha +\beta \right) & I_{CF}\alpha +I_C\left(\alpha +\beta \right) &I_F\alpha +I_C\beta +S_H\alpha\beta  
  \end{bmatrix}}
$
 and \\
 $
 \mathcal{V_{CF}} =
     \tiny{\begin{bmatrix}
    \delta_C+\gamma_C+\mu_H+\frac{\eta }{I_C\zeta +1}-\frac{I_C\eta \zeta }{{\left(I_C\zeta +1\right)}^2} & 0 & 0\\ 
    0 & \delta_F+\gamma_F+\mu_H+\frac{\eta }{I_F\zeta +1}-\frac{I_F\eta \zeta }{{\left(I_F\zeta +1\right)}^2} & 0\\ 
    0 & 0 & \delta_{CF}+\gamma_{CF}+\mu_H+\frac{\eta }{I_{CF}\zeta +1}-\frac{I_{CF}\eta \zeta }{{\left(I_{CF}\,\zeta +1\right)}^2}
    \end{bmatrix}}
$

$
\mathrm{NGM} =
   \begin{bmatrix}
    \frac{\alpha \lambda_H}{\mu_H\left(\delta_C+\eta +\gamma_C+\mu_H\right)} & 0 & -\frac{\alpha \lambda_H\left(\beta -1\right)}{\mu_H\left(\delta_{CF}+\eta +\gamma_{CF}+\mu_H\right)}\\ 
    0 & \frac{\beta \lambda_H}{\mu_H\left(\delta_F+\eta +\gamma_F+\mu_H\right)} & -\frac{\beta \lambda_H\left(\alpha -1\right)}{\mu_H\left(\delta_{CF}+\eta +\gamma_{CF}+\mu_H\right)}\\ 
    0 & 0 & \frac{\alpha\beta\lambda_H}{\mu_H \left(\delta_{CF}+\eta +\gamma_{CF}+\mu_H\right)} 
   \end{bmatrix}
    $
    
According to \cite{VANDENDRIESSCHE200229}, the basic reproduction number $R_0$ is defined be the spectral radius of NGM at DFE.
So,
\begin{equation}\label{BRNModel}
R_0 = \max{ \{ R_{0_{C}}, R_{0_{F}}, R_{0_{CF}} \} }, 
\end{equation}
where $R_{0_{C}}$, $R_{0_{F}}$ and $R_{0_{CF}}$ are defined to be the basic reproduction  numbers associated with Covid-19 only, Influenza like illness (ILI) and co-infected with Covid-19 and ILI.
The expressions for $R_{0_{F}}$ and $R_{0_{CF}}$ are derived as :

$R_{0_{F}}   = \frac{\beta \lambda_H}{\mu_H (\mu_H+\delta_F+\gamma_F+\eta)}$ and $R_{0_{CF}}  = \frac{\alpha\beta \lambda_H}{\mu_H (\mu_H+\delta_{CF}+\gamma_{CF}+\eta)}$.

\section{Stability Analysis}
  $
  \mathrm{J_{CF}} =
     \tiny{\begin{bmatrix}
    
    -\mu_H & -\frac{\alpha\lambda_H}{\mu_H} & 0 & -\frac{\beta\lambda_H}{\mu_H} & 0 & \frac{\lambda_H\left(\alpha\left(\beta -1\right)-\alpha\beta +\beta \left(\alpha -1\right)\right)}{\mu_H} & 0\\ 
    0 & \frac{\alpha\lambda_H}{\mu_H}-\eta -\gamma_C-\mu_H-\delta_C & 0 & 0 & 0 & -\frac{\alpha\lambda_H\,\left(\beta -1\right)}{\mu_H} & 0\\ 
    0 & \eta +\gamma_C & -\mu_H & 0 & 0 & 0 & 0\\ 
    0 & 0 & 0 & \frac{\beta \lambda_H}{\mu_H}-\eta -\gamma_F-\mu_H-\delta_F & 0 & -\frac{\beta\lambda_H\left(\alpha -1\right)}{\mu_H} & 0\\ 
    0 & 0 & 0 & \eta +\gamma_F & -\mu_H & 0 & 0\\ 
    0 & 0 & 0 & 0 & 0 & \frac{\alpha \beta \lambda_H}{\mu_H}-\eta -\gamma_{CF}-\mu_H-\delta_{CF} & 0\\ 
    0 & 0 & 0 & 0 & 0 & \eta +\gamma_{CF} & -\mu_H 
    
    \end{bmatrix}}
$

From the above matrix, the eigenvalues $(\lambda_{i}'s)$ corresponding to the above matrix are given by;\\
$\lambda_{1}=-\mu_H < 0,
\lambda_{2}= \frac{\alpha\lambda_H}{\mu_H}-\eta -\gamma_C-\mu_H-\delta_C,
\lambda_{3}=-\mu_H < 0,
\lambda_{4}= \frac{\beta \lambda_H}{\mu_H}-\eta -\gamma_F-\mu_H-\delta_F,
\lambda_{5}= -\mu_H < 0,
\lambda_{6}=\frac{\alpha \beta \lambda_H}{\mu_H}-\eta -\gamma_{CF}-\mu_H-\delta_{CF},
\lambda_{7}=-\mu_H < 0 $.

\noindent Therefore, the following are the requisites of stable DFE, 
$\frac{\alpha\lambda_H}{\mu_H}-\eta -\gamma_C-\mu_H-\delta_C < 0 \Rightarrow R_{0_{C}} < 1$, 
$\frac{\beta \lambda_H}{\mu_H}-\eta -\gamma_F-\mu_H-\delta_F < 0 \Rightarrow R_{0_{F}} < 1$ and 
$\frac{\alpha \beta \lambda_H}{\mu_H}-\eta -\gamma_{CF}-\mu_H-\delta_{CF} < 0  \Rightarrow R_{0_{CF}} < 1$.
Hence, DFE is locally asymptotically stable iff $R_0 = \max{ \{ R_{0_{C}}, R_{0_{F}}, R_{0_{CF}} \}} < 1 $ which proves the result.

\subsubsection{Global Stability of DFE}
The defined system \eqref{Co-infModelSys} can be written as,

\begin{eqnarray}\label{GlobalStab1}
\frac{dX}{dt} &=& F(X, V), \\ \nonumber
\frac{dV}{dt} &=& G(X, V), G(X, 0) = 0,
\end{eqnarray}

where, $X = (S_{H},R_{C},R_{F},R_{CF})$ which defines the total number of uninfected population compartments, $V = (I_{C},I_{F},I_{CF})$ defines the total number of infected population compartments.

Let $ P =(\frac{\lambda_H}{\mu_H},0,0,0,0,0,0)$ be the disease free equilibrium point of the system.

This disease free equilibrium point of the system is globally stable under the following two conditions:
\begin{enumerate}\label{GlobalStab}
\item  $\frac{dX}{dt} = F(X, 0)$, $X^*$ is globally asymptotically stable.\\
\item  $G(X, V ) = \mathbbm{B}V - \hat{G}(X,V)$ for $(X,V) \in \Omega.$
\end{enumerate}

Where, $\mathbbm{B} = D_{1}G(X^*, 0)$ is an M-matrix (having the off diagonal elements as nonnegative)
and $\omega$ is the region with biological meaning.

Following from the results of Castillo-Chavez et al \cite{CA}, we verify the following conditions:

$F(X,0)=
                                                                    \begin{bmatrix}
                                                                       \lambda_{H}-\mu_{H}S_{H}\\                                                                                                                0\\
0\\
0\\
                                                                        \end{bmatrix}$\\

$\mathbbm{B}=
                                                                    \begin{bmatrix}
                                                                       -(\mu_H+\delta_C+\gamma_C) & 0& 0\\                                                                                                                0& -(\mu_H+\delta_F+\gamma_F)& 0&\\
\lambda_{H}\beta & 0 & -(\mu_H+\delta_{CF}+\gamma_{CF})+\alpha\beta\lambda_{H}
\end{bmatrix}$\\

$\widehat{G}(X,V)=
                                                                   \begin{bmatrix}
                                             (\frac{\eta }{1+\xi I_{C}}-\alpha S_{H})I_{C}+(\beta I_{C}-\alpha(1-\beta)S_{H})I_{CF}+\beta I_{F}I_{C}\\
            (\frac{\eta }{1+\xi I_{F}}-\beta S_{H})I_{F}+(\alpha I_{F}-\beta(1-\alpha)S_{H})I_{CF}+\alpha I_{F}I_{C}   \\                                                                                        %-\beta S_{H}I_{F}-\beta(1-\alpha)S_{H}I_{CF}+\alpha(I_{C}+I_{CF})I_{F}+h(I_{F})\\
                                                                   (\frac{\eta }{1+\xi I_{CF}}-\alpha \beta S_{H})I_{CF}-(\alpha+\beta)I_{F}I_{C}-(\alpha I_{F}+\beta I_{C})I_{CF}\\

                                                                        \end{bmatrix}$\\

Therefore, $\widehat{G}(X,V)\geq 0$ subject to the following requisites;

$\frac{\eta}{1+\xi M} > \alpha \lambda_{H}, $ \\

$\frac{\eta}{1+\xi M'} > \beta \lambda_{H} $,\\

$\frac{\eta}{1+\xi M"} > \alpha\beta \lambda_{H} $,\\

\noindent along with the conditions as follows ;\\
$\beta > \frac{\alpha}{1+\alpha}$,  $  \alpha > \beta(1-\alpha)\lambda_{H}$, %$\eta > \alpha \beta \lambda_{H}$
$\lambda _{H}> \frac{\alpha +\beta}{\beta}M$ and $\lambda _{H}> \frac{\alpha M+\beta M'}{\alpha\beta}$ \\ where, $M$, $M'$ and $ M"$ are the upper bounds of $ I_{C}$, $I_{F}$ and $I_{CF}$ respectively.
%Also, it is evident that $X^*$ is a globally asymptotically stable equilibrium of $\frac{dX}{dt} = F(X, 0)$.
Hence, the main result follows from\cite{CA} which is stated below.\\

\begin{theorem}
The disease free equilibrium point given by $(\frac{\lambda_H}{\mu_H},0,0,0,0,0,0)$,
is globally asymptotically stable for the defined system if $R_{0}<1$ and conditions in \ref{GlobalStab} hold true.
\end{theorem}

\section{Invasion reproduction number $R_{0_{\mathrm{Inv}}}$ }
It is to be noted that in \eqref{BRNModel}, the terms represent the basic reproduction numbers corresponding to infection with either Covid-19 or ILI or with the both, respectively. 
If $R_{0_{C}} > 1$, then Covid-19-19 can invade the DFE and if $R_{0_{F}} > 1$, then ILI can invade the DFE. 
In general, if $R_0 > 1 $, then either Covid-19 or ILI or Co-infection can invade the mode system \eqref{Co-infModelSys}.\par

An invasion reproduction number $R_{0_{\mathrm{Inv}}}$, can determine if the model system \eqref{Co-infModelSys} is already infected with a single pathogen and in our case, it is ILI.
We analyse what happens when Covid-19  invades the model system at the ILI equilibrium i.e. when $R_{0_{F}} > 1$

The endemic equilibrium point for $E^0_F = (S^*_H, I^*_F, R^*_F)$ with ILI can be found as $S^*_H = \frac{\lambda_H}{\mu_H+\beta I^*_F}$,
$R^*_F = \frac{\eta I^*_F}{1+\zeta I^*_F}$
and $I^*_{F}$ is a positive root of the quadratic equation,
$I^{*^{2}}_F+c_1I^*_F+c_2 = 0$,
where $c_1 = \mu_H\zeta+\beta+\frac{\eta \beta-\lambda_H\zeta\beta}{\mu_H+\delta_F+\gamma_F}$ and $c_2 = (1-R_{0_{F}})\frac{\mu_H(\mu_H+\delta_F+\gamma_F+\eta)}{\beta(\mu_H+\delta_F+\gamma_F)}$.
For the calculation purpose we shall use the forms expressed as in $E^0_F$.
We use the NGM approach to derive an invasion matrix  $\mathrm{NGM_{F_{\mathrm{Inv}}}}$ associated with Covid-19 and ILI and thereafter we compute $R_{0_{\mathrm{Inv}}}$.
After following the notation in \eqref{R0Eqn1}, we have :
\begin{equation}\label{R0InvEqn2}
\frac{dY_{I_{F_{\mathrm{Inv}}}}}{dt} =  \mathrm{F_{F_{\mathrm{Inv}}}}(Y_S, Y^{*}_{I_F})- \mathrm{V_{F_{\mathrm{Inv}}}}(Y_S, Y^{*}_{I_F})
\end{equation}

$
 \mathrm{F_{F_{\mathrm{Inv}}}} = 
   \begin{bmatrix} 
     S^*_H\alpha -I^*_F\beta  & 0 & -S^*_H\alpha \left(\beta -1\right)\\ 
     -I^*_F\alpha  & S^*_H\beta  & -I^*_F\alpha -S^*_H\beta \left(\alpha -1\right)\\ 
     I^*_F\left(\alpha +\beta \right) & 0 & I^*_F\alpha +S^*_H\alpha \beta  
     \end{bmatrix} 
$

$
 \mathrm{V_{F_{\mathrm{Inv}}}} = 
   \begin{bmatrix} 
    \delta_C+\eta +\gamma_C+\mu_H & 0 & 0\\ 
    0 & \delta_F+\gamma_F+\mu_H+\frac{\eta }{I^*_F\zeta +1}-\frac{I^*_F\eta \zeta }{{\left(I^*_F\zeta +1\right)}^2} & 0\\ 
    0 & 0 & \delta_{CF}+\eta +\gamma_{CF}+\mu_H 
    \end{bmatrix} 
$

 $
  \tiny{ \mathrm{NGM_{F_{\mathrm{Inv}}}} } =
    \tiny{\begin{bmatrix} 
 -\frac{I_F^*\beta -S^*_H\alpha }{\delta_C+\eta +\gamma_C+\mu_H} & 0 & -\frac{S^*_H\alpha \left(\beta -1\right)}{\delta_{CF}+\eta +\gamma_{CF}+\mu_H}\\ 
    
    -\frac{I^*_F\alpha }{\delta_C+\eta +\gamma_C+\mu_H} & \frac{S^*_H\beta{\left(I^*_F\zeta +1\right)}^2}{\delta_F+\eta +\gamma_F+\mu_H+{I^*_F}^2\delta_F\zeta ^2+{I^*_F}^2\gamma_F\zeta ^2+{I^*_F}^2\mu_H\zeta ^2+2I^*_F\delta_F\zeta +2I^*_F\gamma_F\zeta +2I^*_F\mu_H\zeta } & -\frac{I^*_F\alpha +S^*_H\beta \left(\alpha -1\right)}{\delta_{CF}+\eta +\gamma_{CF}+\mu_H}\\ 

    \frac{I^*_F\left(\alpha +\beta \right)}{\delta_C+\eta +\gamma_C+\mu_H} & 0 & \frac{I^*_F\alpha +S^*_H\alpha\beta }{\delta_{CF}+\eta +\gamma_{CF}+\mu_H} 
    
    \end{bmatrix}}
$
\\

According to the authors in \cite{v11121153}, we can define the invasion reproduction number $R_{0_{\mathrm{Inv}}}$ as the spectral radius of the invasion matrix $\mathrm{NGM_{F_{\mathrm{Inv}}}}$.
Therefore, $\boldsymbol{\rho} (\mathrm{NGM_{F_{\mathrm{Inv}}}}) =  \max{ \{ \lambda_1, \lambda_2, \lambda_3 \} }$, where, 
$\lambda_1 = \frac{S^*_H\beta (I^*_F\zeta+1)^2}{(\delta_F+\gamma_F+\mu_H)(1+I^*_F\zeta)^2+\eta}$, $\lambda_{2,3}= \frac{1}{2} \left[a+g \pm \sqrt{(a-g)^2+4ce} \right]$.
The expressions for $a, g, c$ and $e$ are following:
$a= -\frac{I_F^*\beta -S^*_H\alpha }{\delta_C+\eta +\gamma_C+\mu_H}$, $g = \frac{I^*_F\alpha +S^*_H\alpha\beta }{\delta_{CF}+\eta +\gamma_{CF}+\mu_H}$, $c =   -\frac{I^*_F\alpha }{\delta_C+\eta +\gamma_C+\mu_H}$ and $e = -\frac{I^*_F\alpha +S^*_H\beta \left(\alpha -1\right)}{\delta_{CF}+\eta +\gamma_{CF}+\mu_H}$.

The epidemiological metric $R_{0_{\mathrm{Inv}}}$ can be interpreted as the average number of new infected states that are being produced after an introduction of an average of one Covid-19 infected agent into
the population infected with ILI.

\section{Numerical Simulation}
\subsection{Data Fitting}
We implement our model fitting for an epidemic period starting from when the Indian government formally announced the index cases of Covid-19 for the different states.
We follow the authors in \cite{OKUONGHAE2020110032} while performing the model fitting and the data has been downloaded from \cite{IndiaCOVID-19tracker}.
Collected data contains the days under the lockdown period too and hence the impact of lockdown on the disease dynamics shall inherently capture the transmission rates in the different states of India.
Using the available cumulative number of reported data, we actively seek to estimate the unknown model parameters and fit the infection trajectories of different states of India.
Estimated parameters pertaining to the Covid-19 model are found to be heterogenous in the magnitude and potentially it shows the different infection paths. 
One thing is important to note that as at the time when the first Covid-19 case was declared  in India, it would be extremely difficult to determine the exact number of individuals who were already infected. 
Hence, it might be an aspect that reflects through the heterogeneous distribution of the estimated parameters. 

\begin{figure}[H]
\centering
\subfloat[Subfigure 1 list of figures text][GJ]{
\includegraphics[width=0.28\textwidth]{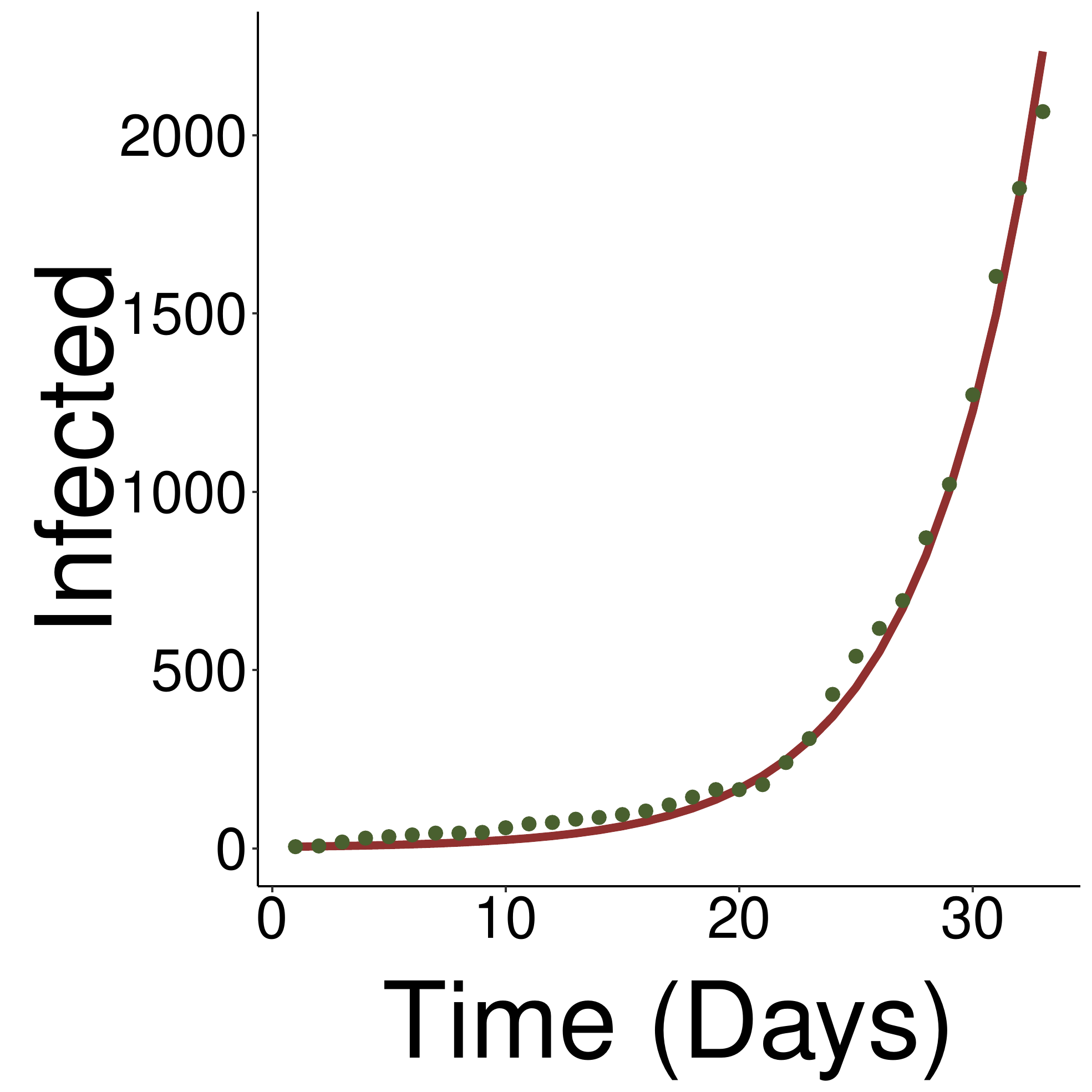}
\label{fig:subfig1}}
%\qquad
\subfloat[Subfigure 2 list of figures text][MH]{
\includegraphics[width=0.3\textwidth]{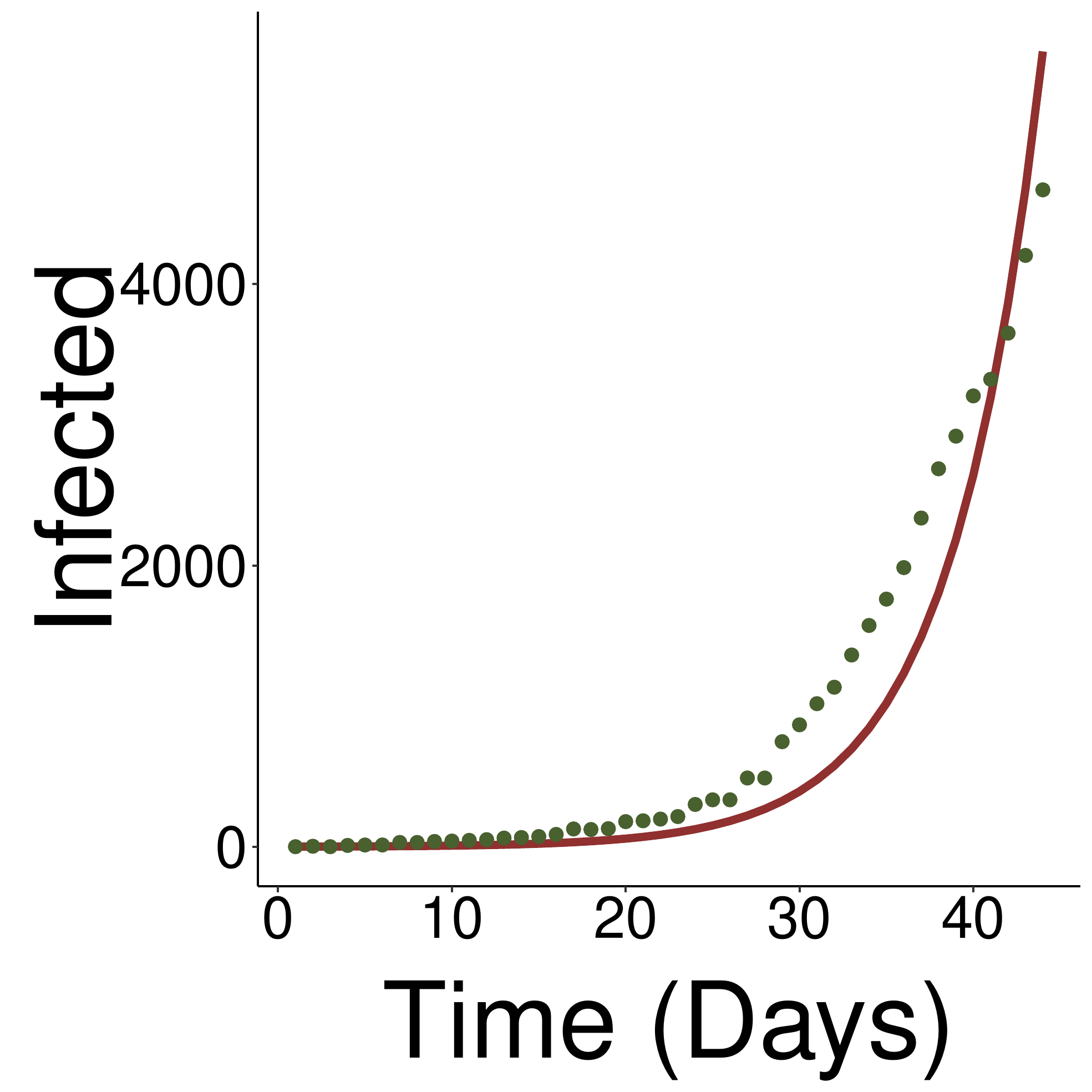}
\label{fig:subfig2}}
\subfloat[Subfigure 3 list of figures text][RJ]{
\includegraphics[width=0.3\textwidth]{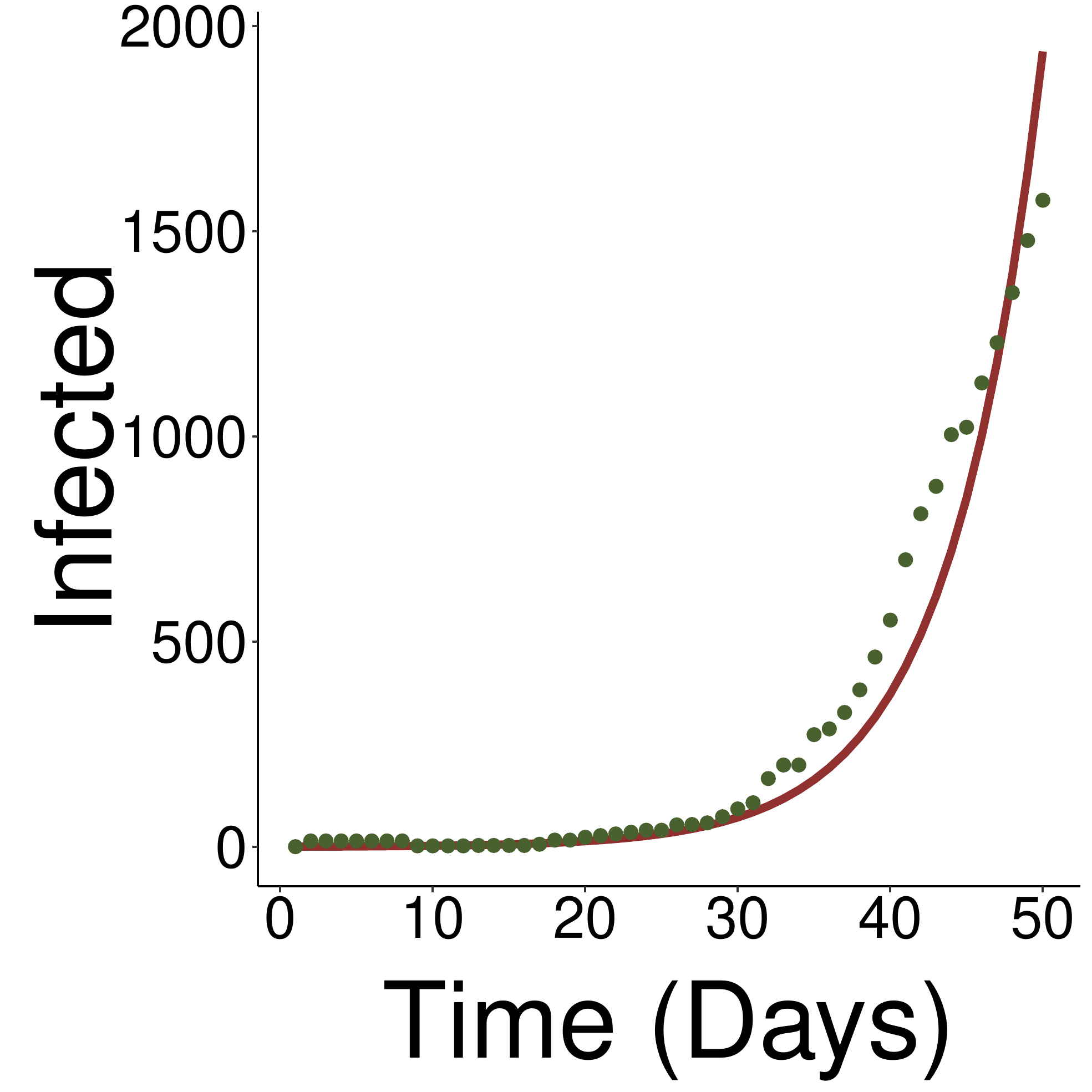}
\label{fig:subfig3}}
\qquad
\subfloat[Subfigure 4 list of figures text][UP]{
\includegraphics[width=0.3\textwidth]{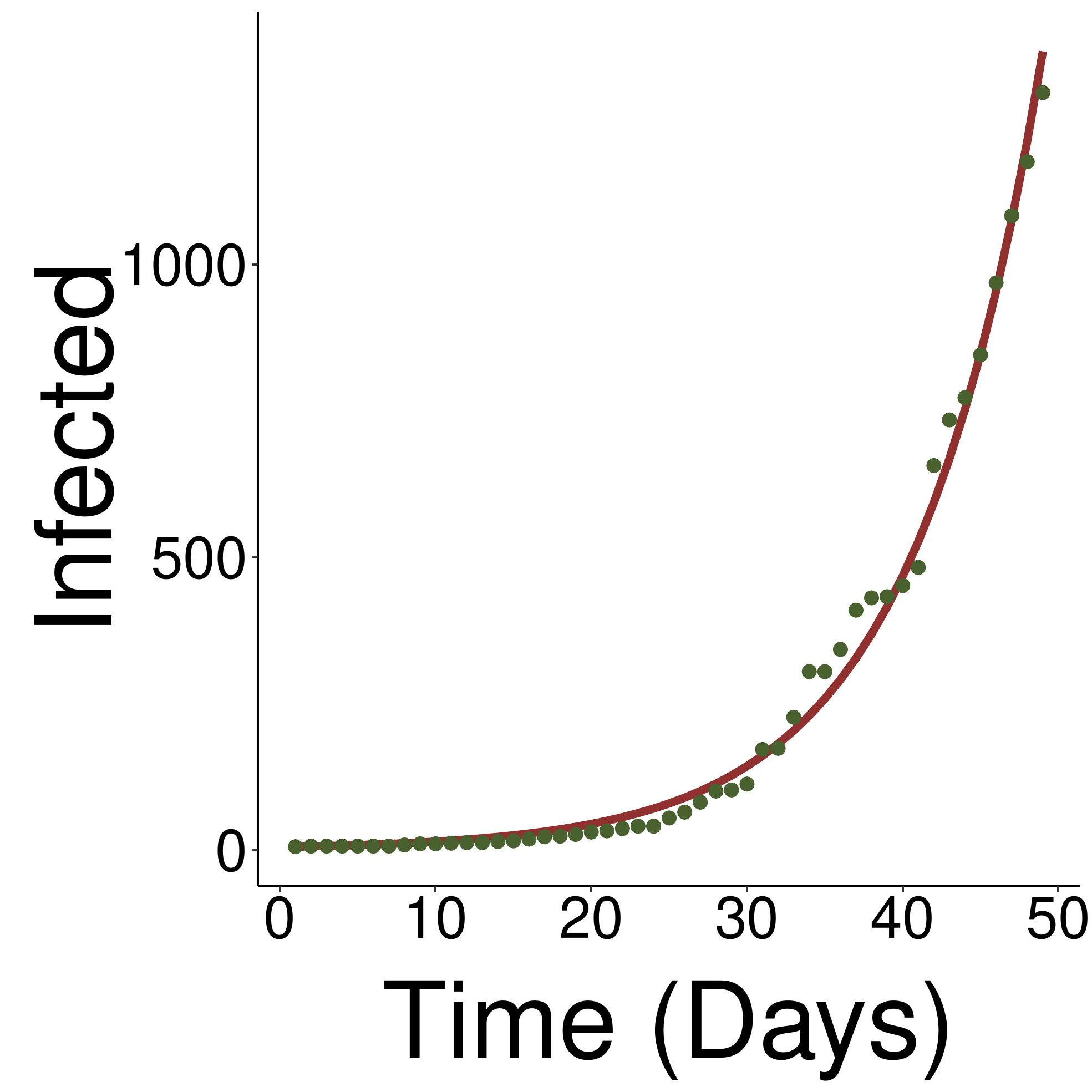}
\label{fig:subfig4}}
\subfloat[Subfigure 5 list of figures text][WB]{
\includegraphics[width=0.3\textwidth]{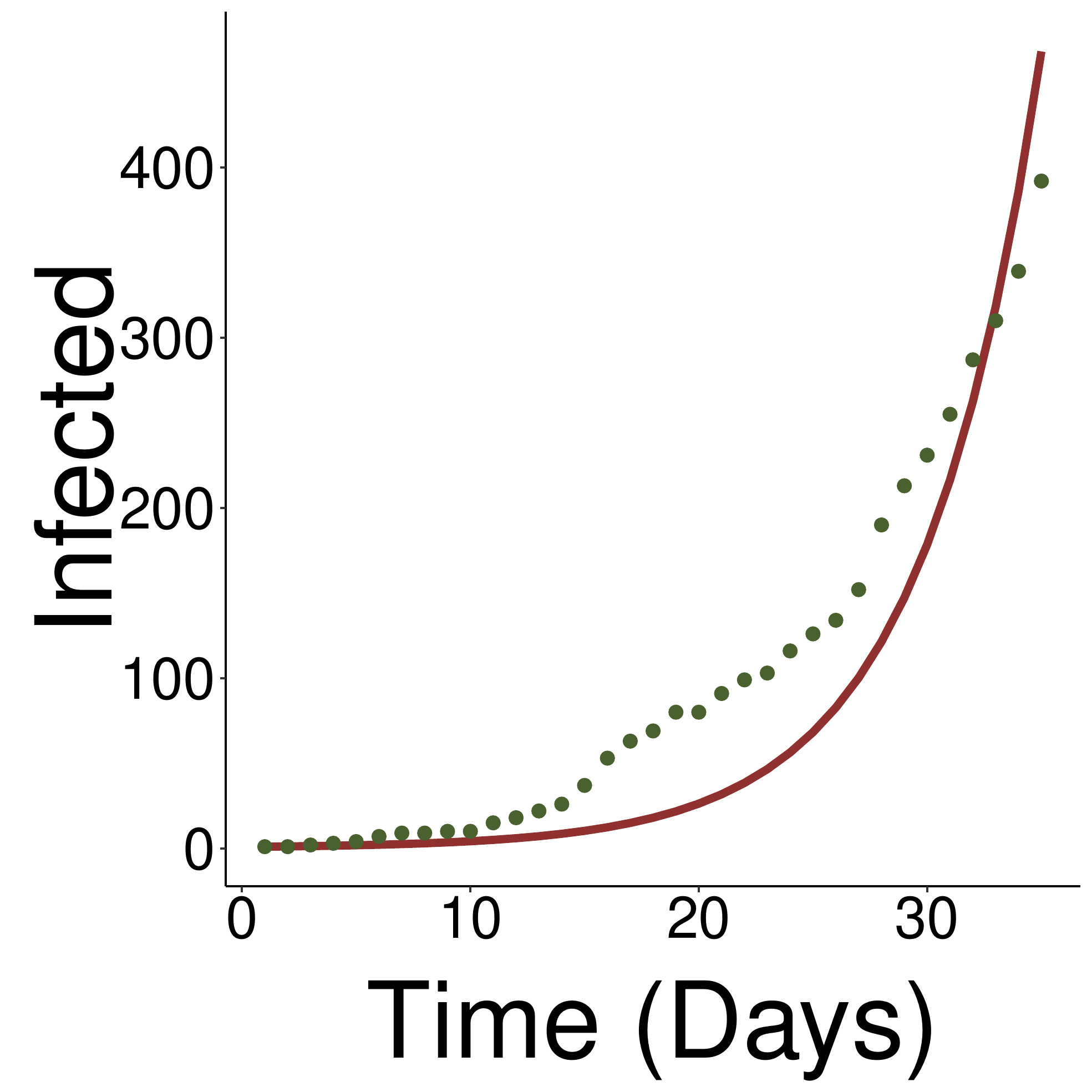}
\label{fig:subfig5}}
\subfloat[Subfigure 6 list of figures text][DL]{
\includegraphics[width=0.3\textwidth]{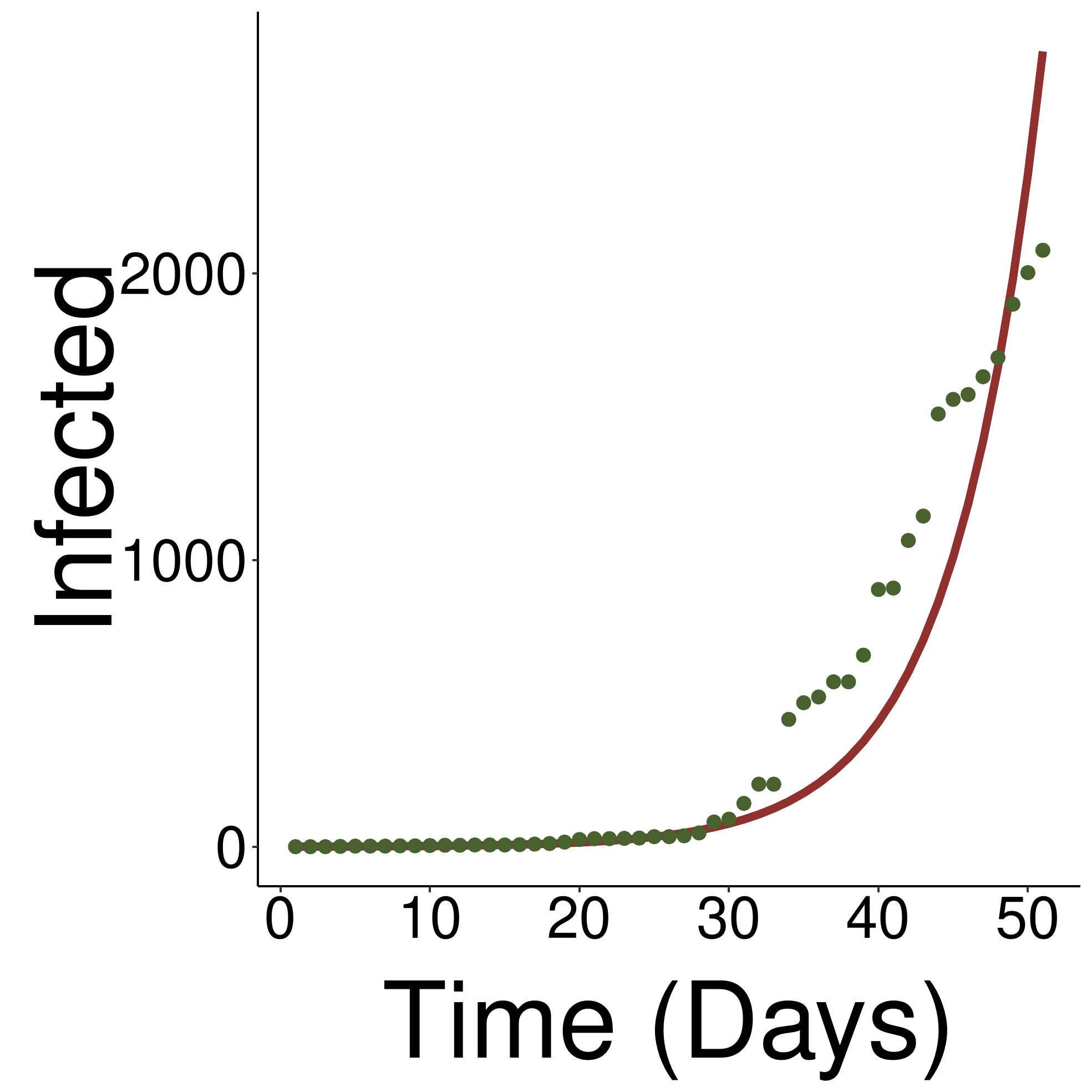}
\label{fig:subfig6}}
\qquad
\subfloat[Subfigure 7 list of figures text][MP]{
\includegraphics[width=0.3\textwidth]{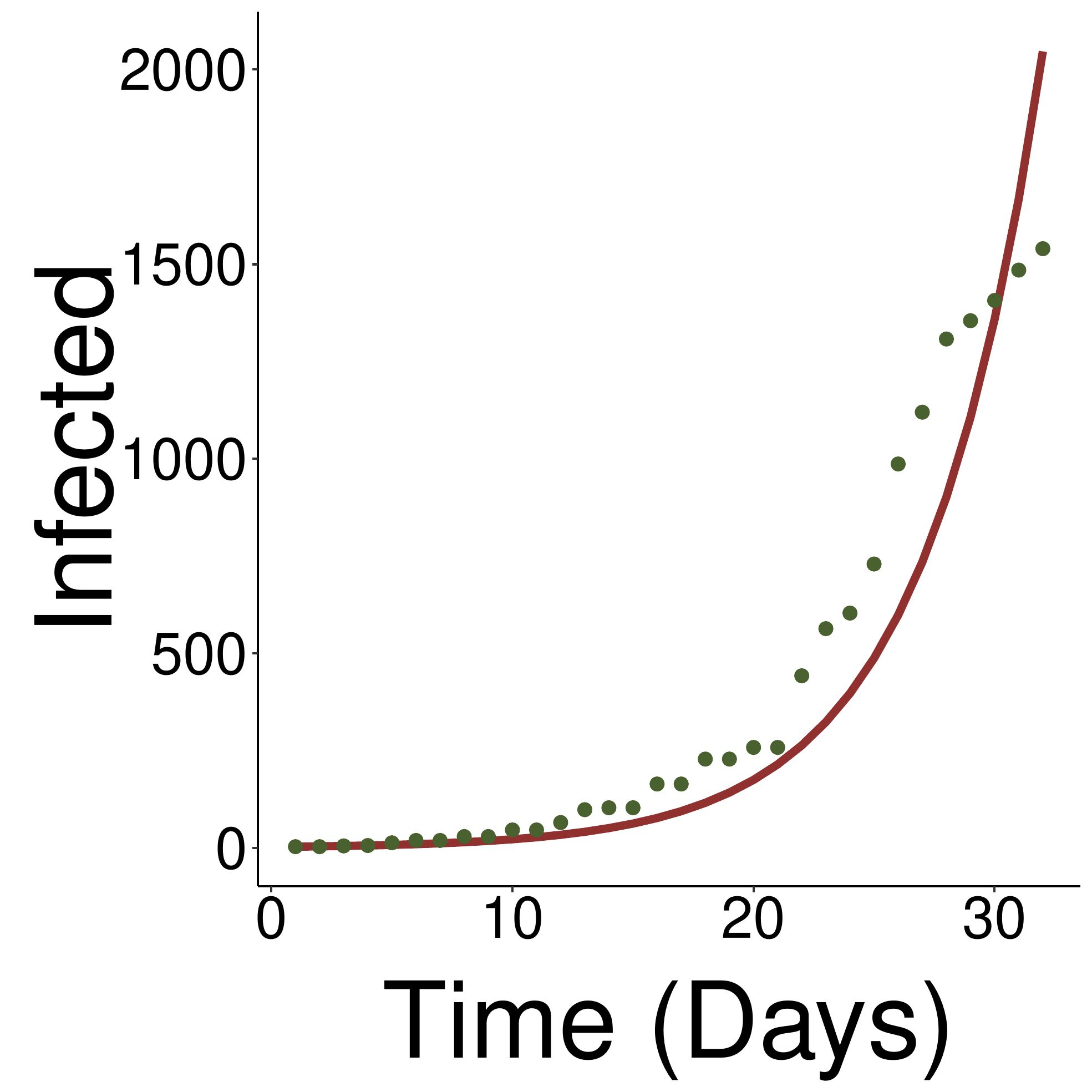}
\label{fig:subfig7}}
\subfloat[Subfigure 8 list of figures text][TS]{
\includegraphics[width=0.3\textwidth]{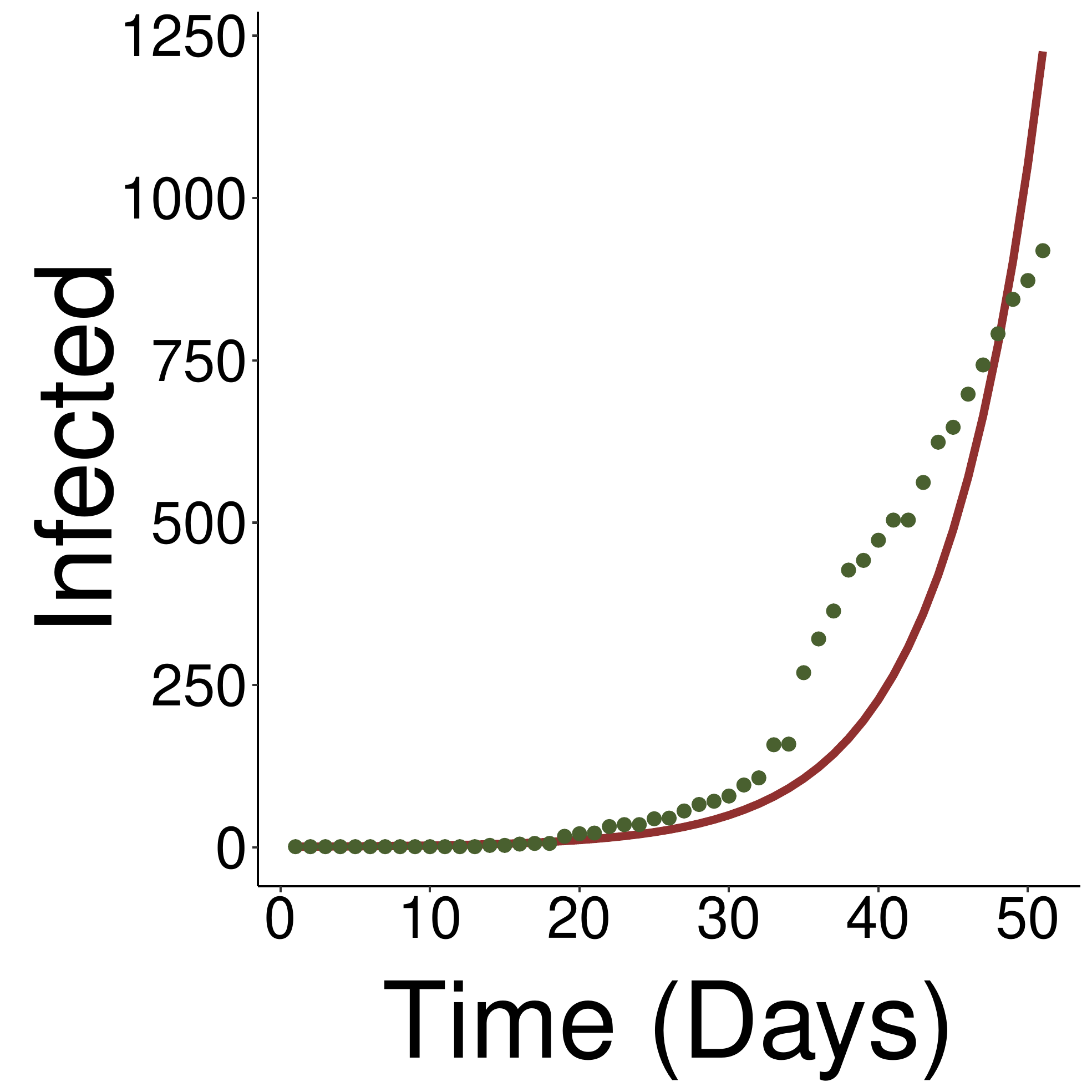}
\label{fig:subfig8}}
\subfloat[Subfigure 9 list of figures text][HR]{
\includegraphics[width=0.3\textwidth]{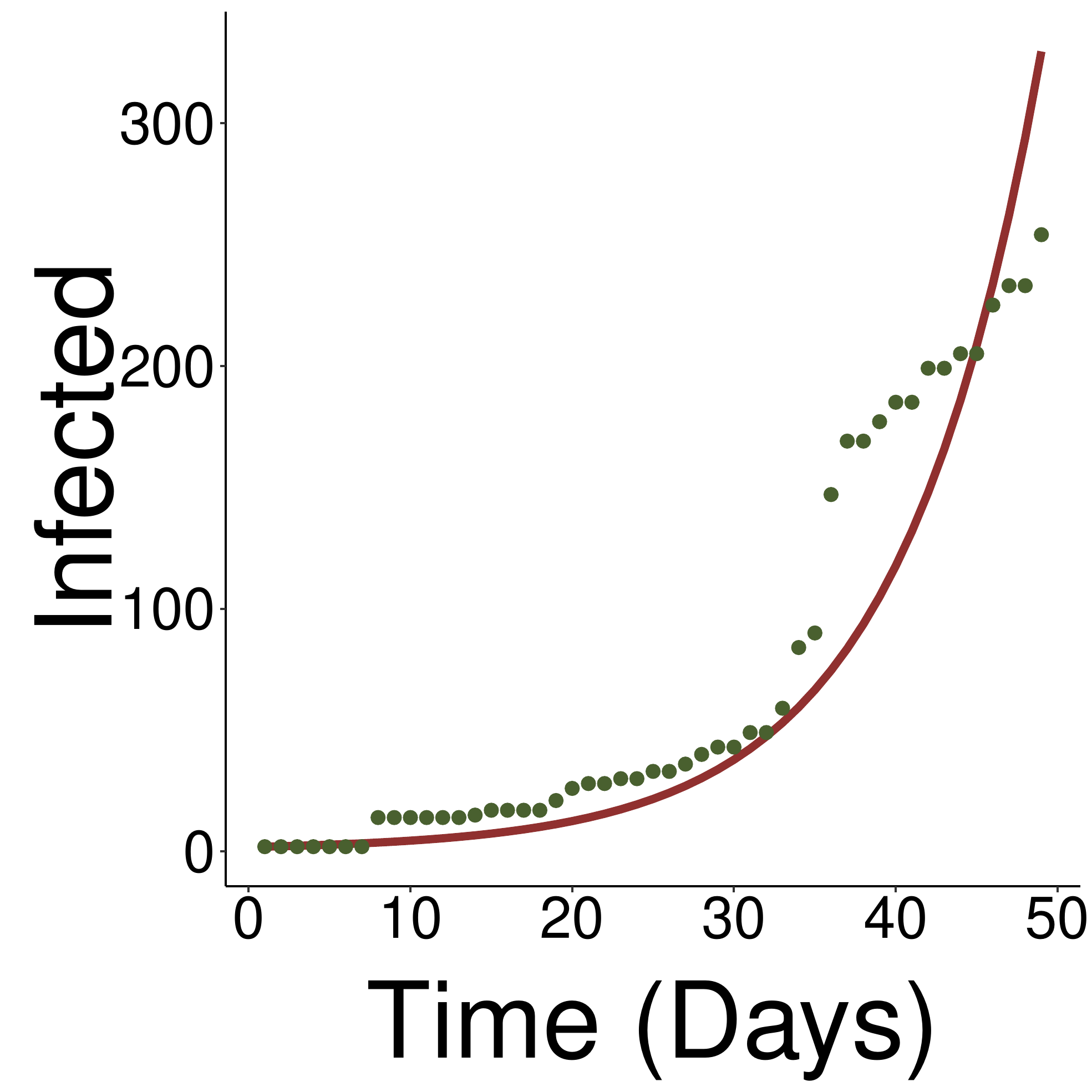}
\label{fig:subfig9}}
\caption{The comparison between the reported Covid-19 cases in  Gujarat, Maharashtra, Rajasthan, Uttar Pradesh, West Bengal, Delhi, Madhya Pradesh, Telengana, Haryana and the simulation of $I_C(t)$ from the model described in \eqref{2}.}
\label{fig:globfig}
\end{figure}

Figure:\ref{fig:globfig} shows that our model \eqref{2} is able to predict the temporal course of daily reported cases of Covid-19 at the national scale where the green dots are the reported cases and the red curves are fitting curves.
It is readily observable from the fitted model that the growth is of exponential type and the influence of lockdown is clearly noticeable.

\subsection{Real time $R_0$ Estimation}

Precise assessment of the parameters that identify infectious transmission is crucial to optimise the different containment strategies. 
An epidemiological metric termed as the time-dependent reproduction $R_t$ has acquired much needed appreciation due to the fact that it can estimate the expected number of secondary cases caused by each infected people in a temporal way \cite{Chowell}.
The magnitude of $R_t$ can provide us the information about the extremity of Covid-19 over different time points.

\begin{figure}[H]
\centering
\subfloat[Subfigure 1 list of figures text][GJ]{
\includegraphics[width=0.3\textwidth]{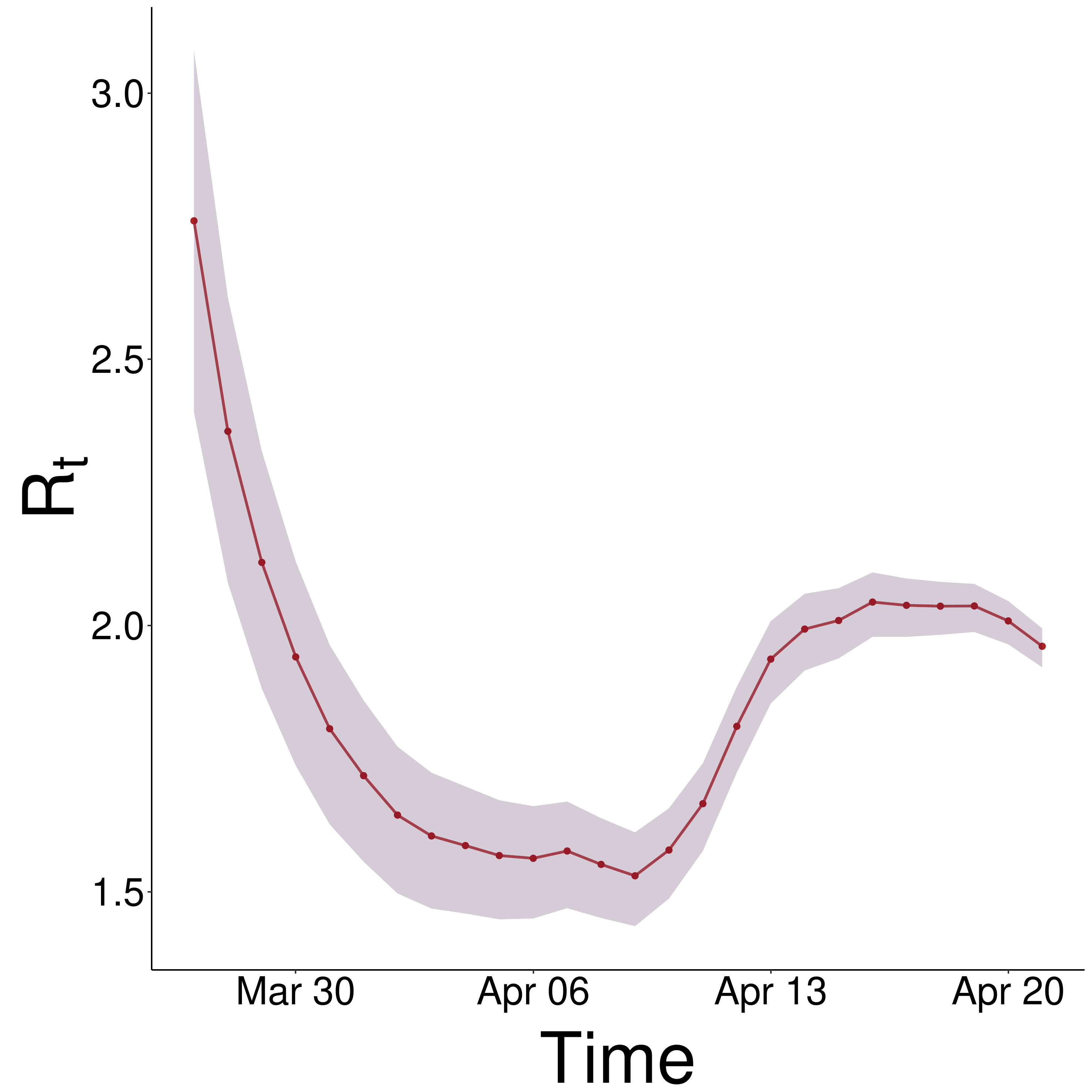}
\label{fig:subfig1Rt}}
%\qquad
\subfloat[Subfigure 2 list of figures text][MH]{
\includegraphics[width=0.3\textwidth]{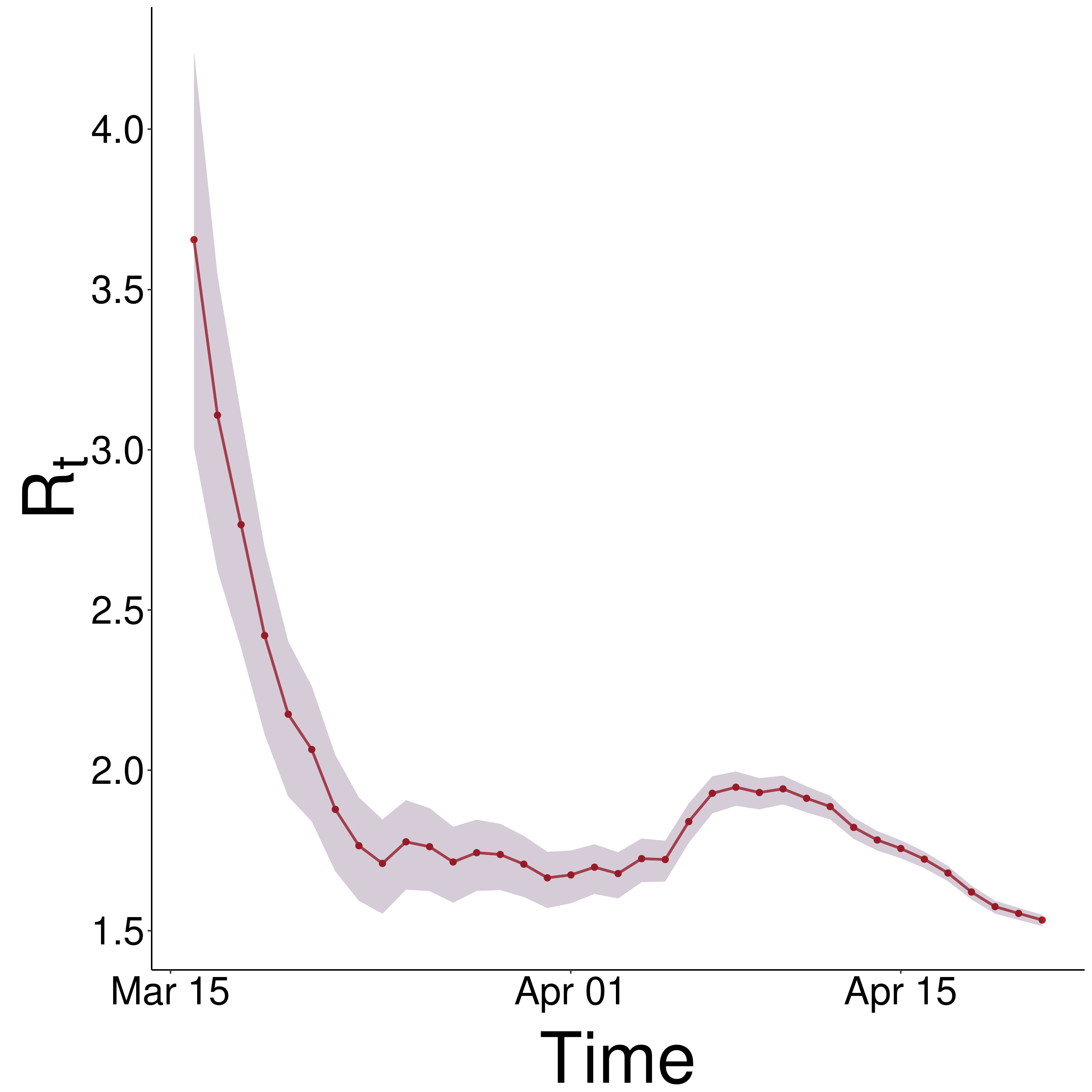}
\label{fig:subfig2Rt}}
\subfloat[Subfigure 3 list of figures text][RJ]{
\includegraphics[width=0.3\textwidth]{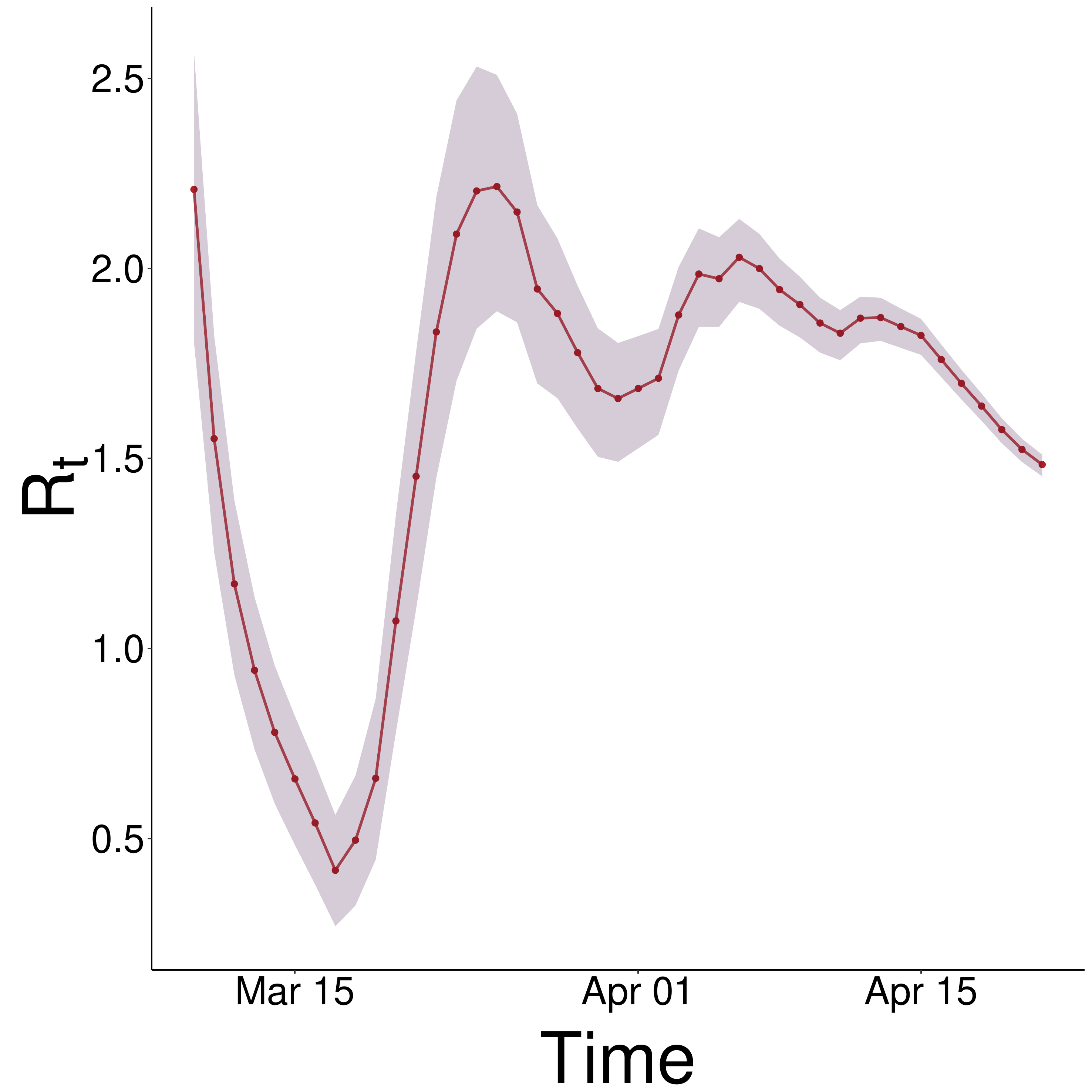}
\label{fig:subfig3Rt}}
\qquad
\subfloat[Subfigure 4 list of figures text][UP]{
\includegraphics[width=0.3\textwidth]{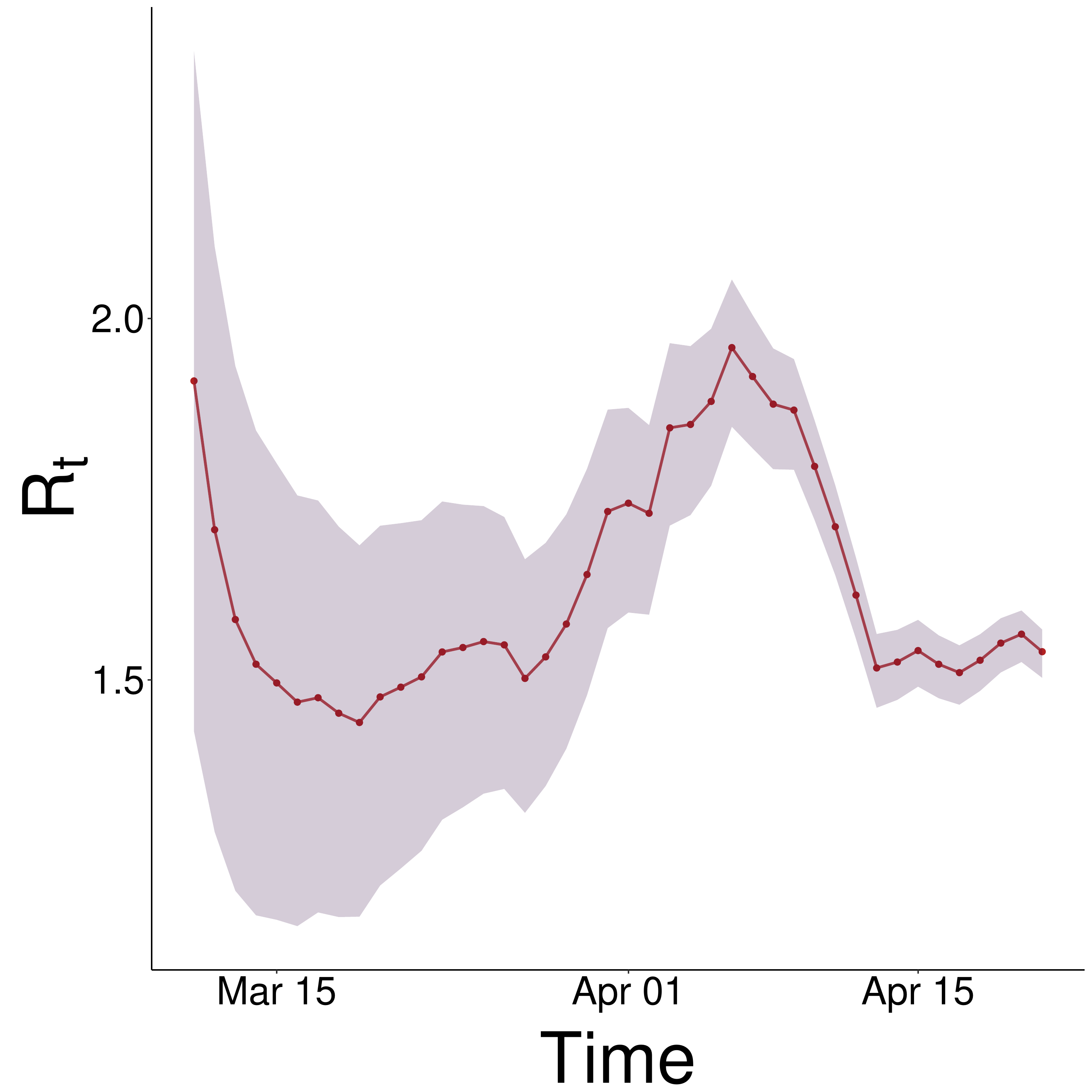}
\label{fig:subfig4Rt}}
\subfloat[Subfigure 5 list of figures text][WB]{
\includegraphics[width=0.3\textwidth]{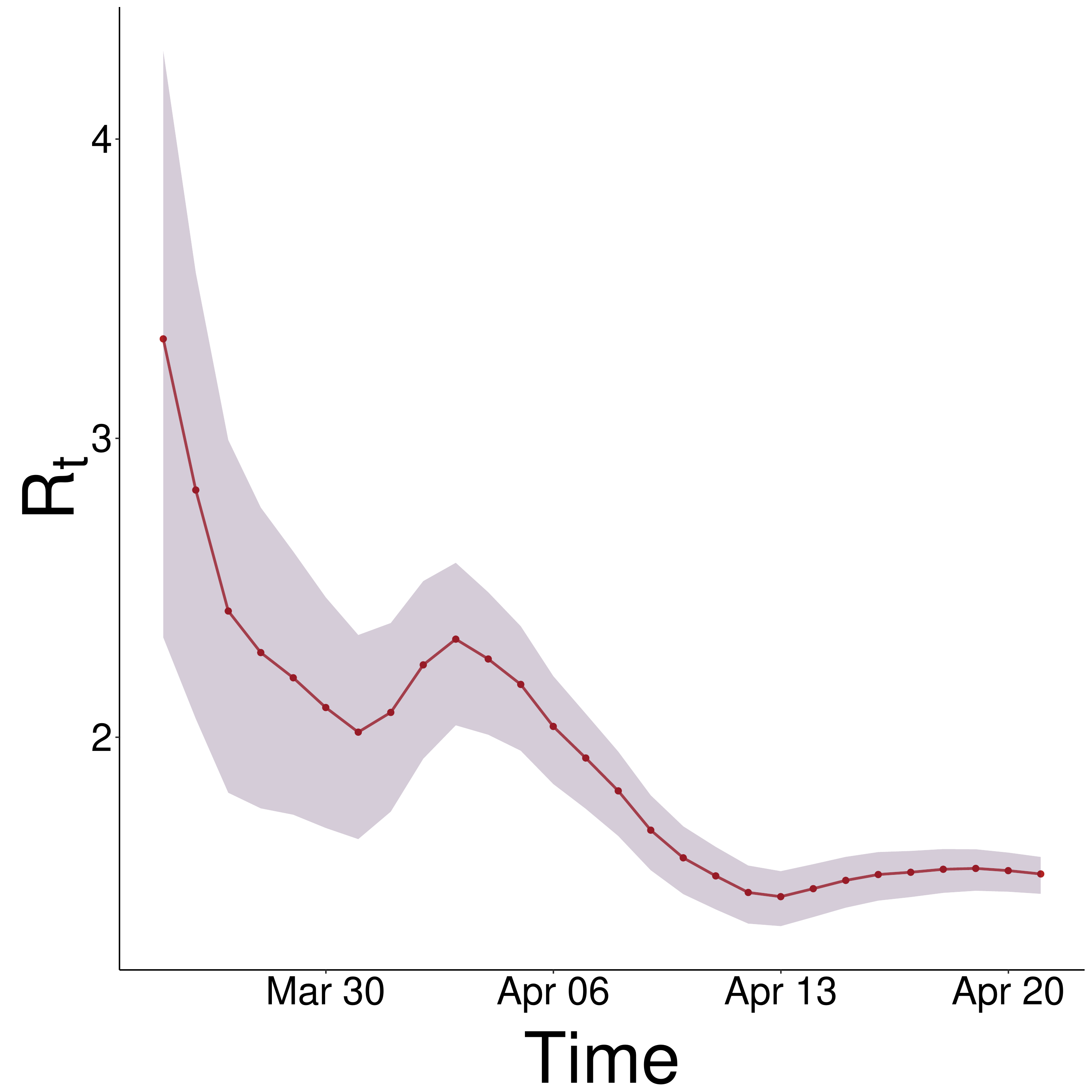}
\label{fig:subfig5Rt}}
\subfloat[Subfigure 6 list of figures text][DL]{
\includegraphics[width=0.3\textwidth]{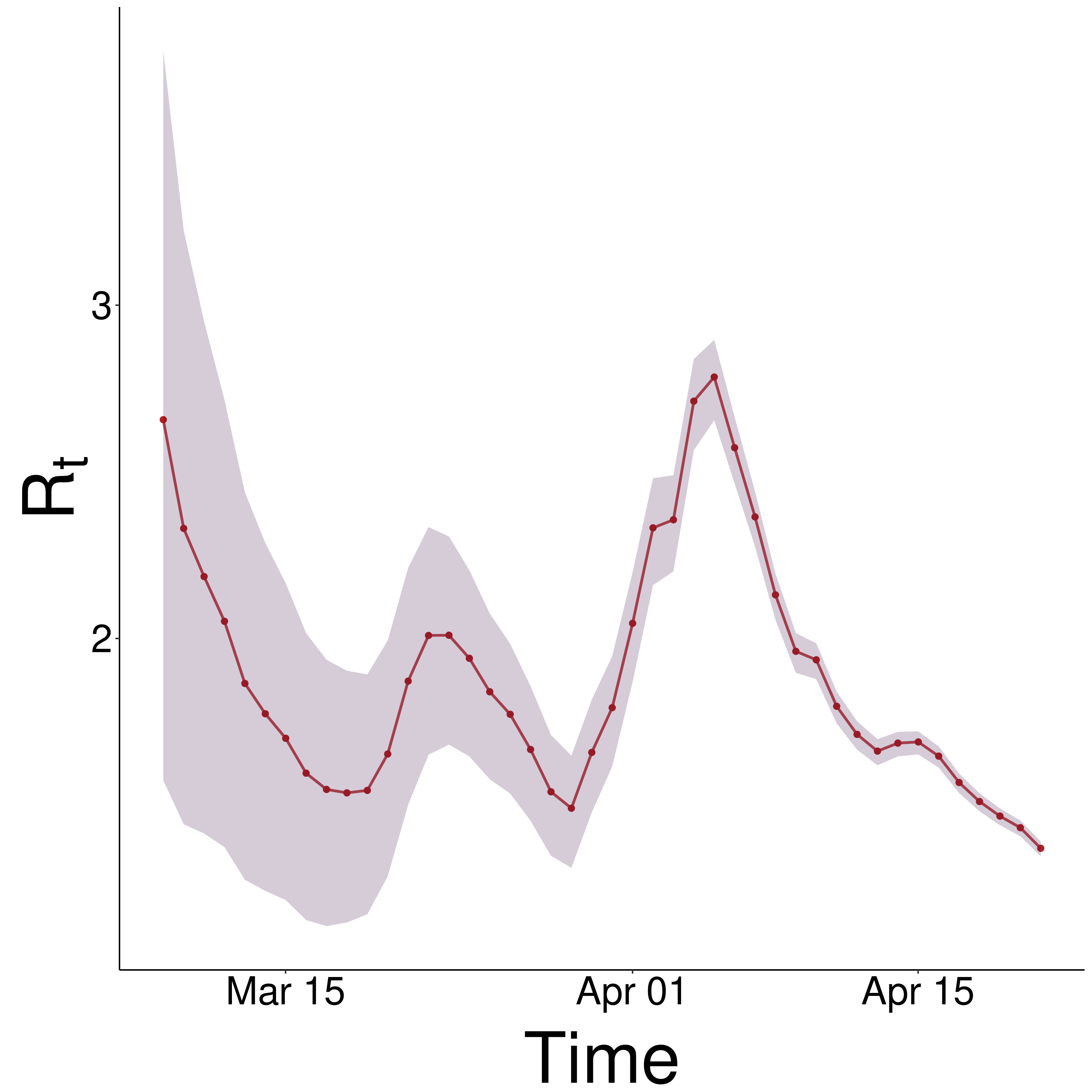}
\label{fig:subfig6Rt}}
\qquad
\subfloat[Subfigure 7 list of figures text][MP]{
\includegraphics[width=0.3\textwidth]{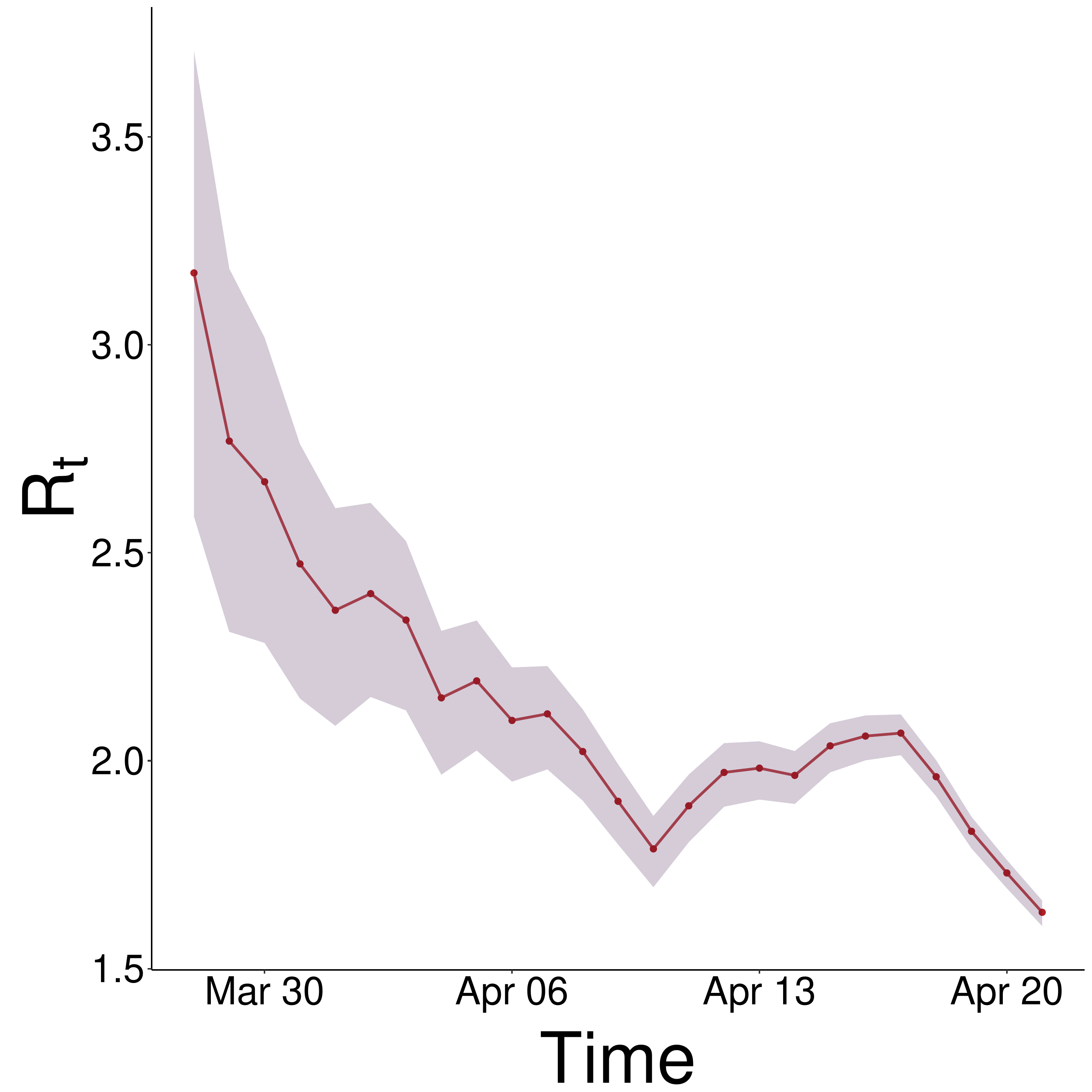}
\label{fig:subfig7Rt}}
\subfloat[Subfigure 8 list of figures text][TS]{
\includegraphics[width=0.3\textwidth]{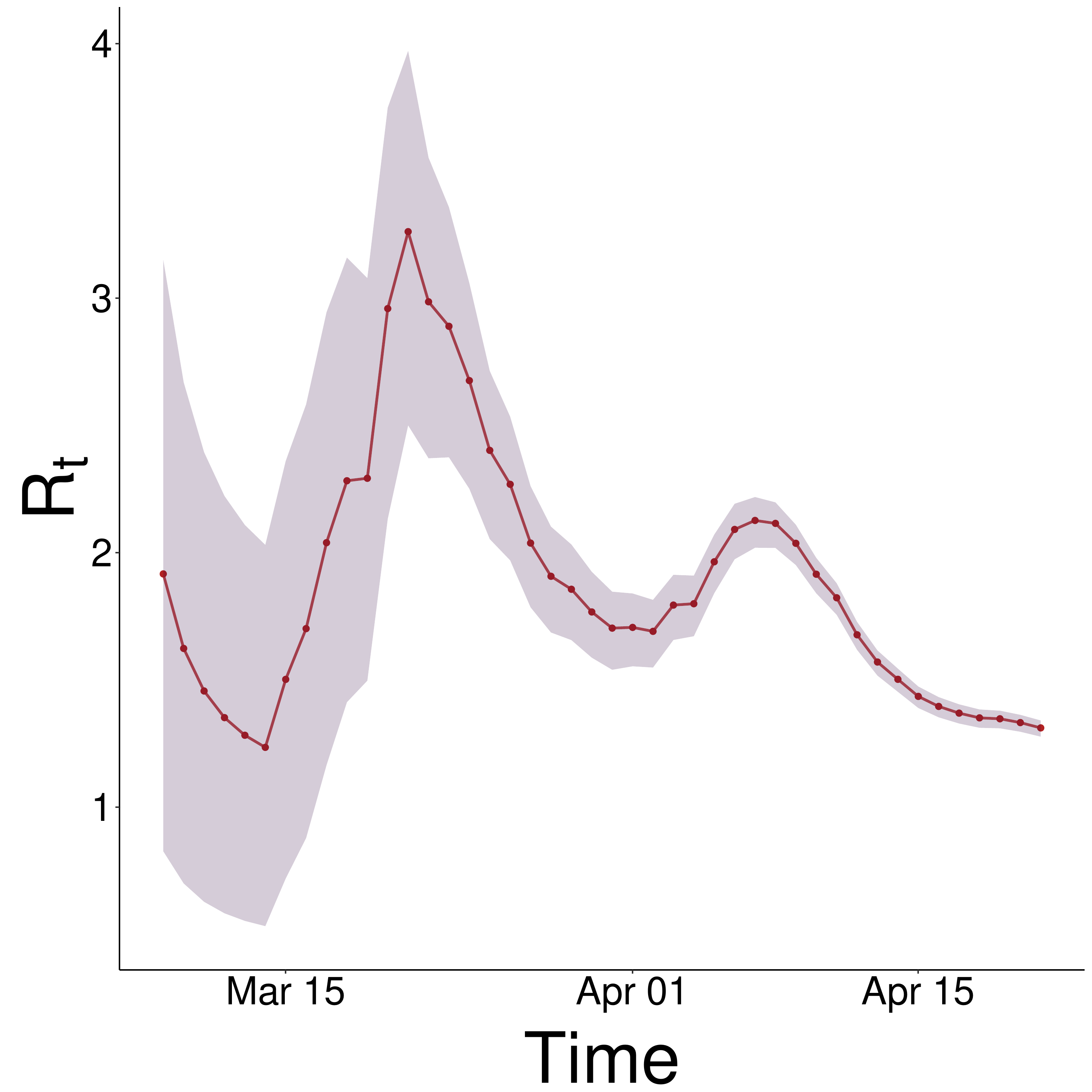}
\label{fig:subfig8Rt}}
\subfloat[Subfigure 9 list of figures text][HR]{
\includegraphics[width=0.3\textwidth]{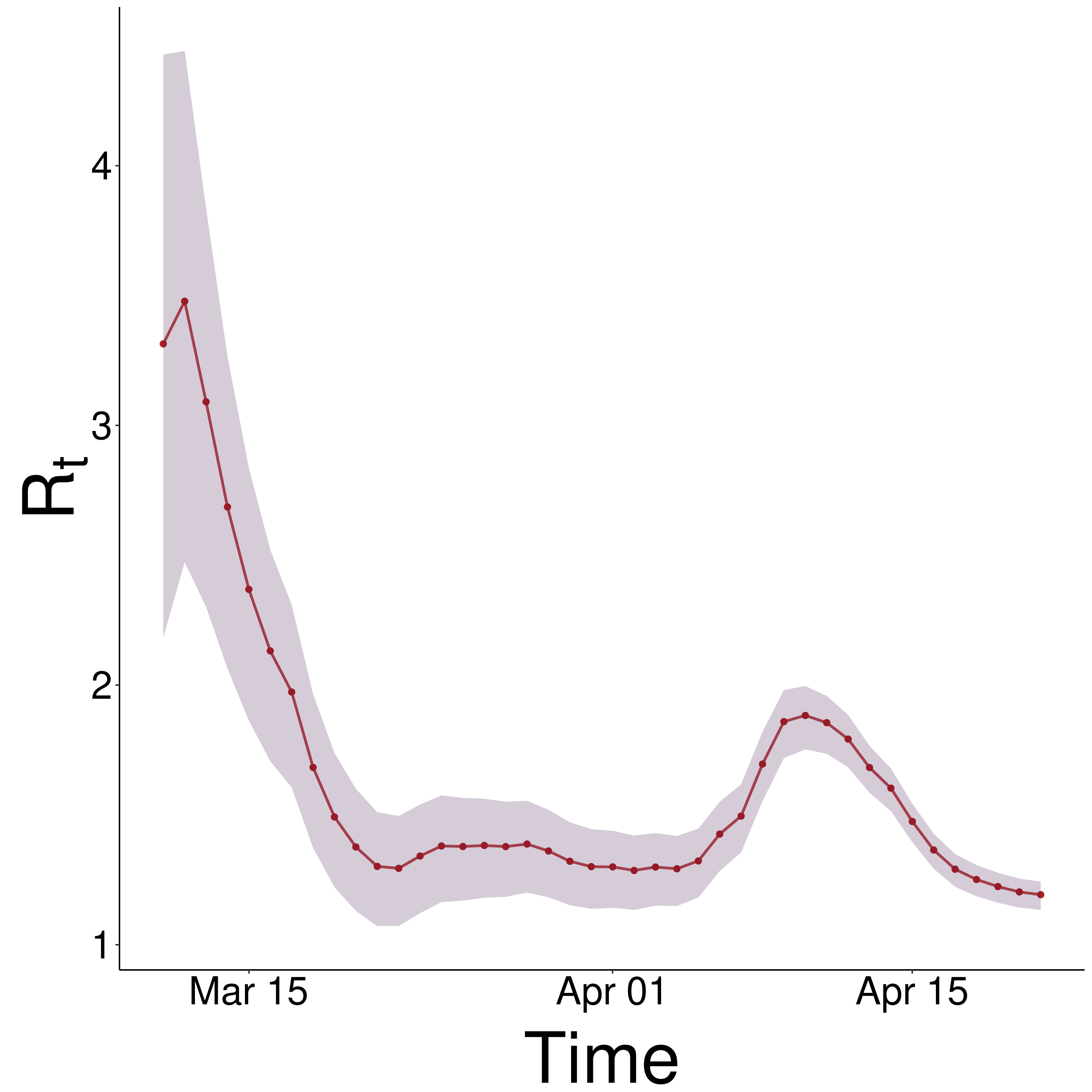}
\label{fig:subfig9Rt}}
\caption{Estimation of the time-dependent reproduction number ($R_t$) for an outbreak of Covid-19 in India.
Estimates of the reproduction number throughout the outbreak (mean (solid-dotted line) and $95\%$ credible interval (shaded area)) obtained from the incidence data shown in the Figure.
To analyse the incidence data of Covid-19  in India, $R_t$ is estimated on sliding windows of width $7$ days.
}
\label{fig:globfigRt}
\end{figure}
To derive the relationship between the daily Covid-19 incidence data and $R_t$, we have employed the \emph{EpiEstim} package \cite{Cori2021}.
We can observe from the Figure: \ref{fig:globfigRt} that the value of $R_t$ ranges between 4.5 to 0.5.
It is also noticeable that the temporal changes in the spread of Covid-19 in India, is highly heterogenous. 
India declared $24^\mathrm{th}$ of March, 2020, a nationwide lockdown for 21 days and its influence can be visible in the Figure: \ref{fig:globfigRt} as the value of $R_t$ decreases.
The magnitude of $R_t$ differed across all the Indian states.
Average $R_t$ is found to be $1.98$ with  $(95\%$ CI $1.93–2.06)$ for the entirety of India during that period of interval and 
in the lockdown phase it is $2.77$ $(95\%$ CI $2.65–2.92)$.
We have observed the geographic differences in the magnitude of $R_t$ across India.
For example, $R_t$ is observed to be greater than $2.5$ for Maharashtra, Delhi, Telengana, whereas it is noticed to be lesser than $2.5$ for UP, West Bengal, Rajasthan.
$R_t$ is found to be greater than $1$ for most of the states during that period.

\subsection{Impact of transmission parameters}
In this section we explore the influence of the transmission parameters ($\alpha$ and $\beta$) on the populations infected with Covid-19, ILI and co-infection cases  ($I_C$, $I_F$ and $I_{CF}$) respectively.
Here, we vary the values of $\alpha$ and $\beta$ and observe the influence on the prevalence on the population.

\begin{figure}[H]
\centering
\subfloat[Subfigure 1 list of figures text][]{
\includegraphics[width=0.32\textwidth]{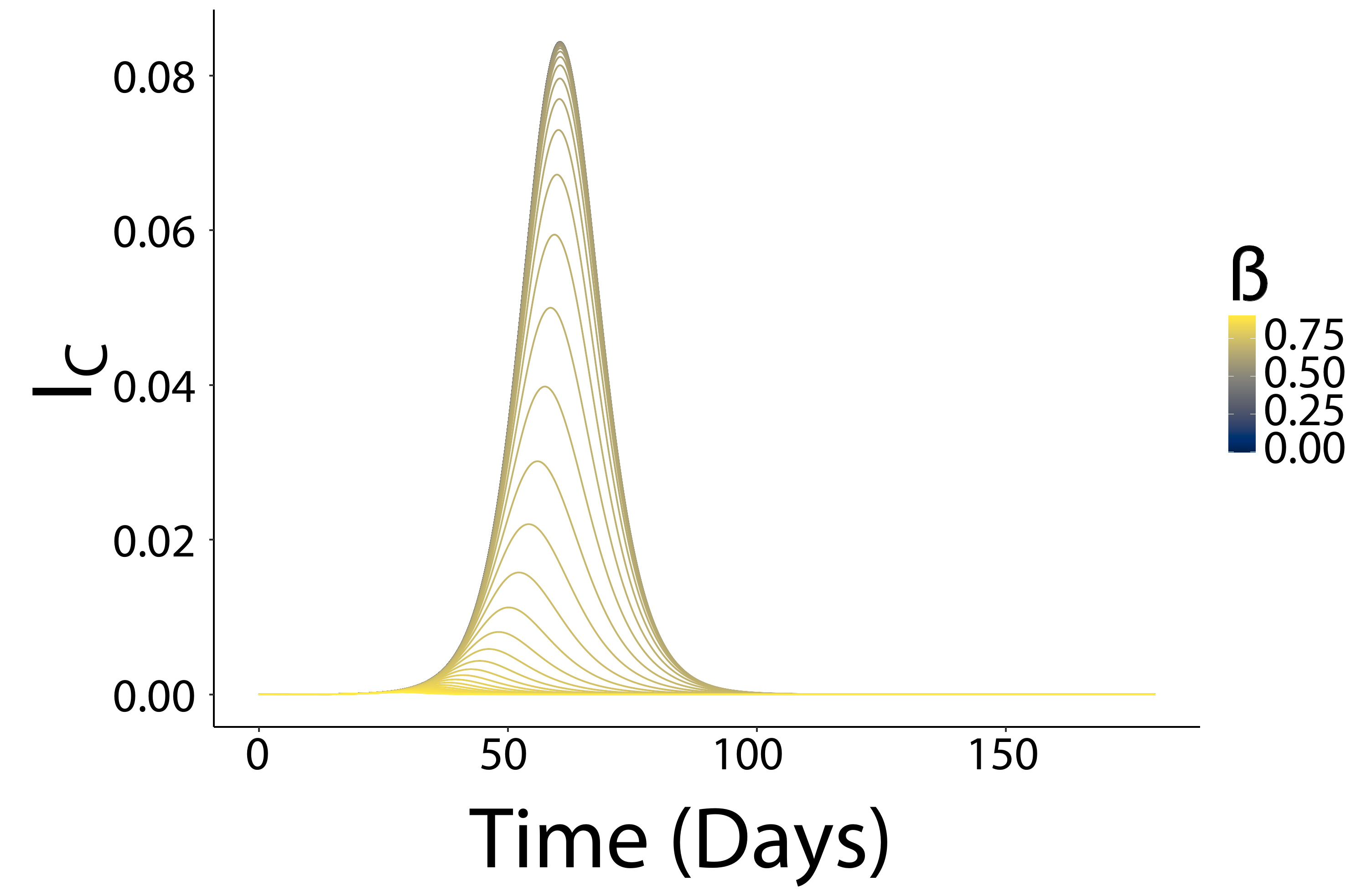}
\label{fig:subfig1ICBeta}}
\subfloat[Subfigure 2 list of figures text][]{
\includegraphics[width=0.32\textwidth]{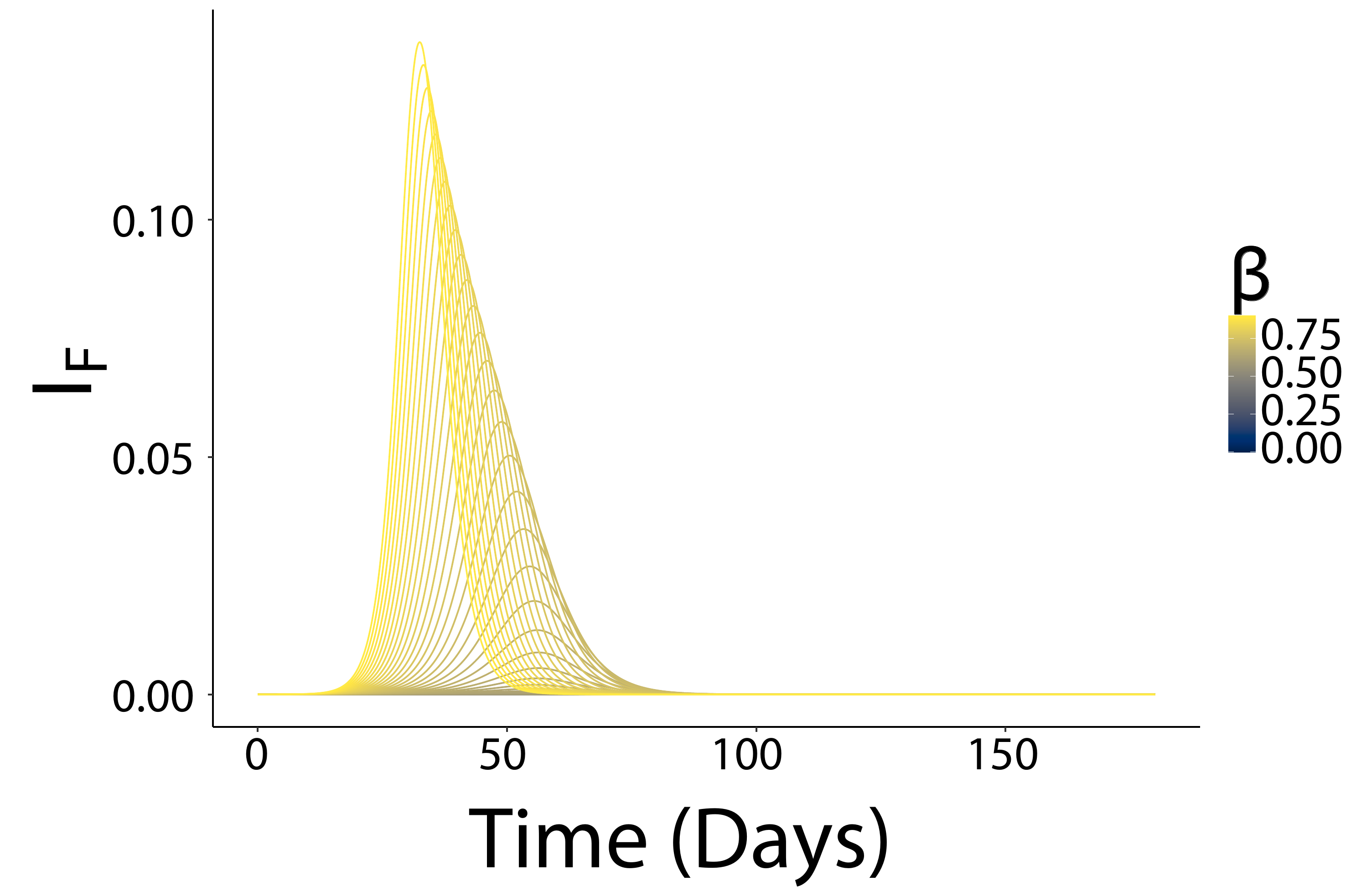}
\label{fig:subfig2IFBeta}}
\subfloat[Subfigure 3 list of figures text][]{
\includegraphics[width=0.32\textwidth]{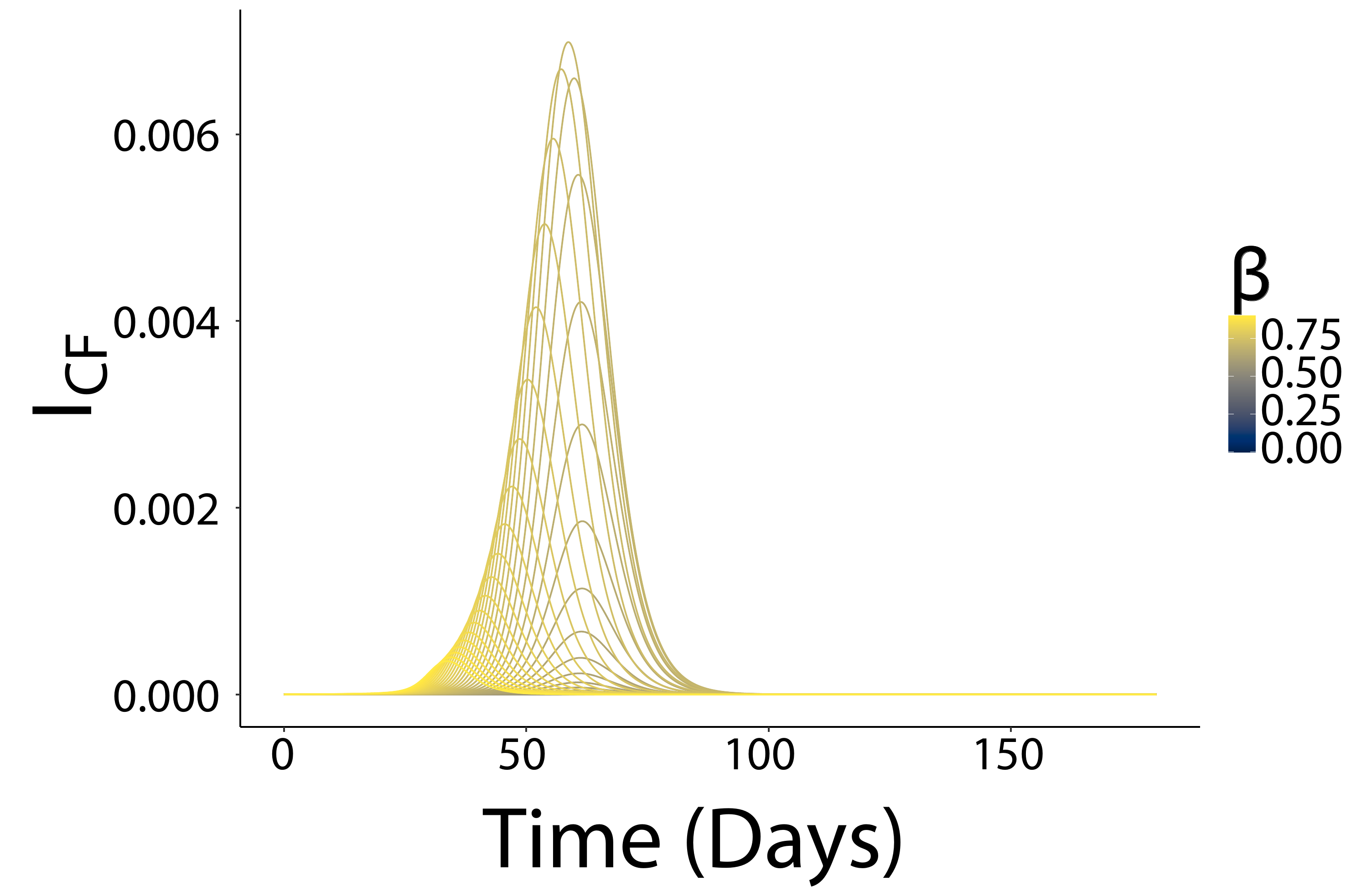}
\label{fig:subfig3ICFBeta}}
\caption{Influence of $\beta$ on the prevalence of Covid-19, ILI and co-infected population.}
\label{fig:globfigbeta}
\end{figure}

\begin{figure}[H]
\centering
\subfloat[Subfigure 1 list of figures text][]{
\includegraphics[width=0.32\textwidth]{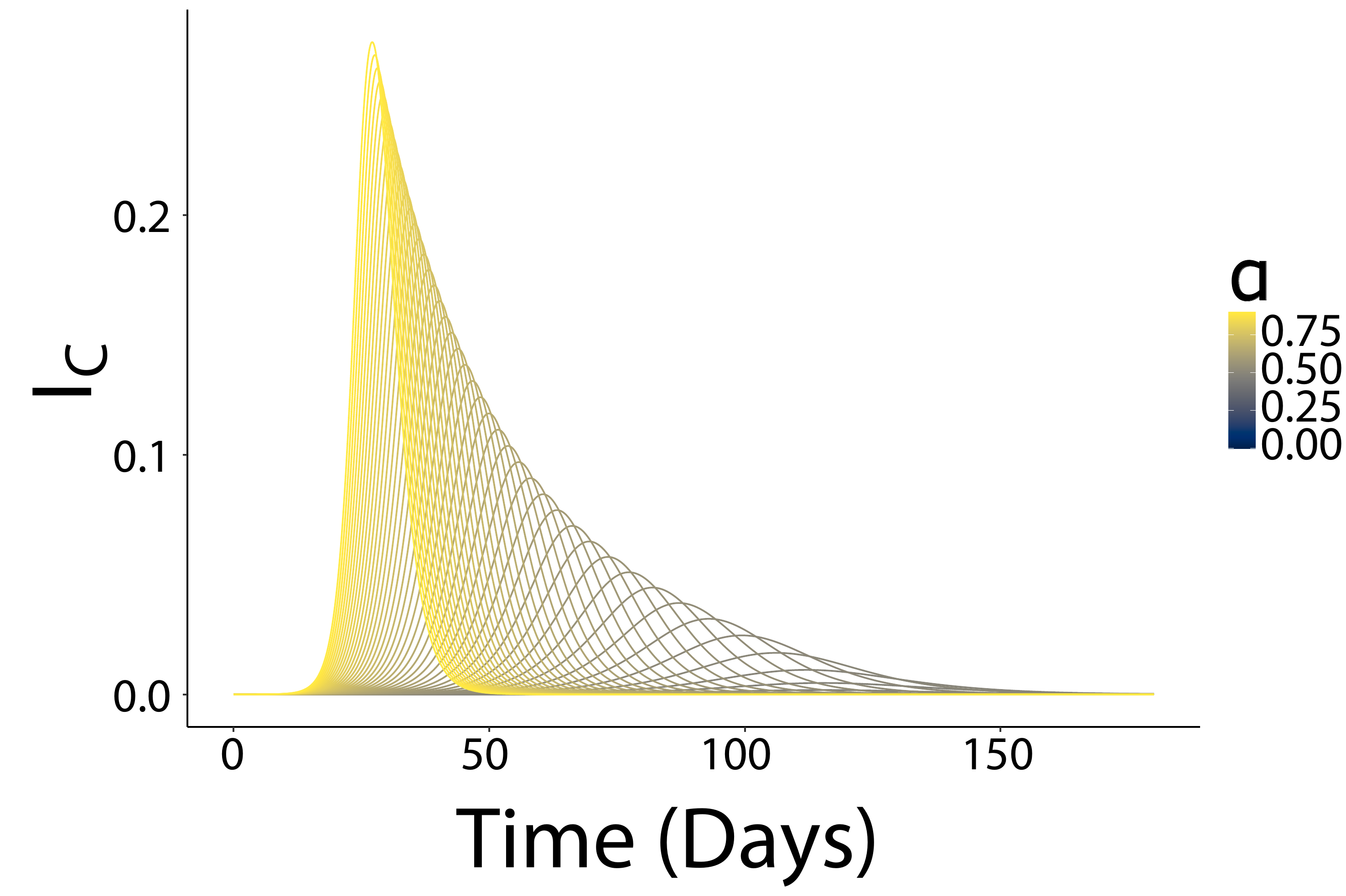}
\label{fig:subfig1ICAlpha}}
\subfloat[Subfigure 2 list of figures text][]{
\includegraphics[width=0.32\textwidth]{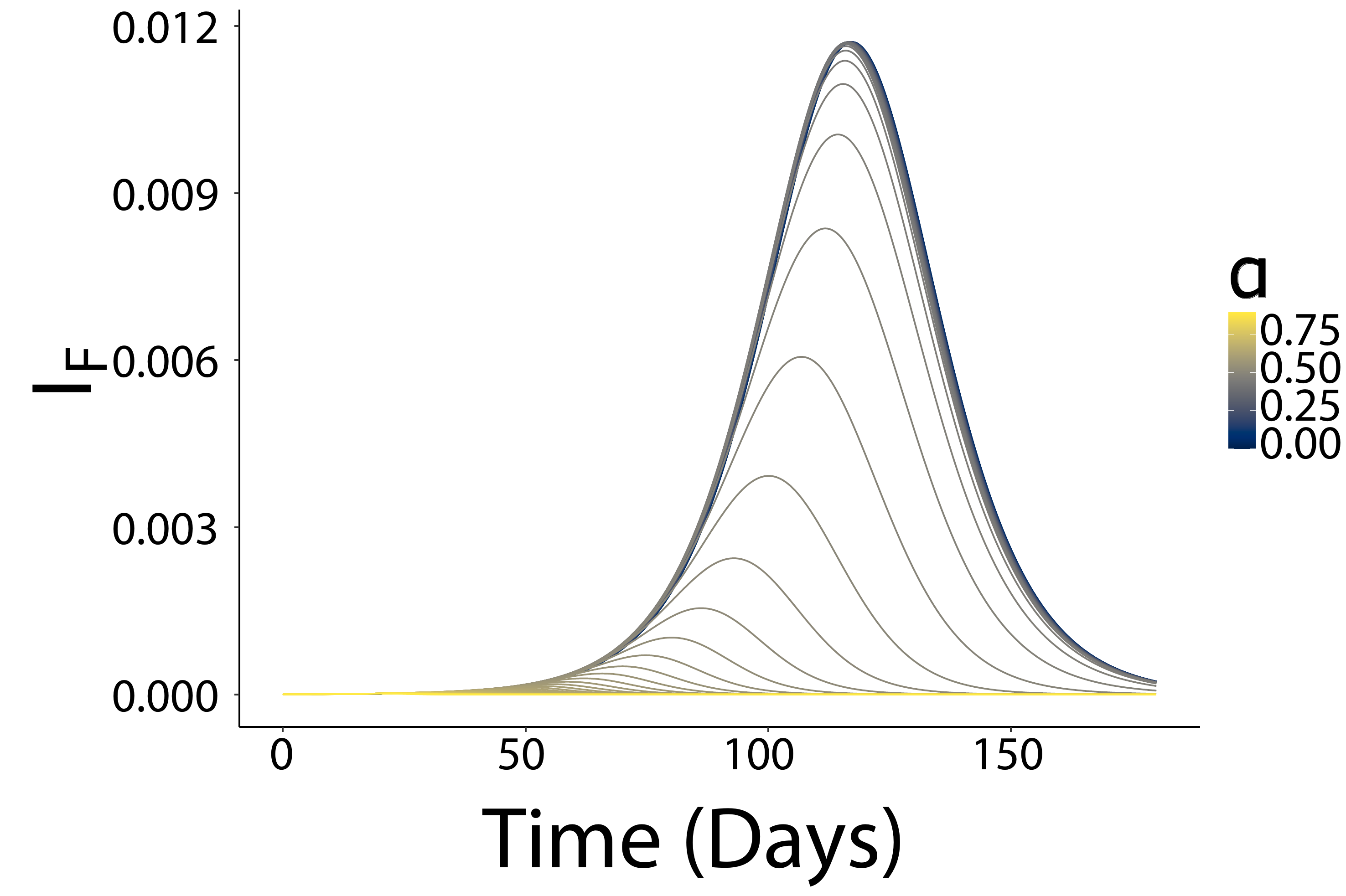}
\label{fig:subfig2IFAlpha}}
\subfloat[Subfigure 3 list of figures text][]{
\includegraphics[width=0.32\textwidth]{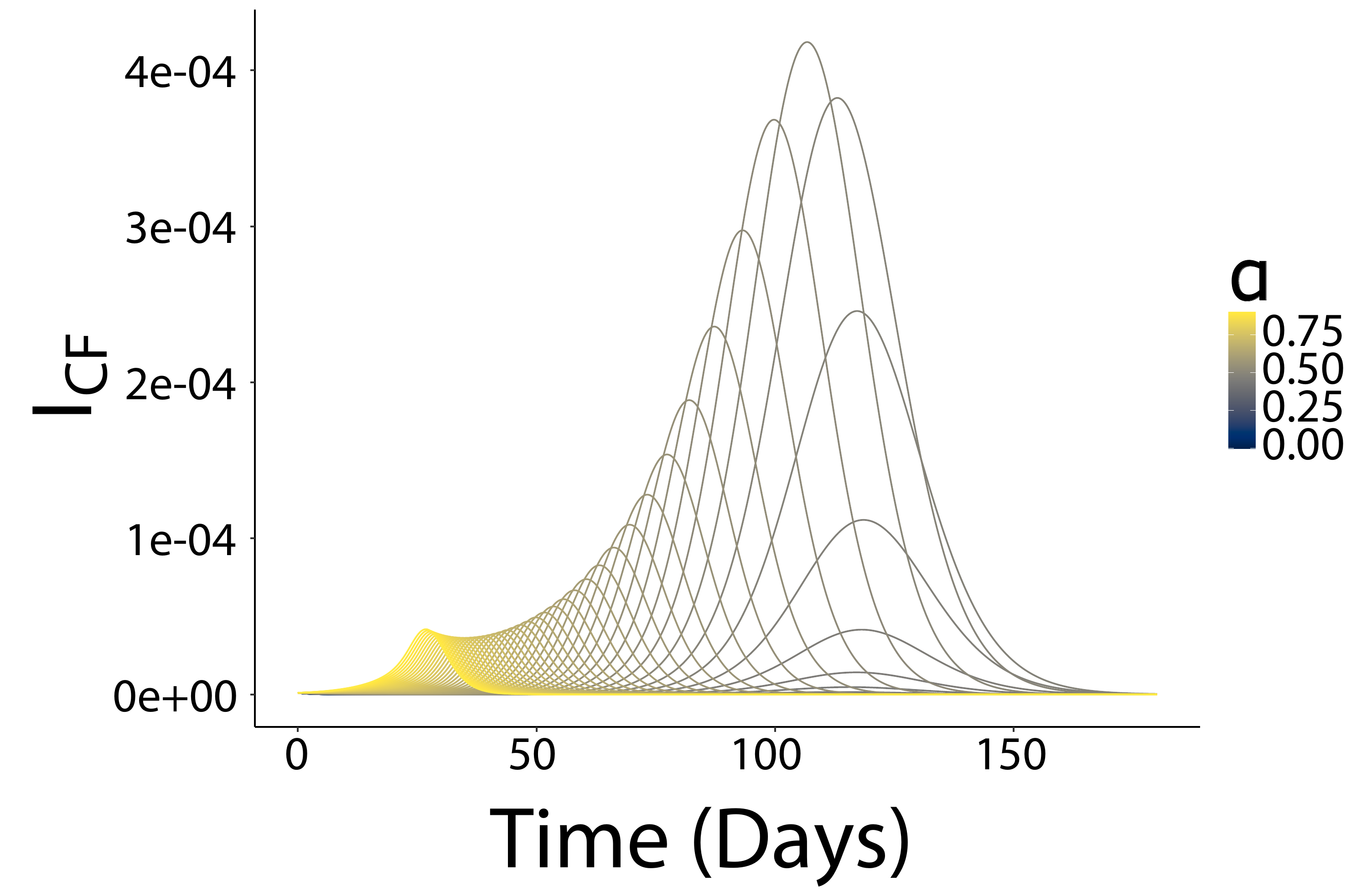}
\label{fig:subfig3ICFAlpha}}
\caption{Influence of $\alpha$ on the prevalence of Covid-19, ILI and co-infected population.}
\label{fig:globfigAlpha}
\end{figure}
For the demonstration purpose, we have chosen the initial conditions to be $S(0) = 995$, $I_C(0) =5$, $I_F(0) = 2$, $I_{CF}(0) = 3$, $R_C(0) = 0$, $R_F(0) = 0$ and $R_{CF} (0) = 0$.
Parameter values are from the Table: (Number).
Infection profiles of the model’s state variables are depicted in the Figure: \ref{fig:globfigbeta} and Figure: \ref{fig:globfigAlpha} wherein it is visible the influence of transmission parameters ($\alpha$ and $\beta$) on the prevalence of Covid-19, ILI and co-infection. 
With the increase in the transmission rates, the sharp surge in infected population is noticeable and similarly with the lower magnitude of transmission rates, reduction in the infected population can give us the assurance of the measurements that are taken account to reduce the spread of Covid-19.
It is interesting to the note that the dynamics of the model is not symmetric (see Figure: \ref{fig:globfigbeta} and Figure: \ref{fig:globfigAlpha}) to the changes performed in two different transmission parameters in spite of being symmetric in the model formulations.
Figure:\ref{fig:globfigAlpha} shows us that higher transmissibility of Covid-19, can create a sudden upsurge of cases of Covid-19 and co-infected people.
The interplay between the transmissibility of Covid-19 and ILI are very interesting to notice and it is being reflected in the dynamics of co-infected population. 
Even the low level of transmissibility of Covid-19, can be a hindrance for the eradication of co-infected cases whereas the influence of transmissibility of ILI is not equally importance.
Therefore, it is of utmost importance to curb the value of $\alpha$ to protect the vulnerable population.

\subsection{Impact of cure rate}
In order to further analyse the model \eqref{Co-infModelSys}, we have chosen to observe the influence of cure rate on compartments that include Covid-19, ILI and co-infected population.

\begin{figure}[H]
\centering
\subfloat[Subfigure 1 list of figures text][]{
\includegraphics[width=0.32\textwidth]{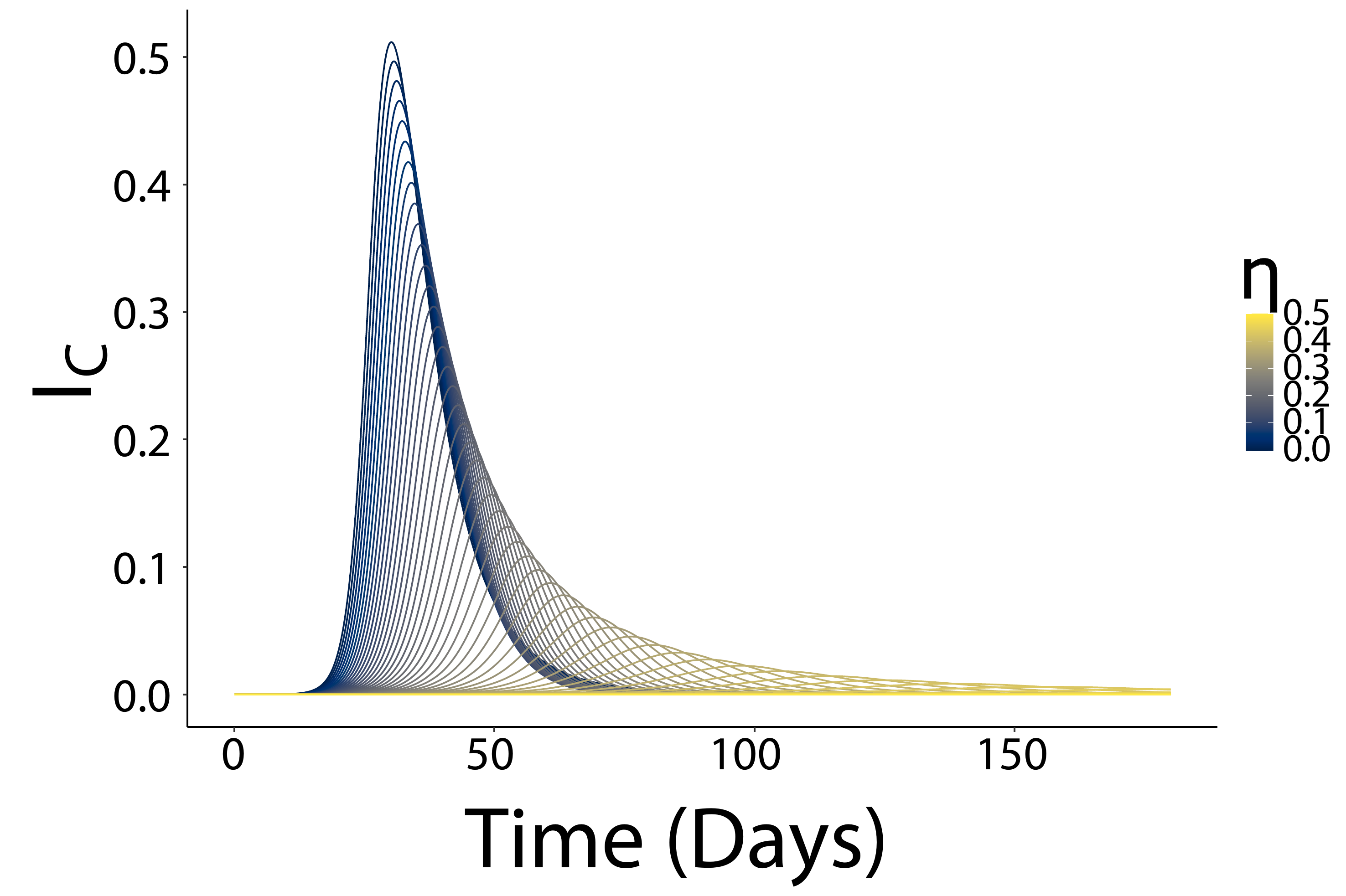}
\label{fig:subfig1ICeta}}
\subfloat[Subfigure 2 list of figures text][]{
\includegraphics[width=0.32\textwidth]{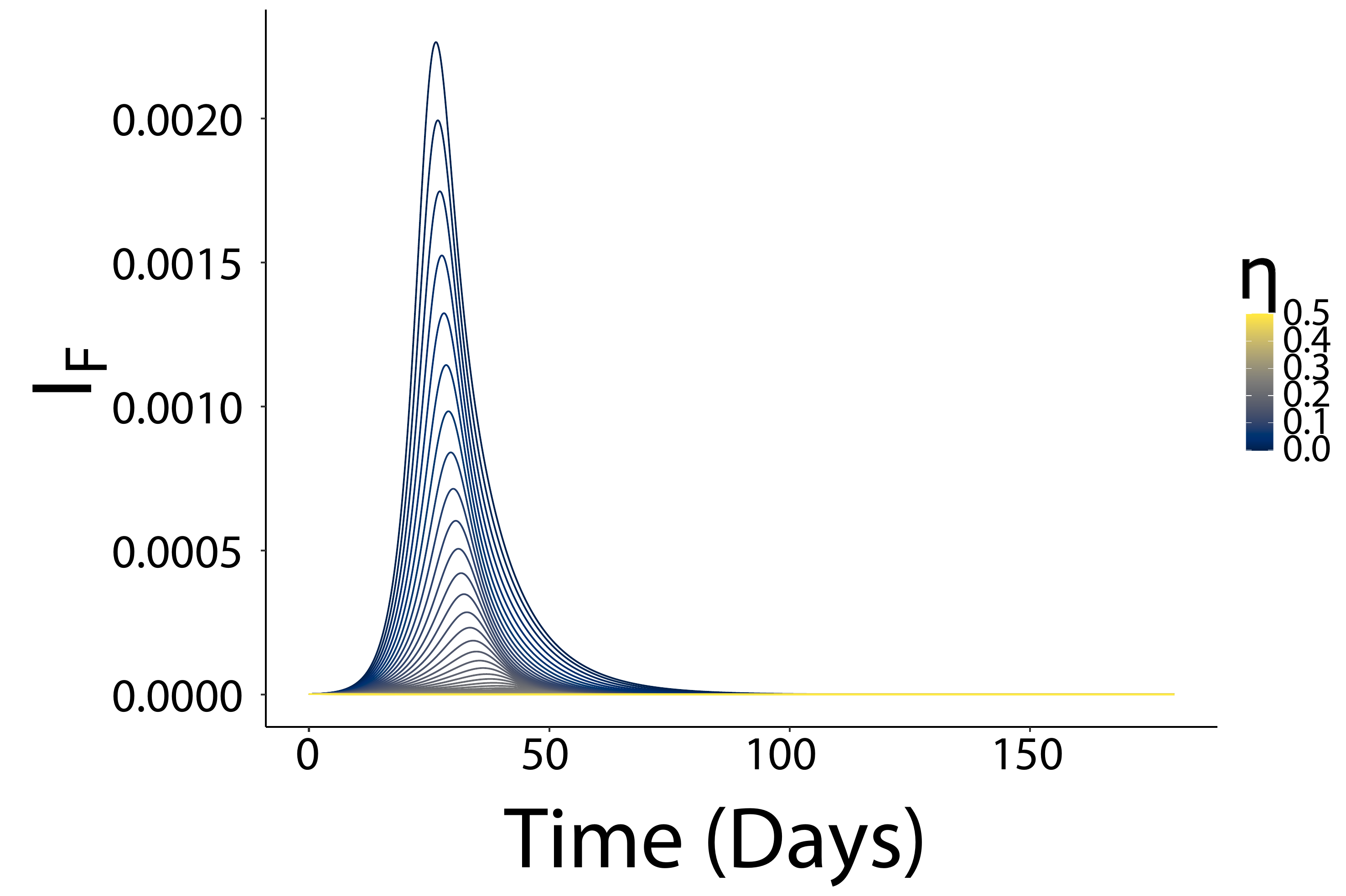}
\label{fig:subfig2IFeta}}
\subfloat[Subfigure 3 list of figures text][]{
\includegraphics[width=0.32\textwidth]{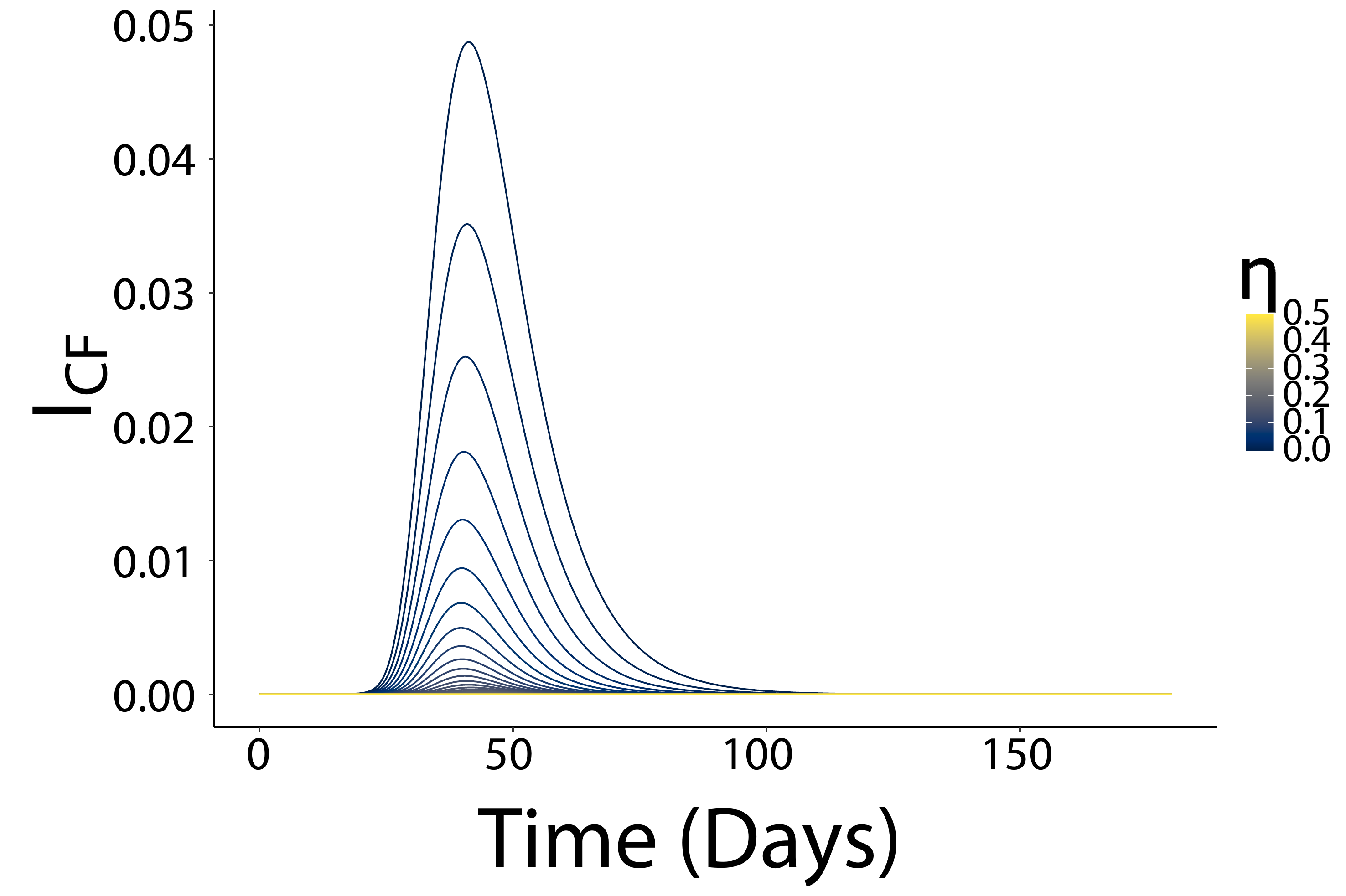}
\label{fig:subfig3ICFeta}}
\caption{The prevalence of Covid-19, ILI and co-infected population.}
\label{fig:globfigEta}
\end{figure}

Dynamical behaviour of these compartments is interesting to note in the constant changes of $\eta$.
From the Figure: \ref{fig:globfigEta}, it is noticeable the changes in the cure rate can bring the cases of Covid-19, ILI down and consequently the cases of co-infection can also be brought down accordingly. 
Finally, as conceived, we can notice that with the increase in  medical facilities like test kits, ventilators, nursing facilities, efficiency of treatment etc, the infected population ($I_C, I_F, I_{CF}$) get recovered and a substantial reduction happens in the infected population.
This observation can lead us to conclude that the better cure rate should be considered as an effective tool to reduce the infection amongst the susceptible population.

\subsection{Time series}
\begin{figure}[H]
\centering
\subfloat[Subfigure 1 list of figures text][]{
\includegraphics[width=0.3\textwidth]{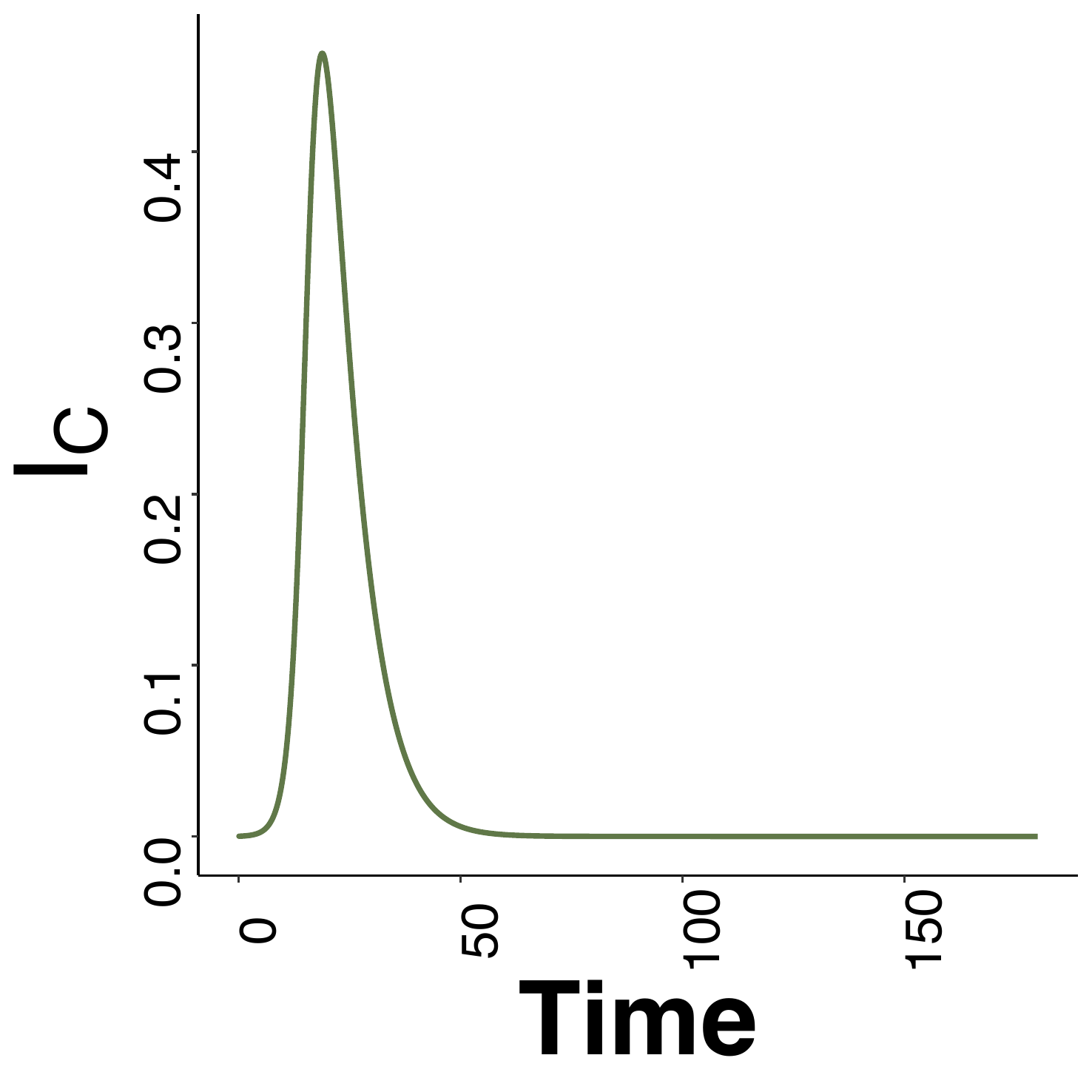}
\label{fig:subfigTmSers1A}}
\subfloat[Subfigure 2 list of figures text][]{
\includegraphics[width=0.3\textwidth]{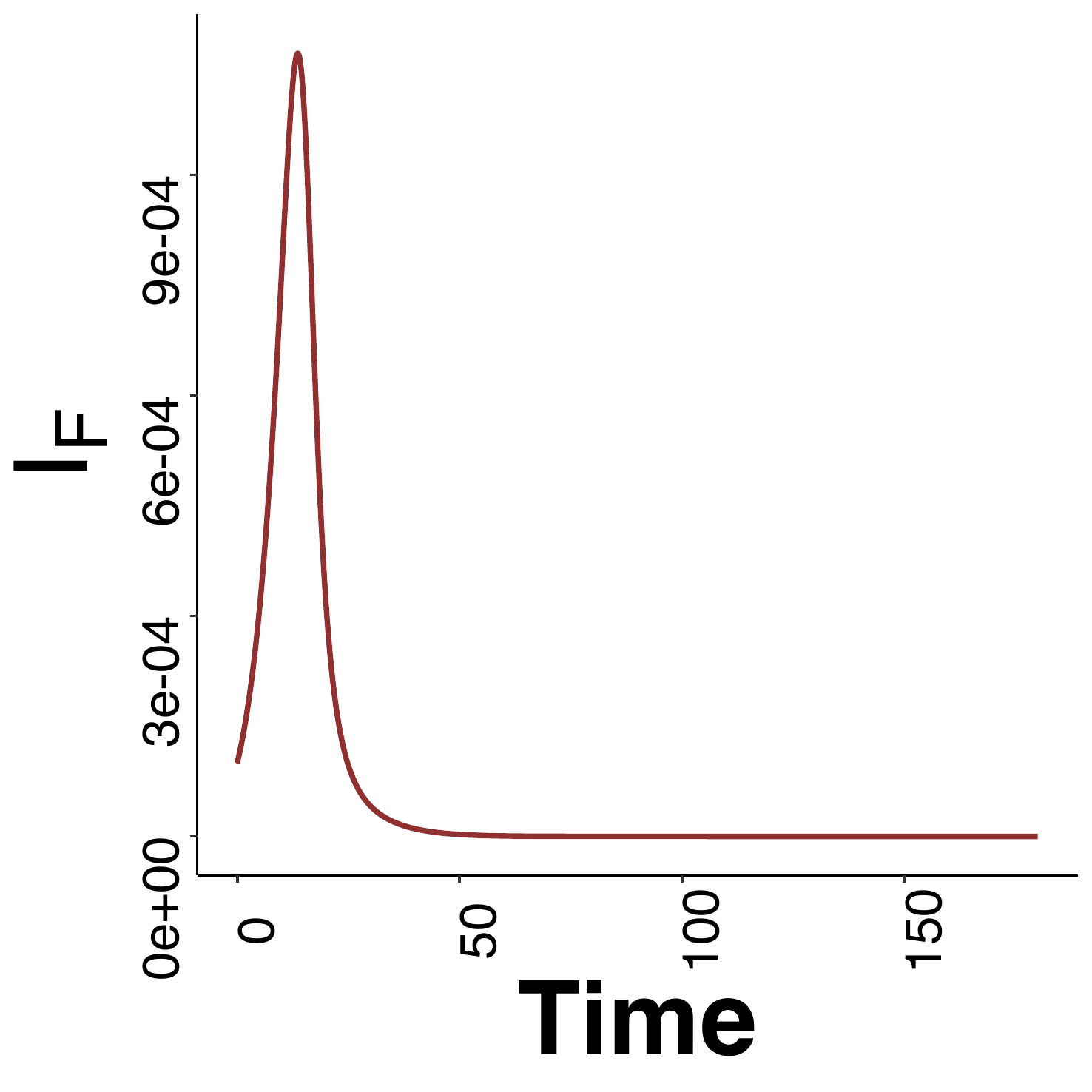}
\label{fig:subfigTmSers2A}}
\subfloat[Subfigure 3 list of figures text][]{
\includegraphics[width=0.3\textwidth]{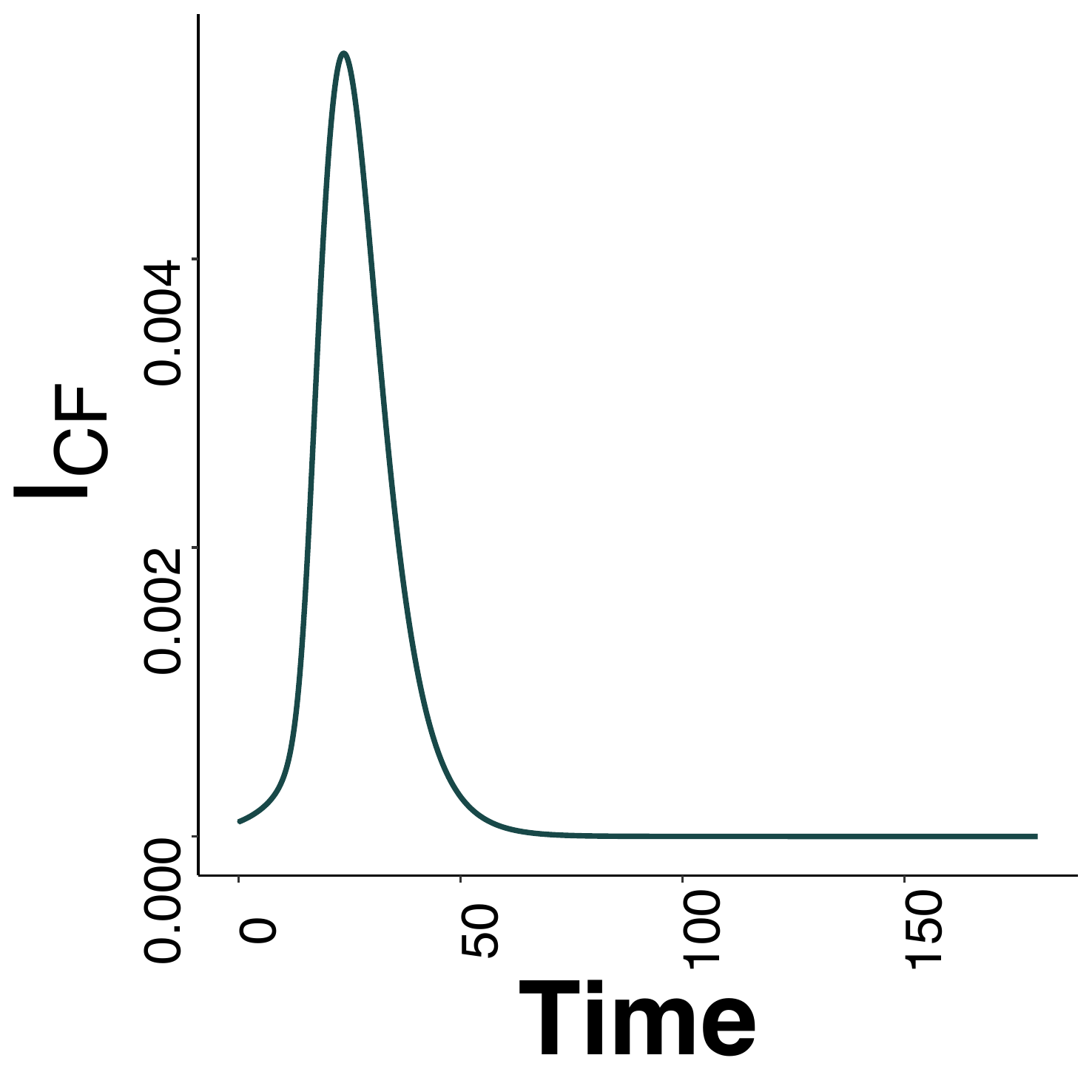}
\label{fig:subfigTmSers3A}}
\caption{The prevalence of Covid-19, ILI and co-infected population.}
\label{fig:TmSers1A}
\end{figure}

\begin{figure}[H]
\centering
\subfloat[Subfigure 1 list of figures text][]{
\includegraphics[width=0.3\textwidth]{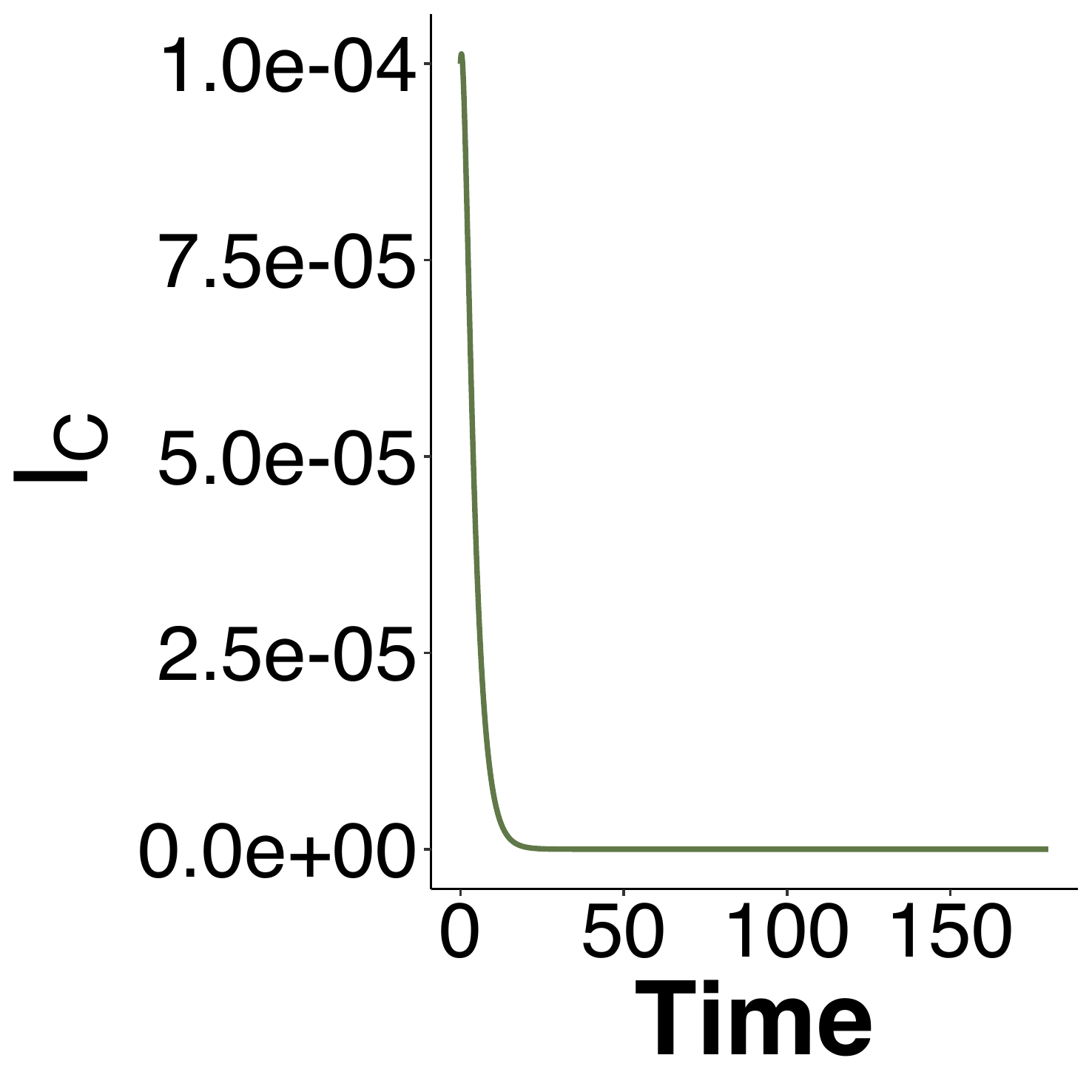}
\label{fig:subfigTmSers1B}}
\subfloat[Subfigure 2 list of figures text][]{
\includegraphics[width=0.3\textwidth]{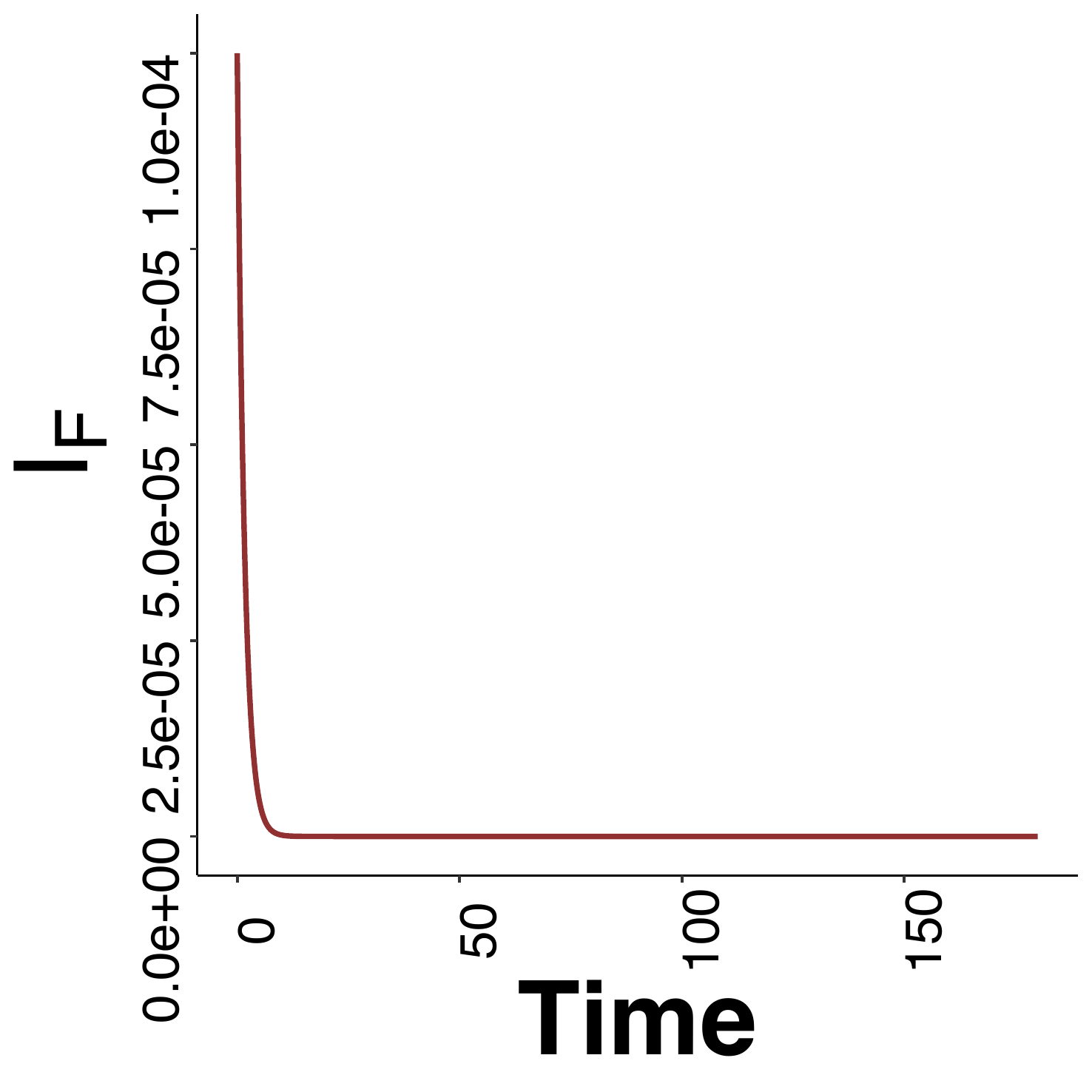}
\label{fig:subfigTmSers2B}}
\subfloat[Subfigure 3 list of figures text][]{
\includegraphics[width=0.3\textwidth]{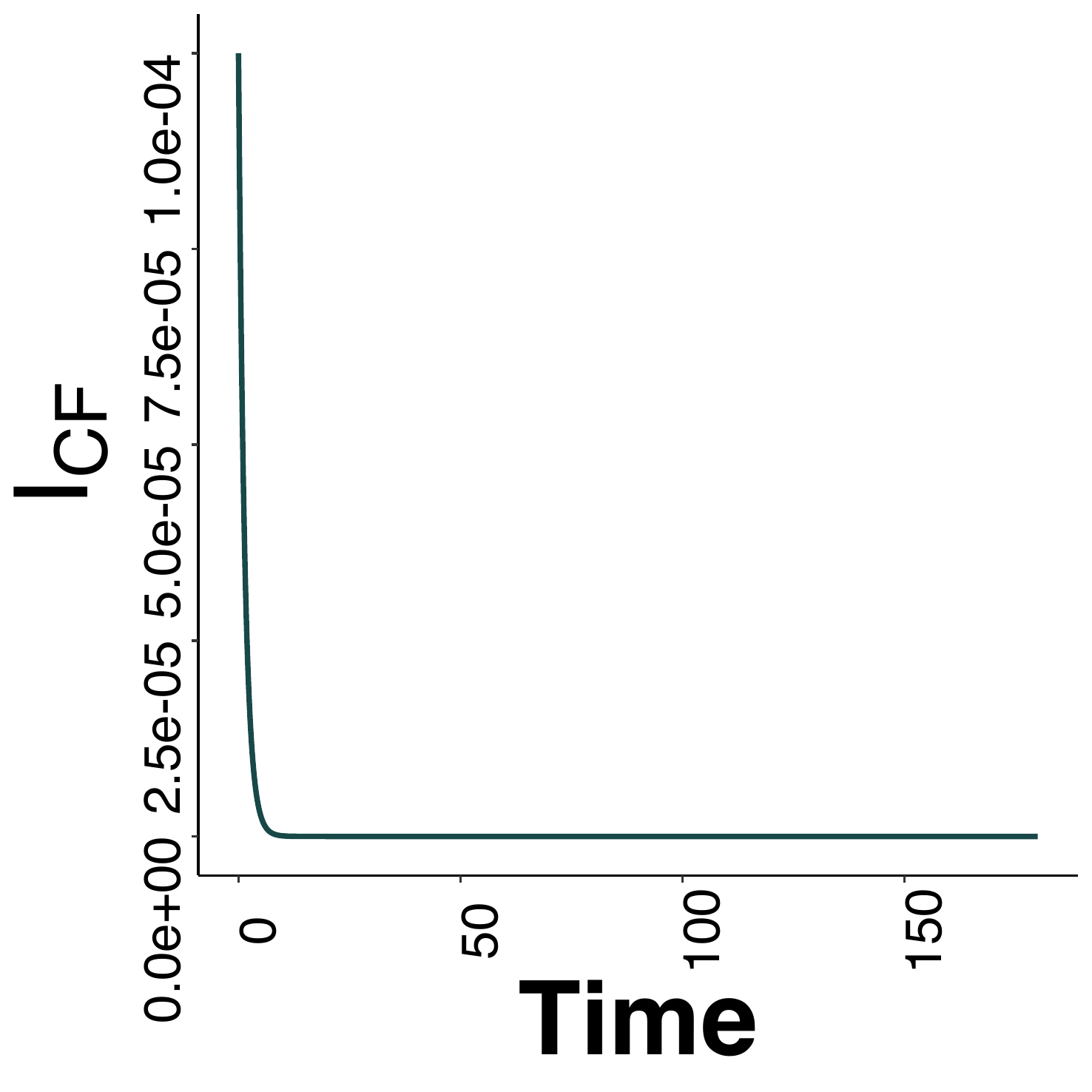}
\label{fig:subfigTmSers3B}}
\caption{The prevalence of Covid-19, ILI and co-infected population.}
\label{fig:TmSers2B}
\end{figure}

Investigations in the time series can  allow us to get further inside in the consequences of the cases when the values of $R_0$ is greater than one or less than one.
The situation when $R_0 < 1$, the disease-free equilibrium point is the only attractor for the system whereas when $R_0 > 1$,
then the endemic equilibrium point. 
Figure: \ref{fig:TmSers2B} depicts the first case and the Figure: \ref{fig:TmSers1A} depicts the latter. 
In this scenario, cases of Covid-19  can be reduced classically by simply increasing the value of $\eta$.

\begin{figure}[H]
\centering
\subfloat[Subfigure 2 list of figures text][]{
\includegraphics[width=0.4\textwidth]{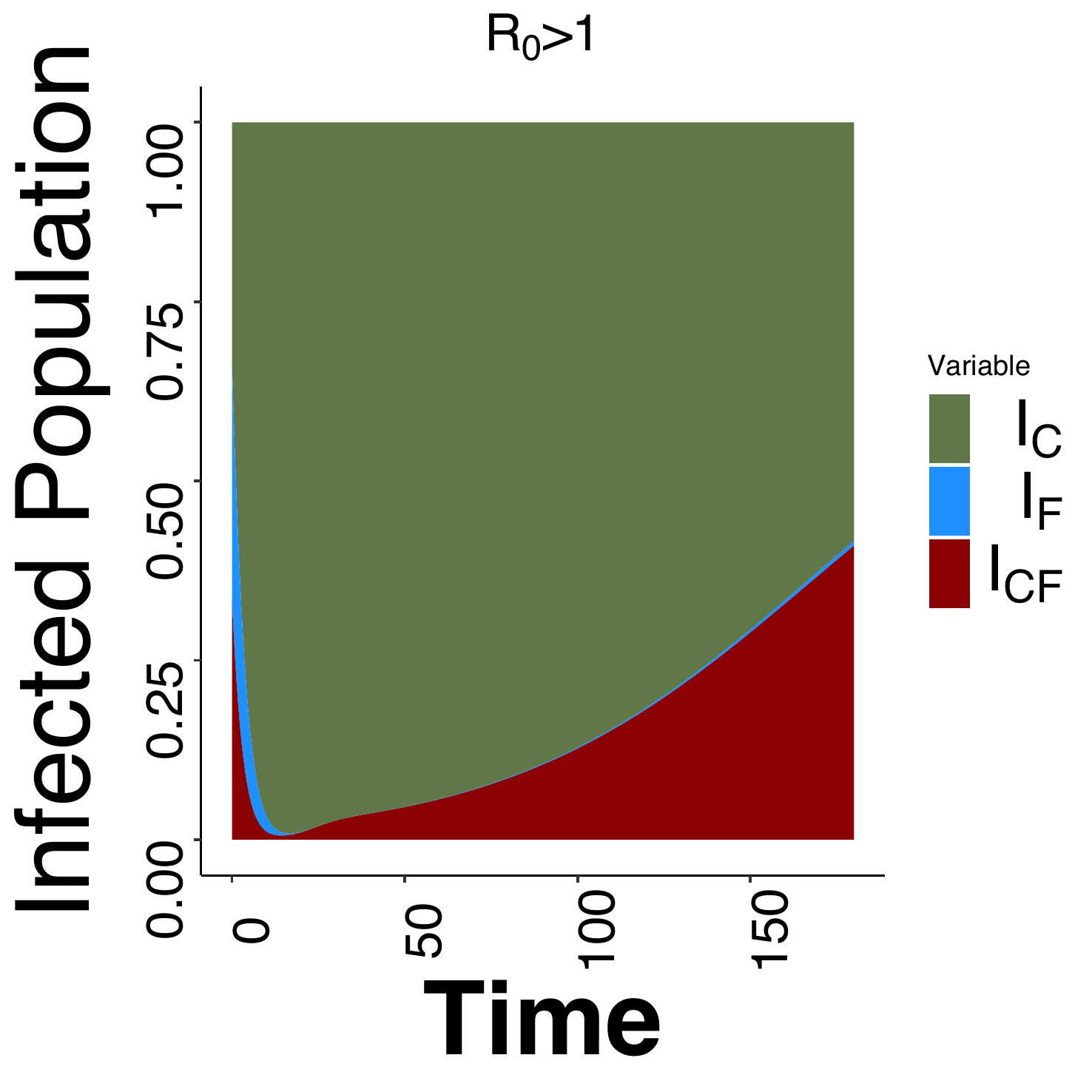}
\label{fig:subfigTmSers2C}}
\subfloat[Subfigure 3 list of figures text][]{
\includegraphics[width=0.4\textwidth]{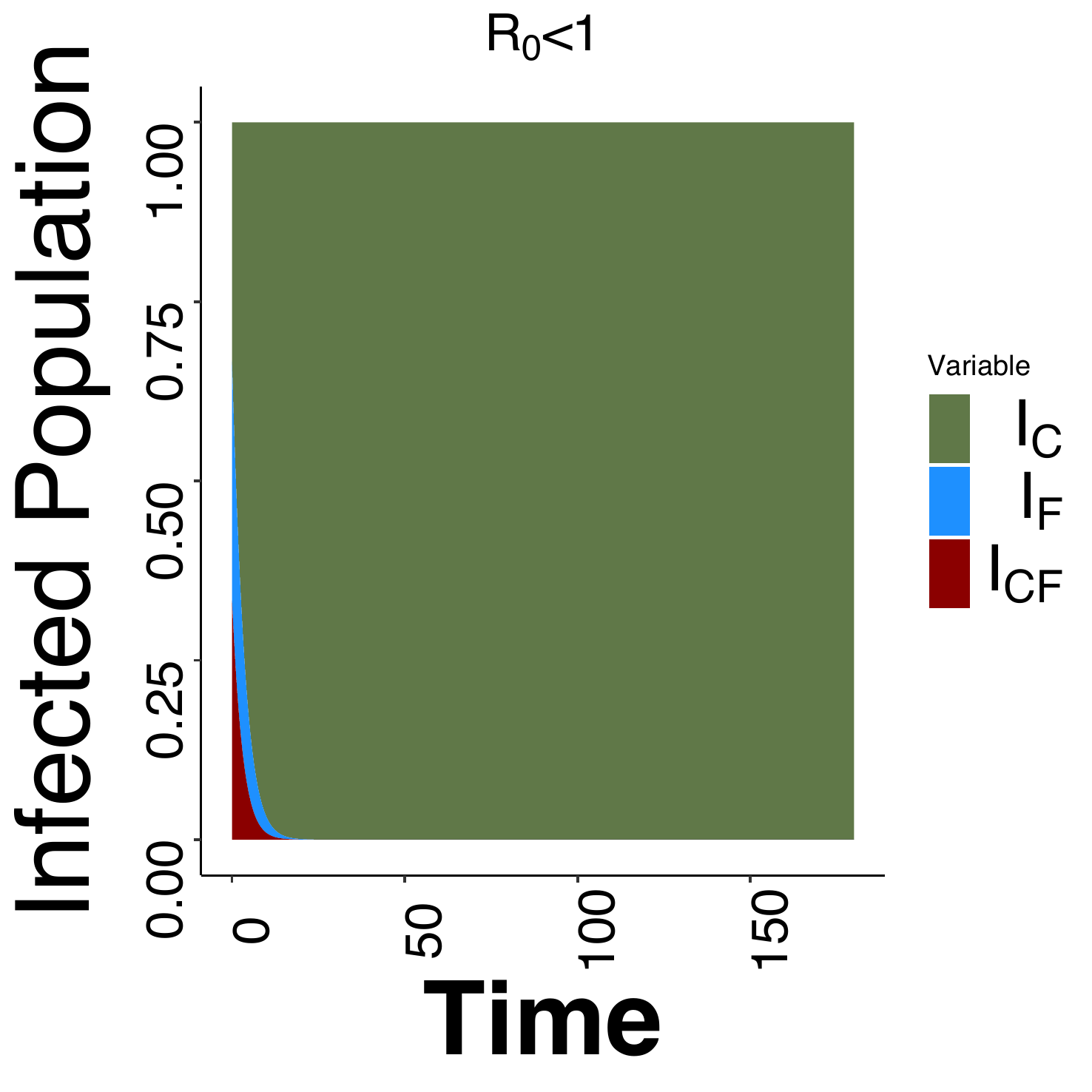}
\label{fig:subfigTmSers3C}}
\caption{The prevalence of Covid-19, ILI and co-infected population.}
\label{fig:TmSers2C}
\end{figure}

Additionally, in Figure: \ref{fig:TmSers2C}, we have shown the prevalence of the infected population with Covid-19, ILI and co-infection under two different situations.
It is clear from the Figure: \ref{fig:TmSers2C} that if  the value of $R_0$ is greater than $1$ then the prevalence of Covid-19, ILI and co-infected people are non-zero and the burden of people infected with Covid-19 and co-infected is higher compared to the people infected with ILI only and the same holds true when we let the value of  $R_0$ is to be less than $1$.

\subsection{Sensitivity Analysis}
Using sensitivity analysis, we can recognise which parameters are important  providing the heterogeneity in the outcome of the basic reproduction number ($R_0$).
In our effort we focus primarily on global sensitivity analysis (GSA) which  investigates the response of model output variables to the model parameters variation. 

\begin{figure}[H]
\centering
\subfloat[Subfigure 1 list of figures text][Sobol indices]{
\includegraphics[width=0.5\textwidth]{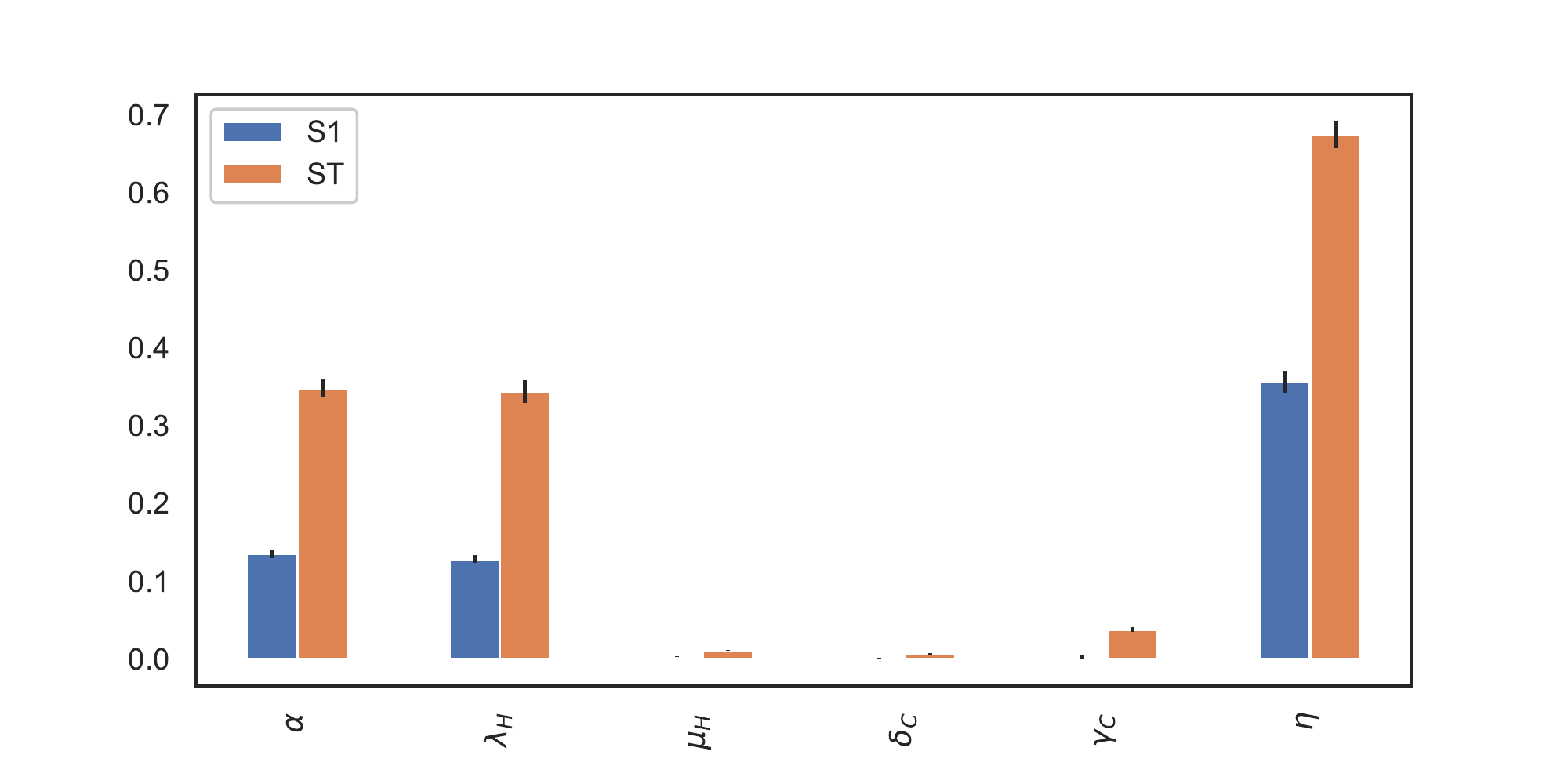}
\label{fig:Sobol1}}
%\qquad
\subfloat[Subfigure 2 list of figures text][Second-order interactions]{
\includegraphics[width=0.4\textwidth]{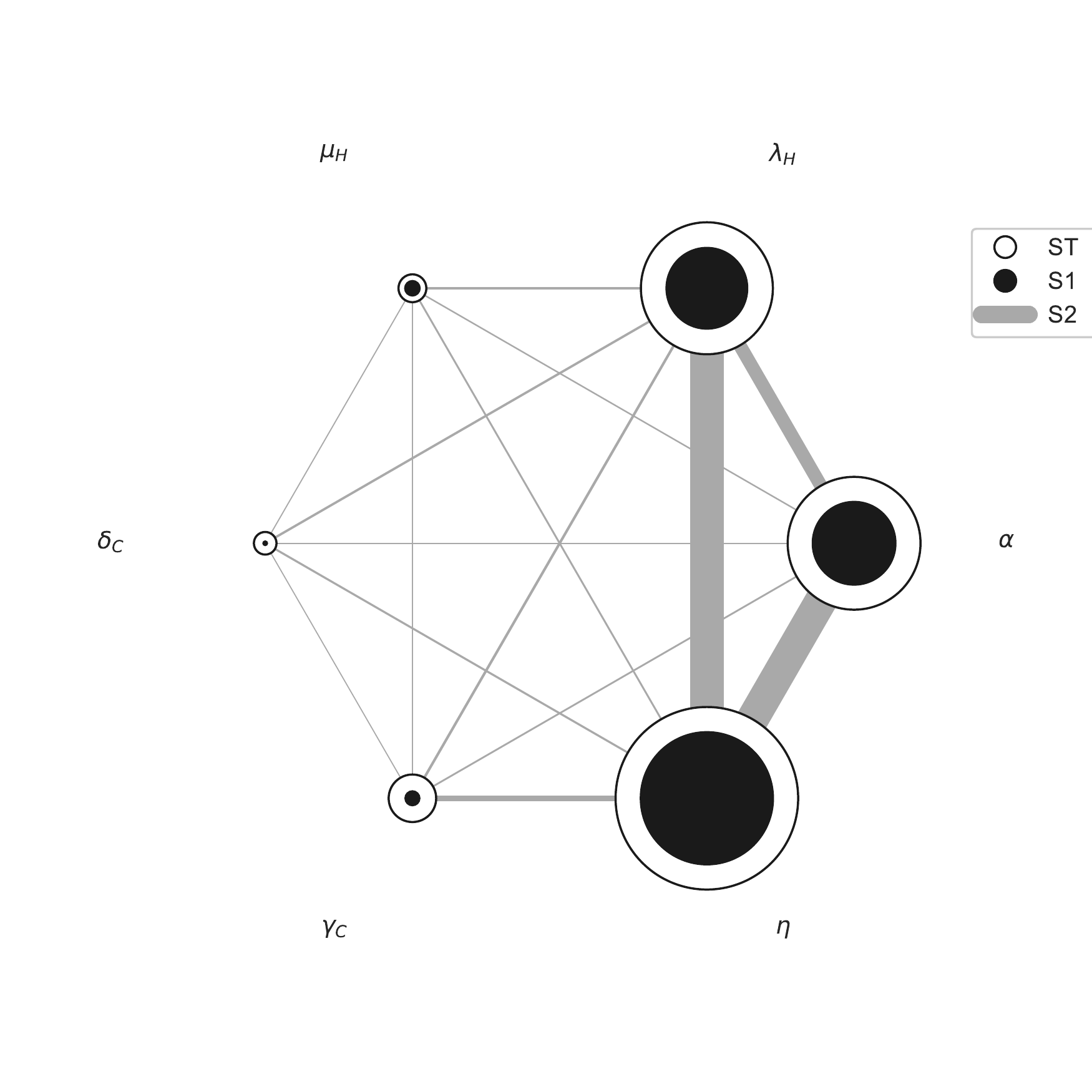}
\label{fig:Sobol2}}
\caption{Sobol' indices}
\label{fig:Sobol}
\end{figure}
We have used Sobol' method as we would like to gather the information about the response of  $R_0$ to a certain parameter while all interactions are taken into the account.
Given the higher value of transmissibility of Covid-19 \cite{Peterson}, we choose to investigate $R_{0_{C}}$ for the GSA in our simulation.
Figure:\ref{fig:Sobol1} reveals that $\eta$ has the highest $\mathrm{ST}$ index and it indicates that it contributes over $65\%$ of $R_{0_{C}}$ variance while accounting for interactions with other parameters, followed by $\alpha$ and $\lambda_H$.
It is to be noted that particularly for the S1 index, it indicates each input’s individual contribution to variance.
We use a more sophisticated visualisation to incorporate the second-order interactions between inputs estimated from the S2 values.
In this case $\eta$ has has strong interactions with $\alpha$ and $\lambda_H$ and then followed by the interaction between 
$\alpha$ and $\lambda_H$.
Here, the size of the $\mathrm{ST}$ and $\mathrm{S1}$ circles correlate with the normalised variable importances.
Figure:\ref{fig:Sobol} delivers a few important messages that shall be the matter of concern in the context of limited medical resources. 
Based on Sobol' sensitivity index, $\mathrm{Si}$, where $i = 1, 2$ three parameters are found to be primary contributors to the variance of transmissibility of Covid-19.
It gives the information about the contributions of $\eta$, $\alpha$ and $\lambda_H$ to the variance of basic reproduction number  and the total effect index, $\mathrm{ST}$ includes both individual contributions and the interaction effects in consideration.
It is evident that the influence of the effective transmission rate of Covid-19 and the treatment facility are very important to lower the value of $R_0$ along with reducing the number of susceptible.
Higher number of susceptible will take toll on the medical resources and henceforth, lack of treatment facility can facilitate higher transmission of Covid-19.

\section{Discussion and Conclusion}

Our work has revealed the co-infection dynamics of Covid-19 and ILI in the presence of non-linear treatment function in different states of India. 
After using our mathematical model with the available data at the initial phase of Covid-19 spread in India, we are able to fit the trajectories of the paths of Covid-19 infection over India and its exponential nature.
By applying numerical simulations, we endeavour to understand the effect of saturated treatment in the course of co-infection. 
After finding the analytical expression of IRN, our current effort build on to the analytical tools which can be employed to understand the multiple-pathogens dynamical system.\par
We have carried out the sensitivity analysis of $R_0$ of the co-infected system with respect to the model parameters.
The outcome identifies three key parameters: the transmission rate of Covid-19: $\alpha$, cure rate: $\eta$ and the recruitment rate: $\pi_H$, play a major role in the co-infection model in the backdrop of a limited medical resource facility.
This suggests the vital roles of lockdown, quarantine, isolation and revamping the medical facilities as the first action shall reduce the number of susceptible people and the second one will make sure that limited medical facilities will not be overwhelmed in India.\par
The decrease in $R_t$ can be attributed to the implementation of public health measures including social distancing, restrictions in mobility, contact tracing etc and a possible improvement in medical care facilities. 
Similar trend of reduction of $R_t$ in India has been observed just like in China or in USA \cite{Yuan, https://doi.org/10.1111/tbed.13868}.
Earlier studies \cite{SENAPATI2021110711, Shil, Nithya} have addressed a varied magnitude of $R_t$ from $1.03$ to $4.18$.
These studies have utilised different data sources and the methods to find $R_t$. 
The time intervals considered in these studies are also different from our endeavour.
So, these estimates can’t be compared readily.\par
One of the highlights of our work is the mathematical formulation of invasion reproduction number \cite{v11121153}, which can mirror the situation whether Covid-19 can invade a susceptible population where another disease is already persistent \cite{Iacobuccim3720, 10.1001/jama.2020.6266, chenz, Burrel, Hazra}, with the aim to investigate the dynamics of co-infection. 
We would like to include that interplay between Covid-19 and ILI has a beneficial impact on the transmission of co-infection in a completely susceptible human population where ILI already exists.
The mathematical expression of  $R_{0_{\mathrm{Inv}}}$ is similar to that of derived in \cite{v11121153}.

\par
It is to be noted that the model considered in this work is a rudimentary mathematical model through which we have focussed to understand the main dynamics of the spread of Covid-19 and ILI in a limited medical facility setting.
Nevertheless, pathogens or the strains of diseases interacting and infecting the hosts may not only promote each other but can also compete or try to be independent.
Therefore, it is interesting to investigate the obstructive effects of co-infection on the Covid-19 transmission after including the evolutionary nature of this virus \cite{doi:10.1098/rspb.1995.0138, Lawton}. 
Furthermore, other factors such as seasonality, spatial heterogeneity will be equally important to address. 

\par
Our modelling foundation is not only tailored to carry out the co-transmission of Covid-19 and ILI, but can be employed to general epidemiological co-transmitting diseases in a limited medical resources.
Predicting the spread of epidemics and developing an early forewarning system for the limited yet demanding healthcare and volume have been put at the foreground of epidemiological modelling. 
It can be achieved through working and planning in a close collaboration with the theoreticians and local authorities. 
As a consequence, it shall bring a unique opportunity to inculcate novel ideas to demanding healthcare options and planning.

\bibliography{BhowmickCo-inf}

\begin{thebibliography}{10}
\expandafter\ifx\csname url\endcsname\relax
  \def\url#1{\texttt{#1}}\fi
\expandafter\ifx\csname urlprefix\endcsname\relax\def\urlprefix{URL }\fi
\expandafter\ifx\csname href\endcsname\relax
  \def\href#1#2{#2} \def\path#1{#1}\fi

\bibitem{doi:10.1164/rccm.2014P7}
W.~G. Carlos, C.~S. Dela~Cruz, B.~Cao, S.~Pasnick, S.~Jamil, Covid-19 disease
  due to sars-cov-2 (novel coronavirus), American Journal of Respiratory and
  Critical Care Medicine 201~(4) (2020) P7--P8.

\bibitem{v12020135}
L.~E. Gralinski, V.~D. Menachery, Return of the coronavirus: 2019-ncov, Viruses
  12~(2) (2020).

\bibitem{LAI2020505}
C.-C. Lai, C.-Y. Wang, P.-R. Hsueh, Co-infections among patients with covid-19:
  The need for combination therapy with non-anti-sars-cov-2 agents?, Journal of
  Microbiology, Immunology and Infection 53~(4) (2020) 505--512.

\bibitem{Cucinotta2020}
D.~Cucinotta, M.~Vanelli, Who declares covid-19 a pandemic, Acta bio-medica :
  Atenei Parmensis 91~(1) (2020) 157--160.

\bibitem{10.1001jama.2020.2565}
Y.~Bai, L.~Yao, T.~Wei, F.~Tian, D.-Y. Jin, L.~Chen, M.~Wang, Presumed
  asymptomatic carrier transmission of covid-19, JAMA 323~(14) (2020)
  1406--1407.

\bibitem{doi:10.1177/2324709620934674}
V.~M. Konala, S.~Adapa, S.~Naramala, A.~Chenna, S.~Lamichhane, P.~R. Garlapati,
  M.~Balla, V.~Gayam, A case series of patients coinfected with influenza and
  covid-19, Journal of Investigative Medicine High Impact Case Reports 8 (2020)
  2324709620934674.

\bibitem{Iacobuccim3720}
G.~Iacobucci, Covid-19: Risk of death more than doubled in people who also had
  flu, english data show, BMJ 370 (2020).

\bibitem{10.1001/jama.2020.6266}
D.~Kim, J.~Quinn, B.~Pinsky, N.~H. Shah, I.~Brown, {Rates of Co-infection
  Between SARS-CoV-2 and Other Respiratory Pathogens}, JAMA 323~(20) (2020)
  2085--2086.

\bibitem{chenz}
N.~Chen, M.~Zhou, X.~Dong, J.~Qu, F.~Gong, Y.~Han, Y.~Qiu, J.~Wang, Y.~Liu,
  Y.~Wei, J.~Xia, T.~Yu, X.~Zhang, L.~Zhang, Epidemiological and clinical
  characteristics of 99 cases of 2019 novel coronavirus pneumonia in wuhan,
  china: a descriptive study, The Lancet 395~(10223) (2020) 507--513.

\bibitem{Hazra}
A.~Hazra, M.~Collison, J.~Pisano, M.~Kumar, C.~Oehler, J.~P. Ridgway,
  Coinfections with sars-cov-2 and other respiratory pathogens, Infection
  control and hospital epidemiology 41~(10) (2020) 1228--1229.

\bibitem{Burrel}
S.~Burrel, P.~Hausfater, M.~Dres, V.~Pourcher, C.-E. Luyt, E.~Teyssou,
  C.~Souli{\'e}, V.~Calvez, A.-G. Marcelin, D.~Boutolleau, Co-infection of
  sars-cov-2 with other respiratory viruses and performance of lower
  respiratory tract samples for the diagnosis of covid-19, International
  Journal of Infectious Diseases 102 (2021) 10--13.

\bibitem{Bandar}
B.~Alosaimi, A.~Naeem, M.~E. Hamed, H.~S. Alkadi, T.~Alanazi, S.~S. Al~Rehily,
  A.~Z. Almutairi, A.~Zafar, Influenza co-infection associated with severity
  and mortality in covid-19 patients, Virology journal 18~(1) (2021) 127--127.

\bibitem{Balraj}
B.~Singh, P.~Kaur, R.-J. Reid, F.~Shamoon, M.~Bikkina, Covid-19 and influenza
  co-infection: Report of three cases, Cureus 12~(8) (2020) e9852--e9852.

\bibitem{10.1093/cid/ciaa1810}
S.~Covin, G.~W. Rutherford, {Coinfection, Severe Acute Respiratory Syndrome
  Coronavirus 2 (SARS-CoV-2), and Influenza: An Evolving Puzzle}, Clinical
  Infectious Diseases 72~(12) (2020) e993--e994.

\bibitem{Belongia1163}
E.~A. Belongia, M.~T. Osterholm, Covid-19 and flu, a perfect storm, Science
  368~(6496) (2020) 1163--1163.

\bibitem{Bai}
L.~Bai, Y.~Zhao, J.~Dong, S.~Liang, M.~Guo, X.~Liu, X.~Wang, Z.~Huang, X.~Sun,
  Z.~Zhang, L.~Dong, Q.~Liu, Y.~Zheng, D.~Niu, M.~Xiang, K.~Song, J.~Ye,
  W.~Zheng, Z.~Tang, M.~Tang, Y.~Zhou, C.~Shen, M.~Dai, L.~Zhou, Y.~Chen,
  H.~Yan, K.~Lan, K.~Xu, Coinfection with influenza a virus enhances sars-cov-2
  infectivity, Cell Research 31~(4) (2021) 395--403.

\bibitem{10.1001/jamahealthforum.2020.0345}
J.~J. Cavallo, D.~A. Donoho, H.~P. Forman, {Hospital Capacity and Operations in
  the Coronavirus Disease 2019 (COVID-19) Pandemic---Planning for the Nth
  Patient}, JAMA Health Forum 1~(3) (2020) e200345--e200345.

\bibitem{sam}
S.~Mirmirani, R.~N. Spivack, Health care system collapse in the united states:
  Capitalist market failure!, De Economist 141~(3) (1993) 419--431.

\bibitem{Gai}
R.~Gai, M.~Tobe, Managing healthcare delivery system to fight the covid-19
  epidemic: experience in japan, Global Health Research and Policy 5~(1) (2020)
  23.

\bibitem{WANG2004775}
W.~Wang, S.~Ruan, Bifurcations in an epidemic model with constant removal rate
  of the infectives, Journal of Mathematical Analysis and Applications 291~(2)
  (2004) 775--793.

\bibitem{Griffinl72}
N.~Griffin, How the lack of government is affecting healthcare in northern
  ireland, BMJ 364 (2019).

\bibitem{ZHOU2012312}
L.~Zhou, M.~Fan, Dynamics of an sir epidemic model with limited medical
  resources revisited, Nonlinear Analysis: Real World Applications 13~(1)
  (2012) 312--324.

\bibitem{Suo}
Z.~Zhonghua, S.~Yaohong, Qualitative analysis of a sir epidemic model with
  saturated treatment rate, Journal of Applied Mathematics and Computing 34~(1)
  (2010) 177--194.

\bibitem{ZHANG2008433}
X.~Zhang, X.~Liu, Backward bifurcation of an epidemic model with saturated
  treatment function, Journal of Mathematical Analysis and Applications 348~(1)
  (2008) 433--443.

\bibitem{Pepita}
P.~Barlow, M.~C. van Schalkwyk, M.~McKee, R.~Labont{\'e}, D.~Stuckler, Covid-19
  and the collapse of global trade: building an effective public health
  response, The Lancet Planetary Health 5~(2) (2021) e102--e107.

\bibitem{Bene}
B.~Armocida, B.~Formenti, S.~Ussai, F.~Palestra, E.~Missoni, The italian health
  system and the covid-19 challenge, The Lancet Public Health 5~(5) (2020)
  e253.

\bibitem{BLYUSS2005177}
K.~B. Blyuss, Y.~N. Kyrychko, On a basic model of a two-disease epidemic,
  Applied Mathematics and Computation 160~(1) (2005) 177--187.

\bibitem{MALLELA2016143}
A.~Mallela, S.~Lenhart, N.~K. Vaidya, Hiv--tb co-infection treatment: Modeling
  and optimal control theory perspectives, Journal of Computational and Applied
  Mathematics 307 (2016) 143--161.

\bibitem{ASADUZZAMAN20151}
S.~M. Asaduzzaman, J.~Ma, P.~van~den Driessche, The coexistence or replacement
  of two subtypes of influenza, Mathematical Biosciences 270 (2015) 1--9.

\bibitem{Til}
G.~T. Tilahun, Modeling co-dynamics of pneumonia and meningitis diseases,
  Advances in Difference Equations 2019~(1) (2019) 149.

\bibitem{GAO2016171}
D.~Gao, T.~C. Porco, S.~Ruan, Coinfection dynamics of two diseases in a single
  host population, Journal of Mathematical Analysis and Applications 442~(1)
  (2016) 171--188.

\bibitem{Tang}
B.~Tang, Y.~Xiao, J.~Wu, Implication of vaccination against dengue for zika
  outbreak, Scientific Reports 6~(1) (2016) 35623.

\bibitem{MERLER2008499}
S.~Merler, P.~Poletti, M.~Ajelli, B.~Caprile, P.~Manfredi, Coinfection can
  trigger multiple pandemic waves, Journal of Theoretical Biology 254~(2)
  (2008) 499--507.

\bibitem{Cacci}
G.~Cacciapaglia, C.~Cot, F.~Sannino, Multiwave pandemic dynamics explained: how
  to tame the next wave of infectious diseases, Scientific Reports 11~(1)
  (2021) 6638.

\bibitem{Fisayo332}
T.~Fisayo, S.~Tsukagoshi, Three waves of the covid-19 pandemic, Postgraduate
  Medical Journal 97~(1147) (2021) 332--332.

\bibitem{doi:10.1098/rsif.2009.0386}
O.~Diekmann, J.~A.~P. Heesterbeek, M.~G. Roberts, The construction of
  next-generation matrices for compartmental epidemic models, Journal of The
  Royal Society Interface 7~(47) (2010) 873--885.

\bibitem{OLAWOYIN201944}
O.~Olawoyin, C.~Kribs, Invasion reproductive numbers for discrete-time models,
  Infectious Disease Modelling 4 (2019) 44--72.

\bibitem{NUNO200720}
M.~Nu{\~n}o, G.~Chowell, X.~Wang, C.~Castillo-Chavez, On the role of
  cross-immunity and vaccines on the survival of less fit flu-strains,
  Theoretical Population Biology 71~(1) (2007) 20--29.

\bibitem{1531-3492_2009_2_279}
C.~M. K.-Z. Britnee~Crawford, The impact of vaccination and coinfection on hpv
  and cervical cancer, Discrete \& Continuous Dynamical Systems - B 12~(2)
  (2009) 279--304.

\bibitem{SARDAR2020110078}
T.~Sardar, S.~S. Nadim, S.~Rana, J.~Chattopadhyay, Assessment of lockdown
  effect in some states and overall india: A predictive mathematical study on
  covid-19 outbreak, Chaos, Solitons \& Fractals 139 (2020) 110078.

\bibitem{MAHAJAN2020110156}
A.~Mahajan, N.~A. Sivadas, R.~Solanki, An epidemic model sipherd and its
  application for prediction of the spread of covid-19 infection in india,
  Chaos, Solitons \& Fractals 140 (2020) 110156.

\bibitem{Blyuss}
K.~Blyuss, C.~W. Kanyiri, K.~Mark, L.~Luboobi, Mathematical analysis of
  influenza a dynamics in the emergence of drug resistance, Computational and
  Mathematical Methods in Medicine 2018 (2018) 2434560.

\bibitem{Kharis_2018}
M.~Kharis, Amidi, Mathematical modeling of avian influenza epidemic with bird
  vaccination in constant population, Journal of Physics: Conference Series 983
  (2018) 012116.

\bibitem{NADIM2020110163}
S.~S. Nadim, J.~Chattopadhyay, Occurrence of backward bifurcation and
  prediction of disease transmission with imperfect lockdown: A case study on
  covid-19, Chaos, Solitons \& Fractals 140 (2020) 110163.

\bibitem{Sezer}
M.~Sezer, I.~M. Wangari, L.~Stone, Analysis of a heroin epidemic model with
  saturated treatment function, Journal of Applied Mathematics 2017 (2017)
  1953036.

\bibitem{VANDENDRIESSCHE200229}
P.~van~den Driessche, J.~Watmough, Reproduction numbers and sub-threshold
  endemic equilibria for compartmental models of disease transmission,
  Mathematical Biosciences 180~(1) (2002) 29 -- 48.

\bibitem{CA}
C.~Castillo-Chavez, Z.~Feng, W.~Huang, On the computation of $R_0$ and its role
  on Global Stability, Mathematical approaches for emerging and reemerging
  infectious diseases: an introduction, Vol. 229-250, Springer-Verlag, New
  York, 2002.

\bibitem{v11121153}
L.~J.~S. Allen, V.~A. Bokil, N.~J. Cunniffe, F.~M. Hamelin, F.~M. Hilker, M.~J.
  Jeger, Modelling vector transmission and epidemiology of co-infecting plant
  viruses, Viruses 11~(12) (2019).

\bibitem{OKUONGHAE2020110032}
D.~Okuonghae, A.~Omame, Analysis of a mathematical model for covid-19
  population dynamics in lagos, nigeria, Chaos, Solitons \& Fractals 139 (2020)
  110032.

\bibitem{IndiaCOVID-19tracker}
I.~C.-. tracker, India covid-19 tracker.

\bibitem{Chowell}
H.~Nishiura, G.~Chowell, The Effective Reproduction Number as a Prelude to
  Statistical Estimation of Time-Dependent Epidemic Trends, Springer
  Netherlands, Dordrecht, 2009, pp. 103--121.

\bibitem{Cori2021}
A.~Cori, Z.~Kamvar, J.~Stockwin, T.~Jombart, E.~Dahlqwist, R.~FitzJohn,
  R.~Thompson, {EpiEstim v2.2-3: A tool to estimate time varying instantaneous
  reproduction number during epidemics},
  \url{https://github.com/mrc-ide/EpiEstim} (2021).

\bibitem{Peterson}
E.~Petersen, M.~Koopmans, U.~Go, D.~H. Hamer, N.~Petrosillo, F.~Castelli,
  M.~Storgaard, S.~Al~Khalili, L.~Simonsen, Comparing sars-cov-2 with sars-cov
  and influenza pandemics, The Lancet Infectious Diseases 20~(9) (2020)
  e238--e244.

\bibitem{Yuan}
J.~Yuan, M.~Li, G.~Lv, Z.~K. Lu, Monitoring transmissibility and mortality of
  covid-19 in europe, International Journal of Infectious Diseases 95 (2020)
  311--315.

\bibitem{https://doi.org/10.1111/tbed.13868}
B.~B. Singh, M.~Lowerison, R.~T. Lewinson, I.~A. Vallerand, R.~Deardon,
  J.~P.~S. Gill, B.~Singh, H.~W. Barkema, Public health interventions slowed
  but did not halt the spread of covid-19 in india, Transboundary and Emerging
  Diseases 68~(4) (2021) 2171--2187.

\bibitem{SENAPATI2021110711}
A.~Senapati, S.~Rana, T.~Das, J.~Chattopadhyay, Impact of intervention on the
  spread of covid-19 in india: A model based study, Journal of Theoretical
  Biology 523 (2021) 110711.

\bibitem{Shil}
P.~Shil, N.~M. Atre, A.~A. Patil, B.~V. Tandale, P.~Abraham, District-wise
  estimation of basic reproduction number (r0) for covid-19 in india in the
  initial phase, Spatial Information Research (2021).

\bibitem{Nithya}
N.~C. Achaiah, S.~B. Subbarajasetty, R.~M. Shetty, R(0) and r(e) of covid-19:
  Can we predict when the pandemic outbreak will be contained?, Indian journal
  of critical care medicine : peer-reviewed, official publication of Indian
  Society of Critical Care Medicine 24~(11) (2020) 1125--1127.

\bibitem{doi:10.1098/rspb.1995.0138}
R.~M. May, M.~A. Nowak, Coinfection and the evolution of parasite virulence,
  Proceedings of the Royal Society of London. Series B: Biological Sciences
  261~(1361) (1995) 209--215.

\bibitem{Lawton}
G.~Lawton, Could co-infection cause coronavirus to evolve?, New scientist
  (1971) 247~(3299) (2020) 10--11.

\end{thebibliography}

\end{document}